\documentclass[12pt,a4paper]{article}
\pdfoutput=1
\usepackage{jheppub}
\usepackage{epstopdf}
\usepackage{graphicx}
\usepackage{epsfig}
\usepackage{dcolumn}  
\usepackage{bm}    
\usepackage{amssymb} 
\usepackage{amsfonts}    
\usepackage{slashed}  
\usepackage[mathscr]{euscript}
\usepackage{epsfig}
\usepackage[position=t,singlelinecheck=off]{subfig}
\usepackage{caption}
\allowdisplaybreaks[3]
\title{Conformal perturbation theory and higher spin entanglement entropy on the torus} 

\author[a]{Shouvik Datta,} 
\author[a,b]{Justin R.~David,} 
\author[c]{and S.~Prem Kumar } 
\affiliation[a]{Centre for High Energy Physics, Indian Institute of Science,\\ C. V. Raman Avenue, Bangalore 560012, India}
\affiliation[b]{
Max-Planck-Institut f\"{u}r Physik (Werner-Heisenberg-Institut), \\
F\"{o}hringer Ring 6, D-80805 Munich, Germany}
\affiliation[c]{Department of Physics, Swansea University,\\Singleton Park, Swansea SA2 8PP, UK.}

\emailAdd{shouvik, justin@cts.iisc.ernet.in}
\emailAdd{\\\qquad\qquad s.p.kumar@swansea.ac.uk}
\abstract{We study the free fermion theory in 1+1 dimensions deformed by chemical potentials for holomorphic, conserved currents at finite temperature and on a spatial circle. For a spin-three chemical potential $\mu$, the deformation is related at high temperatures to a higher spin black hole in hs$[0]$ theory on AdS$_3$ spacetime. 
We calculate the order $\mu^2$ corrections to the single interval R\'enyi and 
entanglement entropies on the torus using the bosonized formulation. A consistent result, satisfying all checks, emerges upon carefully accounting for both perturbative and winding mode contributions in the bosonized language.
The order $\mu^2$ corrections involve integrals that are finite but potentially sensitive to contact term singularities. We propose and apply a prescription for defining such integrals which matches the Hamiltonian picture and  passes several non-trivial checks for both thermal corrections and the R\'enyi entropies at this order. The thermal corrections are given by a weight six quasi-modular form, whilst the R\'enyi entropies are controlled by quasi-elliptic functions of the interval length with modular weight six. We also point out the well known connection between the  perturbative expansion of the partition function in powers of the spin-three chemical potential and the Gross-Taylor genus expansion of large-$N$ Yang-Mills theory on the torus. We note the absence of winding mode contributions in this connection, which suggests  qualitatively different entanglement entropies for the two systems.
}

\begin{document}
 \maketitle 

\def\be{\begin{equation}}
\def\ee{\end{equation}}
\def\bea{\begin{eqnarray}}
\def\eea{\end{eqnarray}}
\def\nn{\nonumber}
\def\pd{\partial}
\def\Re{R\'{e}nyi }
\def\l1{{\text{1-loop}}}
\def\uy{u_y}
\def\ur{u_R}
\def\o{\mathcal{O}}
\def\Cl{{{cl}}}
\def\bz{{\bar{z}}}
\def\by{{\bar{y}}}
\def\bX{\bar{X}}
\def\im{{\text{Im}}}
\def\re{{\text{Re}}}
\def\cn{{\text{cn}}}
\def\sn{{\text{sn}}}
\def\dn{{\text{dn}}}
\def\K{\mathbf{K}}
\def\n1{\Bigg|_{n=1}}
\def\fin{{\text{finite}}}
\def\R{{\mathscr{R}}}
\def\one{{(1)}}
\def\zero{{(0)}}
\def\n{{(n)}}
\def\tr{\text{Tr}}
\def\H{{\cal{H}}}
\def\G{{\cal{G}}}
\def\I{{\cal I}}
\def\TT{\tilde{\mathcal{T}}}
\def\O{{\mathcal{O}}}
\def\cN{{\mathcal{N}}}
\def\P{\Phi}
\def\csch{{\text{cosech}}}
\def\W{{\tilde{W}}}
\def\T{{\tilde{T}}}
\def\by{\bar{y}}
\newcommand*\xbar[1]{%
  \hbox{%
    \vbox{%
      \hrule height 0.5pt % The actual bar
      \kern0.5ex%         % Distance between bar and symbol
      \hbox{%
        \kern-0.1em%      % Shortening on the left side
        \ensuremath{#1}%
        \kern-0.1em%      % Shortening on the right side
      }%
    }%
  }%
} 
\section{Introduction}
The emergence of holographic descriptions \cite{maldacena, witten} of quantum field theories (QFTs), and the subsequent development of the remarkably elegant proposal for holographic entanglement entropy \cite{Ryu:2006bv, Ryu:2006ef}, has led to enormous progress and effort aimed at understanding this fundamental physical observable in QFTs.  A first principles field theoretic approach to calculating entanglement entropy (EE) has largely been limited to free field theories \cite{Casini:2009sr} because the calculation of the reduced density matrix in an interacting QFT presents a significant technical challenge. Conformal field theories (CFTs) in 1+1 dimensions provide tractable examples where this calculation can be performed explicitly by implementation of the replica trick 
\cite{Holzhey:1994we, Calabrese:2004eu, Calabrese:2005in, Calabrese:2009qy}, and where powerful universal results can be inferred using the tool of conformal invariance.
 The effect of interactions on entanglement entropy can then be studied within conformal perturbation theory \cite{Cardy:2010zs}.

The aim of this paper is to study and calculate the effect of perturbing free CFTs in two dimensions, at finite temperature and on a spatial circle, by deformations that decompose into a sum of holomorphic and anti-holomorphic pieces. Such `chiral' deformations of the free fermion CFT have received some attention within the context of the large-$N$ expansion of Yang-Mills theory in two dimensions \cite{douglas, dijkgraaf}. More recently, CFT deformations of this type have been related by holographic duality to higher spin black hole solutions \cite{Gutperle:2011kf, Kraus:2011ds,Gaberdiel:2012yb,Ammon:2012wc,Gaberdiel:2013jca} in Vasiliev's hs$[\lambda]$ theory \cite{Prokushkin:1998bq} in AdS$_3$ spacetime. This holographic connection is the main motivation behind our work, falling within the broader theme of higher spin holography relating CFTs with higher spin currents to higher spin theories of gravity in AdS spacetimes 
\cite{Giombi:2009wh, Klebanov:2002ja, Gaberdiel:2012uj, gg}. However, our results are purely field theoretic and therefore should be of wider interest as analytically calculable corrections to R\'enyi and entanglement entropies  in chiral deformations of CFTs in two dimensions. 

In recent work \cite{paper1, paper2}, we have shown that for any CFT with a ${\cal W}_{\infty}[\lambda]$ symmetry, when formulated on the infinite spatial line, the lowest nontrivial correction to the single interval R\'enyi entropy (and EE) in the presence of a spin-three deformation, is a universal function of temperature and interval size. In \cite{paper1}, explicit calculations within the free boson and free fermion CFTs unexpectedly agreed with each other, accompanied by equally  unexpected agreement  with a holographic proposal for EE due to \cite{deBoer:2013vca, Ammon:2013hba} applied to the higher spin black hole solution of \cite{Gutperle:2011kf}\footnote{See \cite{Datta:2014ypa} and \cite{Castro:2014mza} for some other recent developments on holographic entanglement entropy in higher spin gravity.}. Subsequently in \cite{paper2}, 
universality of the spin-three current correlator on the multi-sheeted cover of the cylinder was proven based only on ${\cal W}_\infty[\lambda]$ symmetries and then used to establish the universal result above for R\'enyi entropy. These results have recently been extended to include deformations by any higher spin current \cite{Long:2014oxa}
\footnote{ Spin-3 corrections to the R\'enyi entropy of two disjoint intervals have been  
studied in \cite{Chen:2013dxa}.}. 
It is natural to ask if the above calculations of EE and RE can be extended to include finite size corrections for the CFT on a spatial circle, yielding concrete predictions for finite-size (non-universal) corrections to EE/RE in higher spin gravity.

Our primary focus in this paper is the free fermion CFT, deformed by the holomorphic (plus anti-holomorphic) spin-three current in this theory. With a view towards extracting all thermal and finite size effects at any given order in conformal perturbation theory, we study the field theory on a circle of circumference $L$, at a temperature $\beta^{-1}$, corresponding to a torus with complex structure parameter $\tau\,=\,i\beta/L$.
The reason for picking this theory is that the branch point twist fields, necessary for implementing the replica trick \cite{Calabrese:2004eu}, have explicit representations in this case. Importantly, this explicit representation is available upon bosonization of the free fermion theory. The free fermion CFT has an infinite number of integer spin conserved currents with spin $s\geq 1$, furnishing a representation of the ${\cal W}_{1+\infty}$ algebra \cite{Pope:1989ew,Bergshoeff:1990yd, Pope:1991ig, Bakas:1990ry}. The dual holographic interpretation in terms of hs$[\lambda]$ theory with $\lambda=0$ requires a truncation of ${\cal W}_{1+\infty}$ to ${\cal W}_{\infty}[0]$ by removing the spin-1 current. We will not address this issue in this work.

As remarked above our approach is to treat the deformation by such currents within conformal perturbation theory in the Lagrangian formulation: 
\be
{\rm I}_{\rm CFT}\to {\rm I}_{\rm CFT}\,+\,\mu\int d^2z\,\left(W(z)\,+\overline W(\bar z)\right)\,.\label{deformation}
\ee
Physically, the deformation can be viewed as a chemical potential for the corresponding spin-three charge in the Hamiltonian language, and the deforming parameter/chemical potential can be thought of as a constant background higher spin gauge field. Whilst one may be tempted to conclude that this picture is completely analogous to the standard situation with a chemical potential for a global U(1) symmetry, there are subtleties associated to chemical potentials for higher spin charges. As pointed out in \cite{dijkgraaf}, and more recently clarified in \cite{deBoer:2014fra} with the holographic picture in mind, passing between Hamiltonian and Lagrangian pictures in the presence of higher spin chiral deformations (and separate anti-chiral ones),  typically involves  redefinition of the deformation parameters (and potential mixing of the two sectors due to contact terms). 

This brings us to one of the important points of our paper. Based on dimension counting, the higher spin deformations (with $s > 2$) are irrelevant operators, and therefore one would expect to see uncontrolled UV divergences in perturbation theory. However, this is not the case; the integrals of holomorphic correlators over the torus, which determine perturbative corrections to the partition function (and RE/EE) for \eqref{deformation}, are finite except for ambiguities involving contact term singularities that require a choice of integration prescription. Although surprising at first sight, the absence of conventional divergences signalling non-renormalizability due to irrelevant perturbations is in line with field theoretic intuition. In particular, one expects that chemical potentials for conserved charges should correspond to altering boundary conditions in some way and should not change the UV divergence structure  of the theory\footnote{On the other hand, the holographic duals to such states which are the spin-three black holes of \cite{Gutperle:2011kf} appear to display an RG flow from an IR AdS$_3$ geometry with 
${\cal W}_\infty[\lambda]$ symmetry 
\cite{Henneaux:2010xg, campoleoni} to a UV AdS$_3$ with different (non-principal) asymptotic ${\cal W}$-algebra symmetry \cite{Afshar:2012hc}. Several works have pointed out both the intriguing consequences of this picture \cite{D-F-K, Ferlaino:2013vga}, and possibly the complete absence of an RG flow interpretation\cite{Compere:2013nba}.}.
We employ a prescription for treating the contact term singularities which is 
essentially an $i\epsilon$-prescription on the torus. The results using this approach exhibit precise, non-trivial agreement with thermal corrections computed using the  canonical or Hamiltonian approach (see e.g. \cite{Kraus:2011ds}). Our prescription differs from the ``covariant'' prescription proposed in \cite{dijkgraaf, douglas} by finite local counterterms involving higher spin currents. The latter picture involves cutting discs around contact points and applying Stokes' theorem. Perhaps the most clear difference between the two prescriptions for computing integrals over the torus is that our prescription yields results holomorphic in $\tau$, whilst the prescription of \cite{dijkgraaf} necessarily introduces non-holomorphic dependence on $\tau$. 

The universal results of  \cite{paper1, paper2} on the finite temperature cylinder, which displayed highly non-trivial agreement with the holographic EE proposal of \cite{deBoer:2013vca, Ammon:2013hba} were based on precisely the same integration prescription that we adopt in the present work on the torus.
It appears that different choices of integration prescriptions accompanied by their specific counterterms are related to different possible thermodynamical interpretations of the higher spin black holes dual to deformed ${\cal W}$-algebra CFTs \cite{deBoer:2014fra, Perez:2012cf, deBoer:2013gz}.

In conformal perturbation theory, the corrections to the thermal partition function are determined by multi-point correlators of $W$-currents on the torus, while for the R\'enyi entropies one needs the correlators in the presence of branch-point twist fields on the torus. In particular, the order $\mu^2$ correction to the partition function for the deformed theory \eqref{deformation} is controlled by the two-point correlator of spin-three currents in the presence of twist fields. The bulk of our work is aimed at correctly pinning down this correlation function. In order to work with an explicit representation for the twist fields we use the bosonized formulation of free fermions. The main source of complication in the computation is that one must separately account for the contributions from both the quantum fluctuations and classical, nonperturbative sectors corresponding to winding modes of the compact bosons on the torus. 

The presence of winding modes ensures that crucial contributions are accounted for and delicate cancellations occur at certain stages. Perhaps most significant amongst these is a complete cancellation of non-hololomorphic dependence on the torus modular parameter $\tau$ which appears (via powers of $1/{\rm Im}(\tau)$) in the intermediate stages of the calculation. The reason for its appearance can be traced to the background uniform charge on the torus required to makes sense of the free boson propagator on the torus. Remarkably, all non-holomorphic dependence originating from the quantum propagators precisely cancels agains identical contributions from the winding mode sectors. Furthermore, the winding mode contributions are crucial for ensuring that the final result for entanglement entropy correctly reproduces the thermal entropy when the interval spans the entire system. In the absence of the winding modes the EE would vanish in this limit.

Our final results for EE and thermal entropy corrections satisfy several nontrivial consistency checks that we describe in detail.  Together they provide evidence in favour of the integration prescription adopted and the relevant correlators on the torus. The R\'enyi entropies are controlled by a quasi-elliptic function of the interval length with quasi-modular weight six, which we determine in closed form.

The results for the order $\mu^2$ corrections to the thermal partition function and the R\'enyi entropies on the torus can be used to extract predictions for finite size effects in the bulk higher spin gravity dual. We find that thermal corrections at this order, both  for  free bosons and free fermions, are determined by weight six quasi modular forms.  

As in the case of the standard BTZ solution in AdS$_3$, the classical thermodynamics of the Gutperle-Kraus black hole yields the high temperature limit $(\beta^{-1}L) \to \infty$ of the CFT partition function. The holographic, classical gravity description which is applicable in the limit of large central charge $c\gg 1$, captures the high temperature Cardy growth of states. Finite size corrections to this emerge from the quantum corrections to the classical gravity saddle point.  In the presence of sources (chemical potential deformations) for higher spin currents, these will be accompanied  by further corrections induced by the deformations, which are non-trivial and yet to be understood.

In the absence of higher spin hair, the  leading 
finite size corrections to entanglement/R\'{e}nyi entropy can be extracted from one-loop determinants evaluated on bulk handlebody solutions which serve as bulk duals for the replicated CFT \cite{Faulkner:2013yia,Barrella:2013wja}. These corrections evaluated from the bulk have been shown to match those of the free fermion and boson CFTs \cite{Datta:2013hba}. It is indeed an interesting future problem to investigate how  handlebody geometries can be realized in the Chern-Simons formulation of higher-spin gravity and how finite size corrections can be evaluated therein.

The paper is organized as follows: In section \ref{two} we calculate free fermion current correlators (for spin-1, spin-2 and spin-3 currents) on the torus (without twist fields) with the goal of using them to compute thermal corrections in conformal perturbation theory. We further cast the correlators in a form that makes direct contact with the bosonized formulation to be employed for RE calculations. In section \ref{three} we explain our prescription for integrating holomorphic current-correlators on the torus. Subsequently, in section \ref{four} the current-correlators are integrated over the torus using the integration prescription. The results are found to be in perfect agreement with exact expectations from other methods for studying deformations of the free fermion theory by the U(1)-current, the stress tensor and the spin three current. We also review and match with the known results of the large-$N$ expansion of Yang-Mills theory in two dimensions.
In section \ref{five} we employ our methods to examine corrections to the thermal partition function of the free boson theory with spin-three deformation, and we demonstrate exact agreement with the results of \cite{Kraus:2011ds}. Sections \ref{six} and \ref{seven} are devoted to the evaluation of the R\'enyi and entanglement entropies for spin-one, spin-two and spin-three deformations along with a number of nontrivial consistency checks along the way. Appendix \ref{walgebra} lists the ${\cal W}_\infty$-algebra OPEs for reference. In appendix \ref{app:elliptic} we collect a large number of useful identities for elliptic functions and modular forms. We also relegate the details of the calculation of correlators in the presence of twist fields from first principles to appendix \ref{appendix:JJVV}. All integrals necessary for evaluating the R\'enyi entropies are derived in detail in appendix \ref{integrals}.

\section{Free fermion current correlators}
\label{two}

The free fermion CFT provides a realization of the ${\cal W}_{1+\infty}$ algebra \cite{Bergshoeff:1990yd, Pope:1991ig} generated by currents with spins $s\geq 1$. Specifically, the theory of $M$ complex free fermions $\{\psi^a\}$ $(a=1,2,\ldots M)$ in the $U(M)$-singlet sector, has an infinite set of integer spin conserved currents. The holomorphic currents $J$, $T$ (the stress tensor) and $W$ with spin one, two and three respectively, are
\bea
&&J\,=\,\psi_a^*\psi^a\,, \qquad T\,=\,\tfrac{1}{2}\left(\partial\psi_a^*\psi^a-\psi_a^*\partial\psi^a\right)\,,\\\nonumber\\\nonumber
&&W\,=\,i\frac{\sqrt{5}}{{12\pi}}\left(\partial^2\psi_a^*\psi^a\,-\,4\partial\psi_a^*\partial\psi^a\,+\,\psi_a^*\partial^2\psi^a\right)
\eea
where the spin-three current has been normalised to match the conventions of \cite{Kraus:2011ds, paper1} for the leading singularity in the $WW$ OPE \eqref{opew3}. The elementary fermions have the OPE 
\be
\psi_a^*(z_1)\psi^b(z_2)\,\sim\,\frac{\delta_a^b}{z_1-z_2}\,,
\ee
and other operator products being trivial.
\subsection{U(1) current correlator}
Current correlators on the torus ${\mathbb C/\Gamma}$ with $\Gamma\,\simeq\, 2\omega_1{\mathbb Z}\,\oplus \,2\omega_2 {\mathbb Z}$, can be readily deduced by making use of the free fermion correlator on the torus \cite{DiFrancesco:1997nk} 
\be
\langle\psi^*_a(z_1)\psi^b(z_2)\rangle\,=\,\frac{\vartheta_\nu\left(u_{12}\right)}{\vartheta_\nu(0)}\frac{\vartheta_1'(0)}{\vartheta_1\left(u_{12}\right)}\,,\qquad u_{12}\,=\,\frac{\pi}{2\omega_1}\left(z_1-z_2\right)\,.\label{fermionprop}
\ee 
Here $\nu =2,3$ for periodic (R) and anti-periodic (NS) boundary conditions, respectively, on the fermions around the spatial circle of the torus. The periods of the torus, $2\omega_1=L$ and $2\omega_2 = i\beta$, are determined by the spatial extent and inverse temperature respectively with $\tau\,=\,\omega_2/\omega_1=i\beta/L$.  In order to avoid cluttering the notation, in all subsequent formulae, unless explicitly specified, theta-functions will be evaluated at vanishing argument.

Using the free fermion correlators on the torus we obtain
\be
\langle J(z_1)J(z_2)\rangle_{{\mathbb T}^2}\,=\,-\sum_{a,b}\delta^a_b\,\left(\langle \psi^*_a(z_1) \psi^b(z_2)\rangle\right)^2\,=\,-\,M\left[\frac{\vartheta_\nu\left(u_{12}\right)}{\vartheta_1\left(u_{12}\right)}\frac{\vartheta_1'}{\vartheta_\nu}\right]^2\,,
\ee
where $\nu=2,3$ depending on the choice of spin structure.  We will now rewrite this formula in a way that will allow us to readily integrate it over the torus. Unlike the fermion Green's function the current correlator is a doubly periodic function on the torus with a second order pole at $z_{12}=0$. Non-trivial identities involving elliptic functions (e.g. \eqref{wpandtheta}) allow us to rewrite the above expression  as 
\bea \label{jj1}
&&\langle J(z_1)J(z_2)\rangle_{{\mathbb T}^2}\,=\,-M\left({\cal H}(z_{12})\,+\,\left(-\frac{i\pi}{L}\right)^2\frac{\vartheta_\nu''}{\vartheta_\nu}\right)\,,\label{jj}\\\nonumber\\\nonumber
&&{\cal H}(z)\,:=\,-\,{\wp}(z;\omega_1,\omega_2)\,-\,\frac{\pi^2}{3 L^2}\,E_2(\tau)\,.
\eea
Here $E_2(\tau)$ is the second Eisenstein series, a holomorphic, {\em almost} modular form with modular weight $2$. Under a general $SL(2, {\mathbb Z})$ transformation on $\tau$, it transforms as
\bea
&& \tau\to \frac{m\,\tau\,+\,n}{p\,\tau\,+\,q}\,,\qquad\qquad m\,q\,-\,p\,n\,=\,1\,,\qquad\qquad p,q,m,n \in {\mathbb Z}\,,\label{modulartau}\\\nonumber
&&E_2(\tau)\to (p\,\tau \,+\,q)^2 \,E_2(\tau)\,+\,\frac{6\,p}{i\pi }\,(p\,\tau\,+\,q)\,.
\eea
We will see subsequently that this representation of the current correlator is natural in the bosonized formulation of the free fermion theory. The first term in \eqref{jj} arises from taking two derivatives of the free (compact) boson propagator on the torus whilst the second term is the outcome of the sum over winding number sectors of the compact boson.
\subsection{Spin-2 and spin-3 current correlators}
The spin-two and spin-three current correlators can also be readily evaluated following Wick contractions and the resulting elliptic functions can be re-expressed in a form that facilitates direct connection to the bosonized approach we will adopt later. The  two-point function for the stress tensor on the torus is
\bea
&&\langle T(z_1)T(z_2)\rangle_{{\mathbb T}^2}\,=\label{tt}\\\nonumber
&&\qquad\qquad\frac{M}{4}\left[2{\cal H}(z_{12})^2\,-\,4{\cal H}(z_{12})\,\frac{\pi^2}{L^2}\frac{\vartheta_\nu''}{\vartheta_\nu}\,+\,\frac{\pi^4}{L^4}\left(\frac{\vartheta_\nu^{(4)}}{\vartheta_\nu}\,-\,\left(\frac{\vartheta_\nu''}{\vartheta_\nu}\right)^2\right)\right]\,. \label{tt1}
\eea
Similarly, the spin-three current correlator is
\bea
&&\langle W(z_1)W(z_2)\rangle_{{\mathbb T}^2}\,=\label{ww}\\\nonumber
&&\qquad\frac{5M}{36\pi^2}
\left[6\,{\cal H}(z_{12})^3\,-\,18\,{\cal H}(z_{12})^2\,\frac{\pi^2}{L^2}\frac{\vartheta_\nu''}{\vartheta_\nu}\,+\,9\,{\cal H}(z_{12})\,\frac{\pi^4}{L^4}\left(\frac{\vartheta_\nu^{(4)}}{\vartheta_\nu}\,+\,\frac{2E_2}{3}\frac{\vartheta_\nu^{''}}{\vartheta_\nu}\,+\,\frac{E_2^2}{9}\right)\right.\\\nonumber
&&\left.\qquad\qquad\qquad-\,\frac{\pi^6}{L^6}\left(\frac{\vartheta_\nu^{(6)}}{\vartheta_\nu}\,+\,2E_2\frac{\vartheta_\nu^{(4)}}{\vartheta_\nu}\,+\,E_2^2\frac{\vartheta_\nu^{''}}{\vartheta_\nu}\right)\right]\,. \label{ww1}
\eea
The correlation functions have the correct leading singularity structure dictated by conformal invariance and the ${\cal W}$-algebra OPEs. 
\subsection{High temperature limits: ${\cal W}_{1+\infty}$ vs. ${\cal W}_{\infty}$}
\label{subsection:highT}
It is a very instructive excercise to verify how the correct high temperature limit $\tau \to i\,0^+$ of the correlators \eqref{jj}, \eqref{tt} and \eqref{ww} is obtained
\footnote{Modular properties of 
correlators involving 
spin-3 currents for  CFTs admitting ${\cal W}$ symmetry were recently 
studied in \cite{Iles:2014gra}.}.
The modular S-transformation, $\tau\to -1/\tau$, relates the limit to the low temperature behaviour as $\tau\to i\infty$. Under the S-transformation the Weierstrass $\wp$-function has weight two:
\be
\wp\left(\frac{z}{\tau};\,\omega_1,\,-\frac{1}{\tau}\omega_1\right)\,=\,\tau^2\,\wp\left(z;\,\omega_1,\,{\tau}\omega_1\right)\,.
\ee
Using the expansion \eqref{plowt} of the $\wp$-function and similar ones for the theta-functions, and the modular property \eqref{modulartau} of $E_2$, discarding exponentially suppressed ${\cal O}(e^{-2\pi L/\beta})$ high temperature corrections, we find,
\bea
&&\langle J(z_1)J(z_2)\rangle\big|_{\tau\to i0^+}\,=\,
\frac{M\,\pi^2}{\beta^2\,\sinh^2\left(\frac{\pi}{\beta} (z_1-z_2)\right)}\label{larget}\\\nonumber\\\nonumber
&&\langle T(z_1)T(z_2)\rangle\big|_{\tau\to i0^+}\,=\,
\frac{M\,\pi^4}{2\beta^4\,\sinh^4\left(\frac{\pi}{\beta} (z_1-z_2)\right)}\\\nonumber\\\nonumber
&&\langle W(z_1)W(z_2)\rangle\big|_{\tau\to i0^+}\,=\,-\frac{5M}{36\pi^2}\left[\frac{6\,\pi^6}{\beta^6\,\sinh^6\left(\tfrac{\pi}{\beta}(z_1 -z_2)\right)}\,+\,\frac{\pi^6}{\beta^6\,\sinh^2\left(\tfrac{\pi}{\beta}(z_1 -z_2)\right)}\right]\,.
\eea
Expectedly, both the stress tensor and U(1)-current correlators, in this limit, are fixed completely by conformal invariance 
and the map from the complex plane to the cylinder (see e.g. \cite{paper1}). However this does not appear to be the case for the spin-three current. In fact its high temperature limit indicates a certain overlap with the U(1)-current. This may be traced to the fact that the free fermion CFT has the ${\cal W}_{1+\infty}$ algebra with a non-trivial OPE between the $W$-current and the U(1)-current. Under the conformal map from the complex plane to the cylinder, it can be shown that the spin-three current does not transform as a tensor unless $J=0$. In particular, a conformal transformation $z(w)$ from the complex-$w$ plane to some Riemann surface (e.g. the cylinder) acts on the spin-three current as 
\be
W(z)\,=\,w'(z)^3 \,W(w)\,+\,\frac{\gamma}{2\sqrt{6}}\,w'(z)\, J(w)\,\{w,z\}\,,
\,\label{Wtransform}
\ee
where $\{w,z\}$ is the Schwarzian derivative, and the normalisation $\gamma$ is chosen to match the $WW$ OPE:
\be
W(z)W(0)\sim \frac{\gamma^2}{z^6}\,,\qquad\gamma\,=\,i\sqrt{\frac{5}{6\pi^2}}\,.
\ee 
The transformation rule \eqref{Wtransform} can be derived by working in the bosonized language (in terms of a chiral boson $\varphi$) on the plane so that $W\,=\,-i\frac{\gamma}{\sqrt 6} :(\partial\varphi)^3:$ and $J\,=\,i\partial\varphi$. Under the coordinate transformation $w(z)$, using $\partial\varphi(z)\,=\,w'(z)\partial_w\varphi(w)$, and defining the current $W$ on both the $z$-plane and the $w$-plane using point-splitting regularization of operator products, one directly obtains the result \eqref{Wtransform}. 
Taking account  of this mixing with the U(1)-current, for the map to the cylinder $z(w) = \beta\ln w/(2\pi)$, we obtain eq.\eqref{larget} for the two point function of the spin-three current.

At high temperatures, it therefore appears that the spin-three current correlator, appropriate for the ${\cal W}_\infty$ theory is   
\bea
\langle W(z_1) W(z_2)\rangle_{{\cal W}_\infty}\to&&\langle W(z_1) W(z_2)\rangle-
\frac{\gamma^2}{6}\left[\frac{6}{M}\langle T\rangle\right]^2
\langle J(z_1) J(z_2)\rangle\,\Big|_{\tau\to i0^+}\\\nonumber\\\nonumber
&& =\,M\gamma^2\frac{\pi^6}{\beta^6\sinh^6\left(\tfrac{\pi}{\beta}(z_1-z_2)\right)}\,.
\eea
The one-point function for the stress tensor on the cylinder is determined by the Schwarzian
\be
\langle T\rangle\Big|_{\tau\to i0^+}\,=\,\frac{M}{12}\{w,z\}\,=\,-M\frac{\pi^2}{6\beta^2}\,,
\ee
which can also be obtained by taking the coincidence limit of the U(1) current correlator \eqref{jj} at high temperature. This manner of extracting ${\cal W}_\infty$ answers from the ${\cal W}_{1+\infty}$ theory is clearly {\it ad hoc}. Indeed, while there is a natural and automatic generalisation of the above subtraction to the torus, we do not have independent evidence to suggest that the resulting expression is the appropriate one for the ${\cal W}_\infty[1]$ theory on the torus. A systematic method along the lines proposed by \cite{Gaberdiel:2013jpa} and attempted in 
\cite{paper1} would be to extract correlation functions from the free fermion theory with the U(1)-current projected out by a cosetting procedure. We leave this for future work and focus attention primarily on the ${\cal W}_{1+\infty}$ theory.

\section{Integrating over the torus}
\label{three}

\begin{figure}
%[!htbp]
\begin{center}
\includegraphics[width=5.2in]{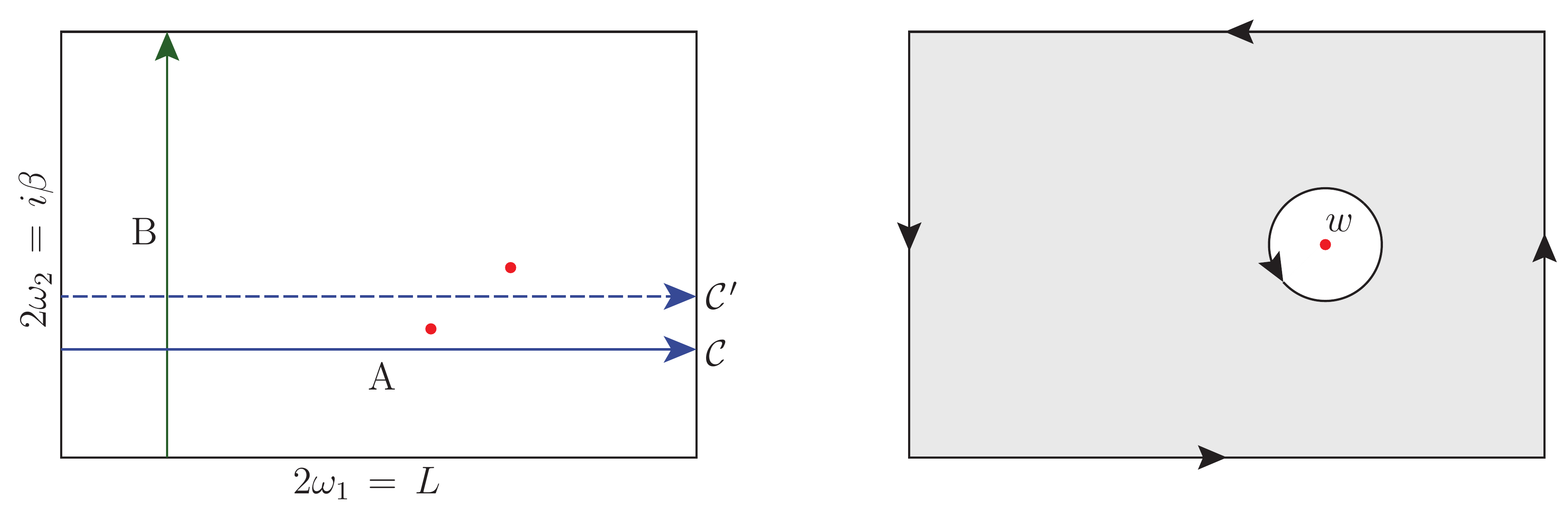}
\end{center}
\caption{\small {\bf Left}: Using prescription-1, integration along the A-cycle is performed first, avoiding any poles along the integration path. The integral along ${\cal C}$ is finite and independent of any smooth deformations of the path. The integral along ${\cal C}$ and ${\cal C}'$, differ by any non-zero residue around a pole (red dot).
{\bf Right}: Using prescription-2, we apply Stokes' theorem to the integral over the torus after excising disks $D_2$ centred around singular contact points (in red).}
\label{torusfig}
\end{figure}

The first non-trivial correction to the thermal partition function of the CFT perturbed by a chemical potential for one of the currents above, is obtained by integrating the correlators over the torus (eqs. \eqref{jj1}, \eqref{tt1} and \eqref{ww1}). Since the correlators contain singularities, we must have a definite prescription for dealing with these. For the spin-three current in particular,
\bea
&&\ln {\cal Z}\,=\ln {\cal Z}_{\rm CFT}\,-\,\beta F^{(2)}\,+\ldots\\\nonumber
&&-\,\beta F^{(2)}\,=\,\tfrac{1}{2}\mu^2\int_{{\mathbb T}^2} d^2 z_1\int_{{\mathbb T}^2} d^2z_2\,\langle W(z_1) W(z_2)\rangle\,\, +\,{\rm h.c.}
\eea
where $F^{(2)}$ is the order $\mu^2$ correction. The integrals are actually tractable and {\em finite} because the integrand is holomorphic. However, the calculation is not without subtleties.
It is well appreciated that such integrals are potentially ambiguous due to singularities from  contact terms \cite{Gaberdiel:2013jca, douglas, dijkgraaf, paper1, paper2} when $z_1\to z_2$\footnote{Note that we have tacitly omitted correlators between the holomorphic and anti-holomorphic currents $\langle W(z_1) \overline W(\bar z_2)\rangle$. These are zero up to contact singularities \cite{deBoer:2014fra} which can be dealt with using the prescriptions that we discuss. The net result is that contact term singularities between holomorphic and anti-holomorphic currents do not contribute since the prescriptions always excise the contact points from the integration domain.}. We will evaluate them using two different prescriptions:
\begin{itemize}
\item{{\bf Prescription-1}: Perform the integrals along Re$(z)$ first, followed by Im$(z)$, or the A- and B-cycles  of the square torus respectively, and in that order (see fig.\eqref{torusfig}). Singularities from contact points along the integration path are avoided using what is effectively an $i\epsilon$-prescription. This prescription matches the correction to the canonical thermal partition function.}

\item{{\bf Prescription-2}: Apply Stokes' theorem to the torus with infinitesimal disks centred around the contact points excised out. This method was also used in \cite{dijkgraaf, douglas}, and whilst yielding a result differing from  the correction to the canonical partition function, the result is closely related to that of prescription-1, and exhibits nice modular properties.}
\end{itemize}
In order to integrate an elliptic (holomorphic) function along a cycle of the torus, one only needs to know the coefficients of the simple and double poles of the elliptic function. This immediately follows from a theorem \eqref{zetaexp} (see \cite{ww} for details) which states that every elliptic (doubly periodic) function can be expanded as a sum involving the Weierstrass $\zeta$-functions and its derivatives, where the coefficients in the expansion are determined from the Laurent series around each of the singularities.

The Weierstrass $\zeta$-function is a quasi-periodic function on the torus satisfying  
$\zeta'(z)\,=\,-\,\wp(z)$ (see appendix\eqref{app:elliptic}). Any elliptic  function with poles of order $r_j$ at $z=a_j$, ($j=1,2\ldots$) can be written as 
\bea
&&f(z)\,=\\\nonumber
&&C\,+\,\sum_{j}\left(c_{j,1}\,\zeta(z-a_j)\,-\,c_{j,2}\,\zeta'(z-a_j)\,+\ldots(-1)^{r_j-1}\frac{c_{j,r_j}\,\zeta^{(r_j-1)}(z-a_j)}{(r_j-1)!}\right)\,.
\eea
Crucially, ellipticity requires that the sum of simple pole residues vanishes
\be
\sum_j c_{j,1}\,=\,0\,.
\ee
The main advantage of this representation is that the function may be immediately integrated, and since all $\zeta^{(s)}(z)$ with $s>1$ integrate to give doubly periodic functions, only the simple and double poles of $f(z)$ influence its integral over 
a cycle of the torus.
Therefore, if the integration contour is chosen to avoid the poles, keeping all simple poles to one side of the contour, then integration along the A(B)-cycle yields,
\bea
\oint_{A,B}\,f(z)
\,=\,2\omega_{1,2}\,C \,-\,2\,\zeta(\omega_{1,2})\sum_k c_{k,1}\,a_k\,-\,2\zeta(\omega_{1,2})\sum_{k} c_{k,2}\,,
\eea
where we have used \eqref{zetasigmadef}. When simple poles lie both sides of the contour, we must keep track of any additional contributions from the residues at these poles. Simple poles do not arise for thermal corrections at the order we work in, whilst for the R\'enyi entropy they appear due to twist operator insertions on the real (spatial) axis.

\subsection{Integration on ${\mathbb T}^2$: Prescription-1}
We now perform the integral over the torus by first integrating along the A-cycle (or spatial axis) and subsequently along the B-cycle. For simplicity, we take all simple poles to lie on the real axis, to avoid issues arising from the contours threading between poles, and we find
\bea
I_{AB}\,=\,\int_0^{|2\omega_2|}dt&&\int_0^{2\omega_1}d\sigma\,f(\sigma+i t)\,=\,\\\nonumber
&&4|\omega_2|\left(\omega_{1}\,C\,-\,\zeta(\omega_1)\sum_{k}c_{k,1}\,a_k\,-\,\zeta(\omega_1)\sum_{k} c_{k,2}\right)\,.
\eea
One may check that performing the integrals in the reverse order yields a different result.
Before attempting to understand the physical interpretation of this prescription, let us recast the expressions in terms of $L, \beta$ and $\tau$. For this we use the identity,
\be
\zeta(\omega_1)\,=\,\frac{\pi^2}{6L}E_2(\tau)\,,
\ee
to obtain
\bea
&&{I}_{AB}\,=\,\beta\,L\,C\,-\,\frac{\pi^2}{3}{\rm Im}(\tau)\,E_2(\tau)\,\left(\sum_{k=1}^nc_{k,1}\,a_k\,+\,\sum_{k=1}^n c_{k,2}\right)\,.\label{IAB1}
\eea
\paragraph{Interpretation:} The prescription for the calculation we have adopted above can be understood as follows. The current-current correlator $\langle W(z_1) W(z_2)\rangle$ only has a singularity when $z_1=z_2$. This cannot be a simple pole on the torus, since an elliptic function cannot have only one simple pole. Our prescription requires performing the spatial integrals first, taking care to avoid any pole(s) along the integration path ($z_{1,2}\,=\,\sigma_{1,2}\,+\,it_{1,2}$): 
\bea
&&\int_{{\mathbb T}^2}d^2z_2\int_{{\mathbb T}^2} d^2z_1 \langle W(z_1) W(z_2)\rangle=\,\\\nonumber\\\nonumber
&&\lim_{\epsilon\to 0}\int_0^{|2\omega_2|}dt_2\int_0^{2\omega_1}d\sigma_2
\left(\int_0^{t_2-\epsilon}dt_1\,+
\int_{t_2+\epsilon}^{|2\omega_2|}dt_1\right)\int_0^{2\omega_1}d\sigma_1
\langle W(z_1)W(z_2)\rangle\,.
\eea
We can now use the definition of the conserved charge
\be
Q(t)\,\equiv\,\int_0^{\beta} d\sigma\,W(z)\,
\ee
which is actually independent of time, to rewrite the integral as
\bea
\int_{{\mathbb T}^2}d^2z_2\int_{{\mathbb T}^2} d^2z_1 &&\langle W(z_1) W(z_2)\rangle\,=\label{firstmethod}\\\nonumber\\\nonumber
&&\lim_{\epsilon\to 0}\int_0^\beta dt_2\left(\int_{t_2+\epsilon}^\beta dt_1\,+\,\int_0^{t_2-\epsilon} dt_1\right)\langle Q(t_1)\,Q(t_2)\rangle
\,=\,\beta^2\langle Q^2\rangle\,.
\eea
This is precisely the ${\cal O}(\mu^2)$ contribution to the {\em canonical} partition function. The interpretation in terms of conserved charge expectation values is also supported by the fact that upon doing the $\sigma_1$ integral first we get a result independent of the Euclidean times, $t_1$ and $t_2$. The same prescription has already been used to calculate the integrals relevant for thermal and R\'enyi entropies on the thermal cylinder ($\tau\to i 0^+$) in \cite{paper1, paper2}.
%Therefore
%\be
%\int_{{\mathbb T}_2}{\cal I}_{AB}\,=\,\beta^2\,\langle \oint_A W \oint_A W\rangle\,=%\,\beta^2\,\langle Q\,Q\rangle\,.
%\ee
% The application of the same method  with the roles of the A- and B-cycles exchanged %yields
%\be
%\int_{{\mathbb T}_2}{\cal I}_{BA}\,=\,L^2\langle\oint_B W\oint_B W\rangle\,,
%\ee
%whose physical interpretation is less obvious.

\subsection{Integration on ${\mathbb T}^2$: Prescription-2}
We now discuss the second method based on the prescription given in \cite{dijkgraaf, douglas}. For this method we first note that the deformation by a holomorphic operator can be viewed as the integral of an exact 2-form on the torus:
\be
\frac{1}{2i}\int_{{\mathbb T}^2-{\sum_k D_2(a_k)}} dz\wedge d\bar z\, \,{W(z)\ldots}\,
%=\,-\frac{1}{2i}\int d\left(\bar z\,W(z)\,dz\right)
\,=\,\int_{{\mathbb T}^2-{\sum_k D_2(a_k)}} d\left(\frac{z-\bar z}{2i}\,W(z)\,dz\right)\ldots\,,
\ee
where $D_2(a_k)$ is an infinitesimal disk centred at $z=a_k$ and $\{a_1,a_2,\ldots a_n\}$ is the set of poles in the operator product of $W(z)$ with any other insertions (denoted as $``\ldots"$).
Applying Stokes' theorem, we reduce the integral to a line-integral of the one-form 
\be
\omega\,\equiv\,\frac{z-\bar z}{2i}\,W(z)\,dz\,,\label{omegachoice}
\ee
along the boundary of the torus (with the excised discs). Here $\omega$ is defined only up to an additive holomorphic one-form $g(z)dz$. This  choice was originally made by Douglas in \cite{douglas} and it leads to a nice physical interpretation, related to the  prescription-1 discussed above. When applied to the two-point correlator of the $W$-currents, noting that the only singularity of the two-point function of the two $W$-currents is when $z_1-z_2=0$, we find that the boundary integral (fig.\eqref{torusfig}) simplifies to
\bea
\int_{{\mathbb T}^2-{D_2(z_2)}}&&d^2z_1\,\langle W(z_1)W(z_2)\rangle\,=\,\\\nonumber
\\\nonumber
&&\beta\int_0^{2\omega_1=L}d\sigma_1\,\langle W(\sigma_1) W(z_2)\rangle +\oint_{|z_1-z_2|=\epsilon}dz_1\,\frac{z_1-\bar z_1}{2i}\,\langle W(z_1) W(z_2)\rangle\,.
\eea
The first term on the right hand side is the correlator integrated along the A-cycle whilst the second is a contour integral around the infinitesimal circle centred at $z_1=z_2$. The result of the first term (since $\sigma_1\neq z_2$) is identical to that  obtained using prescription-1. The contour integral in the second term involves a non-holomorphic function so one should be careful in evaluating it. It can receive non-vanishing contributions from both simple and double poles in the current-current correlator. However since an elliptic function cannot have a solitary simple pole, the  only contribution from the excised disc is from the double pole at $z_1=z_2$. In particular,
\be
\langle W(z_1) W(z_2)\rangle\,\sim\,\sum_{s=2}^6\frac{c_s}{(z_1-z_2)^s}\,,
\ee
where the coefficients $c_s$ are the expectation values of operators contained in the 
OPE of two $W$-currents \eqref{opew3}. Therefore, we obtain
\be
\int d^2z_2\int d^2z_1\langle W(z_1)W(z_2)\rangle\,=\,\beta^2\langle Q\,Q\rangle\,+\pi\beta\,L\,c_2\,.\label{secondmethod}
\ee
This matches the general prescription quoted in \cite{dijkgraaf}. 

Therefore, we have shown explicitly via eqs. \eqref{firstmethod} and \eqref{secondmethod} that the  results for the lowest order correction to the free energy depend on the integration prescription as:
\be
-\beta F^{(2)}_{\rm presc-2}\,=\, -\beta F^{(2)}_{\rm presc-1}\,+\,\frac{\mu^2}{2} (\pi \beta\,L c_2 )\,+\,{\rm h.c.}
\ee
But $c_2$ is the coefficient of the second order pole in the OPE of two spin-three currents. Explicitly, 
\be
W(z_1) W(z_2)\,\sim\,-\frac{5c}{6\pi^2\, z_{12}^6}\,+\ldots -\,
\frac{1}{\pi^2\,z_{12}^2}\left(5 U + \tfrac{3}{4}\,T''\right)\,+\ldots
\ee
where $U$ is the spin-four current of the ${\cal W}$-algebra, and we have used the OPE  \eqref{opew3} along with the normalisation adopted in the body of this paper. Hence the two prescriptions we have discussed differ by the one-point function of the spin-four (and dimension four) operator appearing in the above OPE. Therefore the difference between the two prescriptions (at order $\mu^2$) can be viewed as a local counterterm proportional to the spin-four operator: 
\bea
&&-\beta F^{(2)}_{\rm presc-2}\,=\, -\beta F^{(2)}_{\rm presc-1}\,-\,\frac{\mu^2}{2\pi}\int d^2z \langle \Lambda^{(4)}(z)\rangle\,+\,{\rm h.c.}\,,
\label{relateboth}\\\nonumber\\\nonumber
&&\Lambda^{(4)}(z)\,=\,5U(z) \,+\tfrac{3}{4}T''(z)\,,
\eea
It is important to note that this local counterterm is finite and is uniquely determined for the two specific prescriptions discussed in this paper. For the reasons we have already described above, prescription-1 is automatically  related to the canonical partition function whilst prescription-2 can be now be thought of as a deformation of the partition function by both  spin-three and spin-four chemical potentials and where the spin-four chemical potential is proportional to $\mu^2$ according to eq. \eqref{relateboth}. At higher orders in the $\mu$-expansion, further differences between the two prescriptions can be explained by additional (fixed) finite counterterms proportional to successively higher spin (and higher dimension) currents. Generalizing the picture, one can obtain a dictionary between generic higher spin chemical potential parameters in the two prescriptions.
For a detailed and elegant explanation of this relationship between prescription-2 and the canonical picture we refer the reader to \cite{dijkgraaf}.

The final result for the free energy correction using the second prescription can be written compactly in terms of the non-holomorphic but modular form $\widehat E_2$ which is obtained by a slight modification of $E_2$:
\bea
&&\,-\beta F^{(2)}_{\rm presc-2}\,=\,2\beta L\left(\beta L C - \frac{\pi^2}{3}{\rm Im}(\tau)\widehat E_2(\tau,\bar \tau)\,c_2\right)\,,\\\nonumber\\
&&\widehat E_2(\tau,\bar\tau)\,\equiv\,E_2(\tau)-\frac{3}{\pi\,{\rm Im}(\tau)}\,.
\eea
 Remarkably, the effect of prescription-2 is to simply replace the explicit factor of $E_2(\tau)$ in eq.\eqref{IAB1} with $\widehat E_2(\tau,\bar\tau)$, ensuring that the result has good modular properties. 

In the calculation above there were only double poles contributing to the final result. Simple poles in operator products become relevant when considering three and four-point functions and therefore affect the correction terms at higher orders in $\mu$ and other physical observables such as the R\'enyi and entanglement entropies.

\section{Thermal corrections for free fermion theory}
\label{four}

We now apply the integration prescriptions discussed above to calculate the 
corrections to the thermal partition function for the free fermion theory in perturbation theory. The current-current correlator \eqref{ww} for $M$ free fermions is expressed in terms of powers of the function ${\cal H}(z_{12})$ which is the ``perturbative'' contribution to the two-point function of the U(1) current \eqref{jj}.
Here we use the term ``perturbative'' with reference to the bosonized formulation of free fermions which will be reviewed at a later stage. In the bosonized language, free fermion current correlators have both perturbative and non-perturbative or winding mode contributions, and in this sense ${\cal H}$ is the perturbative piece. 
 The correlation function is a doubly periodic function of  $z_{12}\,=\,z_1-z_2$ as expected. In order to integrate it over the torus, as explained above, we must first use the  Laurent series \eqref{laurent} for the $\wp$-function and apply the expansion of elliptic functions in terms of the Weierstrass $\zeta$-function and its derivatives \eqref{zetaexp}. The two non-trivial expansions we require are for ${\cal H}^2$ and ${\cal H}^3$ which appear in the correlators for the stress tensor and the spin-three current:
\be
{\cal H}(z)^2\,=\,\tfrac{\pi^4}{9L^4}\left(E_2^2+E_4\right)
\,-\,\tfrac{2\pi^2}{3L^2} E_2\,\zeta'(z)\,-\,\tfrac{1}{6}\zeta^{(3)}(z)\label{h^2}
\ee
and
\be
-{\cal H}^3(z)\,=\,C\,-\,c_2\,\zeta'(z)\,-\,\frac{c_4}{3!}\,\zeta^{(3)}(z)\,-\,\frac{1}{5!}\,\zeta^{(5)}(z)\,,\label{h^3}
\ee
with
\bea
C=\tfrac{\pi^6 }{135\,L^6}\left(4E_6\,+\,15 E_4 E_2\,+\,5 E_2^3\right)\,,
\,\quad c_2= \tfrac{\pi^4}{15 L^4}\left(5 E_2^{2}\,+\,3E_4\right)\,,
%\nonumber\\
%&&\qquad\qquad 
\quad c_4= \tfrac{\pi^2}{ L^2}E_2\,.\nonumber
%,\qquad c_6=1\,.
\eea
It is both a useful exercise and a valuable check to first focus attention on the simple examples involving deformations by the U(1) current and the stress-tensor. In either case, since we can deduce the exact partition sum we can compare these expectations with the perturbative expansion employing the integration prescription introduced above.
\subsection{Perturbation by the U(1)-current}
Let us first consider the well understood  example of the CFT with a chemical potential for the U(1) current $J(z)$:
\be
{\cal L}_{\rm CFT} \to {\cal L}_{\rm CFT}\,+\,\varepsilon\int d^2z (J(z)\,+\,\bar J(\bar z))\,.
\ee
The first correction to the free energy (assuming that $\langle J\rangle\,=\,\langle \bar J\rangle\,=\,0$ at the conformal point) appearing at quadratic order in $\varepsilon^2$ is 
\be
-\beta F^{(2)}\,=\,\frac{\varepsilon^2}{2}\,\int d^2 z_1\int d^2 z_2 \,\langle J(z_1) J(z_2)\rangle_{{\mathbb T}^2}\,+\,{\rm h.c.}
\ee
Using the expression \eqref{jj} for the current correlator and employing our prescription-1 to perform the integrals, we find
\be
-\beta F^{(2)}\,=\,M\pi^2\varepsilon^2\beta^2\,\frac{\vartheta_\nu''}{\vartheta_\nu}\,.
\label{u1f2}
\ee 
We may now check this against the exact grand canonical partition sum for the free fermion CFT with a U(1) chemical potential. 
Following the relevant steps laid out in \cite{Kraus:2011ds}, the  partition function for the deformed CFT is (with periodic boundary conditions around the spatial circle),
\bea
&&\ln{\cal Z}\,=\label{u1partition}\\\nonumber
&&M\ln\cos^2(\pi\varepsilon\beta)\,+\, 2M\sum_{m=1}^\infty\left[\ln\left(1\,+\,e^{2\pi i \tau m \,+\,2\pi i \varepsilon\beta}\right)\,+\,\ln\left(1\,+\,e^{2\pi i \tau m \,-\,2\pi i\varepsilon\beta}\right)\right]\,,
\eea
where the $m=0$ modes have been separately accounted for as we are in the Ramond sector. The exercise can be repeated, with half-integer moding, in the NS sector. Both partition sums then lead to the closed form expression,
\be
{\cal Z}\,=\,\left|\frac{\vartheta_\nu(\pi\varepsilon\beta|\tau)}{\eta(\tau)}\right|^{2M}\,.
\ee
The R and NS sectors correspond to $\nu=2,3$ respectively. This matches our perturbative result upon expanding to  order 
$\varepsilon^2$. We have used the fact that $\vartheta_\nu'/\vartheta_\nu=0$ and that the partition function in each sector is real valued for $\tau = i\beta/L$. 
%Note that the U(1) deformation could be interpreted as a continuous deformation of the spin structure.

\subsection{Perturbation by the stress tensor}
Perturbing the CFT by the stress tensor should have the effect of simply rescaling the background temperature (whilst keeping the spatial circle fixed). Let us now see how this works out,
\be
{\cal L}_{\rm CFT} \to {\cal L}_{\rm CFT}\,+\,\varepsilon\int d^2z\, (T(z)\,+\,\overline T(\bar z))\,.\label{stressdef}
\ee
This situation is interesting because the stress tensor also has a non-vanishing one-point function on the torus. Therefore we have two potential checks to satisfy. The corrections to the free energy at order $\varepsilon$ and $\varepsilon^2$ are respectively given by
\bea
&&-\beta F^{(1)}\,=\,2\varepsilon\int d^2z\,\langle T(z) \rangle_{{\mathbb T}^2}\,,
\label{stressperts}\\\nonumber
\\\nonumber
&&-\beta F^{(2)}\,=\,\varepsilon^2\int d^2z_1\int d^2 z_2\, \langle T(z_1)T(z_2) \rangle_{{\mathbb T}^2}\,.
\eea
Here we have taken account of the fact that the holomorphic and anti-holomorphic sectors contribute equally.
The one-point function for the stress tensor can be computed using the derivative of the fermion propagator \eqref{fermionprop} and taking a coincidence limit for the two points
\be
-\beta F^{(1)}\,=\,-\,\varepsilon\,M(\beta L)\frac{\pi^2}{L^2}\left(\frac{1}{3}E_2\,+\,\frac{\vartheta_\nu''}{\vartheta_\nu}\right)\,.\label{f1tt}
\ee
Let us now compare this with the change in the free fermion partition sum upon infinitesimally changing the temperature,
\be
\beta \to \beta \,+\,\delta\beta\,,\qquad \delta\tau\,=\,i\frac{\delta\beta}{L}\,.
\ee
In a sector with fixed spin structure, we then have
\bea
&&{\ln {\cal Z(\tau + \delta\tau)}}\,=\, \ln{{\cal Z}(\tau)}\,+\,
\delta\tau\,\partial_\tau\ln{{\cal Z}(\tau)}\,+\,\frac{1}{2}\delta\tau^2\,\partial_\tau^2\ln{{\cal Z}(\tau)}\,+\ldots\label{varbeta}\\\nonumber\\\nonumber
&& {\cal Z}(\tau)\,=\,\left[\frac{\vartheta_\nu(\tau)}{\eta(\tau)}\right]^{2M}\,.
\eea
For a rectangular torus with $\tau = i\beta/L$ all functions above are real valued.
The heat equation \eqref{heateq} for theta-functions and the identity $24\,\partial_\tau\ln \eta\,=\,2\pi i E_2$ together then yield
\be
\delta\left({\ln {\cal Z}}\right)\,=\,M\delta\beta\frac{\pi}{4 L}\left(\frac{1}{3}E_2\,+\,\frac{\vartheta_\nu''}{\vartheta_\nu}\right)\,+ {\cal O}(\delta\beta^2)\,.
\ee
This matches eq.\eqref{f1tt} and the temperature variation can be identified as $\delta\beta\,=\,-2\pi\varepsilon\beta$. Therefore, perturbing the CFT by the stress tensor can be viewed as a marginal deformation that rescales the temperature. 

We now turn to the second order correction to the partition function. The stress tensor correlator \eqref{tt} along with the expansion \eqref{h^2} of ${\cal H}^2$ can be used to readily evaluate the necessary integrals so that
\be
-\beta F^{(2)}\,=\,\varepsilon^2\frac{M\beta^2\pi^4}{4L^2}\left[\frac{2}{9}(E_4-E_2^2)\,+\,\frac{\vartheta_\nu^{(4)}}{\vartheta_\nu}\,-\,\left(\frac{\vartheta_\nu''}{\vartheta_\nu}\right)^2\right]\,.
\ee
Once again, making use of the heat equation and the Ramanujan identity $E_2'(\tau)\,=\,2\pi i(E_2^2-E_4)/12$, it immediately follows that this is precisely the correction at order $(\delta\beta)^2$ to the free fermion  partition function in eq.\eqref{varbeta}, as a consequence of an infinitesimal rescaling of the inverse temperature $\beta$.

\subsection{Spin-three chemical potential}
Having seen how holomorphic perturbation theory on the torus works with our integration prescription (prescription-1), we now apply this to the case with spin-three chemical potential. The order $\mu^2$ correction to the partition function is formally
\be
-\beta F^{(2)}\,=\,\mu^2\int d^2z_1\int d^2 z_2\,\langle W(z_1)W(z_2)\rangle_{{\mathbb T}^2}\,.
\ee
We use the expansions \eqref{h^2} and \eqref{h^3} to evaluate the integrals and find that
\bea
&&-\beta F^{(2)}\,=\label{fermionF}\\\nonumber
&&\,80M\mu^2\beta^2\frac{\pi^4}{L^4}\left[\frac{1}{2^5\cdot 3^4\cdot 5}\left(10 E_2^3\,-\,6 E_4 E_2\,-4E_6\right)\,-\,\frac{1}{2^5\cdot 3^2}(E_4-E_2^2)\frac{\vartheta_\nu''}{\vartheta_\nu}\right.\nonumber\\\nonumber
&&\left.\qquad\qquad\qquad-\,\frac{1}{2^6\cdot 3^2}\left(\frac{\vartheta_\nu^{(6)}}{\vartheta_\nu}
\,+\,2E_2\frac{\vartheta_\nu^{(4)}}{\vartheta_\nu}\,+\,E_2^2\frac{\vartheta_\nu''}{\vartheta_\nu}\right)\right]\,.
\eea
The term in parentheses without any theta-functions, in the first line of the above expression, is a weight six quasi-modular form and was also obtained by \cite{douglas} within the context of 2D Yang-Mills theory. The additional contributions appearing with theta-functions arise from  winding modes (in the bosonized language) which were not relevant for the application considered in \cite{douglas}. We will return to this point shortly. Note that the entire  expression is a quasi-modular form of weight six. Furthermore, the $q$-expansion of the modular form with $q\,=\, e^{2\pi i\tau}$ has positive integer coefficients:
\be
-\beta F^{(2)}\,=\,160M\mu^2\beta^2\frac{\pi^4}{L^4}\left(q\,+\,14q^2\,+\,84q^3\,+\,220 q^4\,+\ldots\right)\,.\label{qexpff}
\ee
Although it is not obvious {\em a priori}, this expansion precisely matches a direct calculation of the grand partition sum closely along the lines of Kraus and Perlmutter \cite{Kraus:2011ds}. The partition function of the free fermion theory with a spin-three chemical potential, and {\em without} any constraint on the U(1) charge, is
\bea
&&\ln{\cal Z}\,=\,\,2M\sum_{m=1}^\infty\left[\ln\left(1\,+\,e^{2\pi i\tau m\, +\,  b \mu m^2}\right)\,+\,\ln\left(1\,+\,e^{2\pi i\tau m\, -\,  b \mu m^2}\right)\right]\,,\\\nonumber 
&&b\,=\,\frac{4\sqrt{6}\,\pi^3}{L}\tau\gamma\,.
\eea
Expanding to order $\mu^2$, the partition sum is
\be
\ln{\cal Z}\,=\,2M\ln\left[\frac{\vartheta_2}{\eta(\tau)}\right]\,+\,160M\mu^2\beta^2\frac{\pi^4}{L^4}\sum_{m=1}^\infty\frac{m^4\,q^m}{(1+q^m)^2}\,+\ldots\label{qexpff1}
\ee
The $q$-expansion of the second term reproduces eq.\eqref{qexpff}. This confirms the validity of our perturbative approach and the associated integration prescription. It also makes  clear that Lagrangian perturbations involving holomorphic currents have an equivalent, simple interpretation in the Hamiltonian framework.

\subsubsection{High temperature limits and ${\cal W}_{1+\infty}$ vs. ${\cal W}_{\infty}[0]$}
\label{section:hightw3}
At zero temperature, when $\tau\to i\infty$, the free energy corrections vanish since $q\to0$. It is important that the correction to the thermal partition function is not exactly a modular form due to the anomalous transformation laws for $E_2$ and the derivatives of $\vartheta_\nu$; this  ensures that the high temperature limit is non-vanishing. The high temperature behaviour, when $\tau \to i0^+$, is obtained as usual by first applying the modular transformation $\tau\to -{\tau}^{-1}$ and subsequently taking the limit  $\tau\to i\infty$. The anomalous modular transformation rules play a crucial role in this limit and we find
\be
-\beta F^{(2)}\Big|_{\beta/L\to 0}\,=\,M\mu^2\frac{7\pi^3 L}{6\,\beta^3}\left(1\,+\,{\cal O}\left(e^{-2\pi L/\beta}\right)\right)\,,\label{F2ff}
\ee
whose leading term does not match the high temperature universal result at order 
$\mu^2$ in \cite{paper1, paper2}. This is because the free fermion theory has a ${\cal W}_{1+\infty}$ symmetry and we have neither cosetted out the U(1)-current (as in \cite{Gaberdiel:2013jpa,paper1}), nor have we worked in the sector with vanishing  U(1)-charge. The method of \cite{Kraus:2011ds} which involved the introduction of a Lagrange multiplier imposing the zero U(1)-charge condition on the ensemble average, is expected to be correct only at high temperatures. 

We can, however, use our observations in section \ref{subsection:highT} and  verify that the high temperature behaviour \eqref{F2ff} differs from the universal result for ${\cal W}_\infty[\lambda]$ theory, precisely due to a shift by the correlator of the $U(1)$ current:
\be
-\beta F^{(2)}\big|_{\tau\to i0^+}\,=\,-\mu^2\frac{5M}{36\pi^2}\int d^2z_1\int d^2 z_2\left[
\frac{6\,\pi^6}{\beta^6\,\sinh^6\left(\tfrac{\pi}{\beta}(z_{12})\right)}\,+\,\frac{\pi^6}{\beta^6\,\sinh^2\left(\tfrac{\pi}{\beta}(z_{12})\right)}\right]\,.\nonumber
\ee
Evaluating these two terms separately using our integration prescription,
\bea
&&-\mu^2\frac{5M}{36\pi^2}\int d^2z_1\int d^2 z_2
\frac{6\,\pi^6}{\beta^6\,\sinh^6\left(\tfrac{\pi}{\beta}(z_{12})\right)}\,=\,
M\mu^2\frac{8\pi^3 L}{9\beta^3}\,,
\label{winftyvs}\\\nonumber\\\nonumber
&&-\mu^2\frac{5M}{36\pi^2}\int d^2z_1\int d^2 z_2
\frac{\pi^6}{\beta^6\,\sinh^2\left(\tfrac{\pi}{\beta}(z_{12})\right)}\,=\,M\mu^2\frac{5\pi^3 L}{18\beta^3}\,.
\eea
The first of these two integrals reproduces the universal high temperature correction for 
${\cal W}_\infty[\lambda]$ theories with any $\lambda$, perturbed by the spin-three chemical potential, and the sum of these two contributions yields the corresponding high temperature result \eqref{F2ff} for the ${\cal W}_{1+\infty}$ theory.
It would be very interesting to understand how to extract the correct torus 
partition function for the ${\cal W}_\infty[0]$ CFT from free fermions at any temperature.

\subsection{Free fermions and the 2D Yang-Mills connection}
\label{subsection:2dYM}

A fascinating aspect of the free fermion CFT on the torus, deformed by the spin-three current, is its relation to the large-$N$ expansion of Yang-Mills theory in two dimensions (YM$_2$). This is well known and there is extensive literature on the subject, e.g. see \cite{douglas, dijkgraaf, Griguolo:2004uz}. Thus our computation of higher spin corrections to the partition function of the ${\cal W}_{1+\infty}$ CFT is related to the QCD-string genus expansion in 2D Yang-Mills theory at large-$N$. For this reason it is interesting to review this connection briefly and to highlight significant differences.

Yang-Mills theory in two dimensions is an exactly solvable system and can be reformulated as a string theory in two dimensions \cite{grosstaylor}. In the 't Hooft large-$N$ limit, the chiral partition function of YM$_2$ with $U(N)$ gauge group, on the square torus with area ${\cal A}$, can be formally expressed as a genus expansion,
\be
\ln{\cal Z}_{{\rm YM}_2}\,=\,\sum_{g=1}^\infty\frac{1}{N^{2g-2}}\,{\cal F}_g(\lambda)\,,
\ee
where $\lambda$ is the dimensionless combination of the 't Hooft coupling and the area ${\cal A}$,
\be
\lambda\,=\,g^2_{\rm YM_2}N\,{\cal A}\,.
\ee
Thus $\lambda$ is essentially the dimensionless area of the torus, with the strong and weak coupling limits corresponding to the large and small area limits, respectively. 
The functions ${\cal F}_g(\lambda)$ are generating functions counting maps without folds from a genus $g$ Riemann surface to the target spacetime torus \cite{grosstaylor}. They have the remarkable property \cite{rudd, dijkgraaf, dijkgraaf2, kaneko}
 that they are quasi-modular functions of 
\be
\tau\,=\,\frac{i\lambda}{2\pi}\,.
\ee
The strong coupling expansion of ${\cal F}_g(\lambda)$ encodes a sum over maps with winding number $n$, with an action given by the exponential of the area
\be
{\cal F}_g(\lambda)\,=\,\sum_{n=1}^\infty\sum_j\omega_g^{n,j}\,e^{-n\lambda}\,\lambda^j\,.
\ee
The power laws $\lambda^j$ arise from the integrating over the positions of branch-points and collapsed handles for these maps. The (non-negative) coefficients $\omega_g^{n,j}$ are the Hurwitz numbers which count the number of maps (without folds) with winding number $n$, and a certain number of branch points and collapsed handles. 

The $g=1$ contribution, from genus one worldsheets, is dictated  by the $c=1$ torus partition function, which counts maps of the torus onto a torus,
\be
{\cal F}_1(\lambda)\,=\,-\ln\eta(\tau)\,,
\ee
where $\eta(\tau)$ is the Dedekind eta function. The absence of theta-functions associated to free fermion spin structures is indicative of the fact that, in the bosonized language, the winding modes are irrelevant for the application to large-$N$ YM$_2$ and can be systematically eliminated as discussed in \cite{douglas}. The $g=2$ term is the correction to the Yang-Mills free energy at order $N^{-2}$. Ignoring all winding mode contributions (in the bosonized language), the second order correction to the free fermion partition function using our integration prescription yields, instead of eq.\eqref{fermionF} 
\bea
-\beta F^{(2)}&&=\,M\,\left(6\pi\gamma\,\mu\frac{2\pi}{L}\right)^2\,\left(2\pi\,{\rm Im}\,\tau\right)^2\times\label{F2}\\\nonumber
&&\frac{1}{2^5\cdot 3^4\cdot 5}
\left(10E_2^3\,-\,6 E_4 E_2\,-\,4E_6\,+\,
\frac{90}{2\pi\,{\rm Im}\,\tau}(E_4-E_2^2)+\frac{1080}{(2\pi\,{\rm Im}\,\tau)^3}\right)
\,.
%\right.
%\\\nonumber\\\nonumber
%&&\left.-\frac{1}{2^5\cdot 3^2}\,\left(E_4\,-\,E_2^2\,+\,\frac{36}{(2\pi\,{\rm Im}\,\tau)^2 }\right)\,
%\frac{\vartheta_\nu''}{\vartheta_\nu}\,+\,\frac{1}{2^5}\frac{1}{(2\pi\,{\rm Im}\,\tau)}\frac{\vartheta_\nu^{(4)}}{\vartheta_\nu}\,-\,\frac{1}{2^6\cdot 3^2}\frac{\vartheta_\nu^{(6)}}{\vartheta_\nu}
%\right]\,.
\eea
Note the appearance of powers of $1/{\rm Im}(\tau)$ which were absent in 
eq.\eqref{fermionF}. The dependence on powers of $1/{\rm Im}(\tau)$ is natural when winding modes in the bosonized picture are ignored, and only the perturbative contributions to the correlation functions are included. We refer the reader to the  discussion in appendix \ref{appendix:JJVV} for further clarification.
The correspondence between free fermion theory with the spin-three deformation and large-$N$ Yang-Mills leads us to make the identifications,
\be
\frac{\lambda}{2\pi}\,\leftrightarrow\, \frac{\beta}{L}\,,\qquad\qquad
\frac{1}{N}\,\leftrightarrow\, \frac{\mu}{L}\,(12\pi^2\gamma)\,,
\ee
where the numerical factors in the second expression originate from the normalization of our spin-three current. With these identifications eq.\eqref{F2} agrees with the corresponding expression in \cite{douglas} for the order $\mu^2 \sim N^{-2}$ correction to the partition function\footnote{Our expression appears to  differ from that of \cite{douglas} by a term $\sim\lambda^{-1}$, but without any exponential factors. The significance of this is unclear at present.}. The strong and weak coupling expansions in YM$_2$ can be viewed as the low  and high temperature expansions, respectively, of the free fermion CFT. 
The low temperature $q$-expansion of \eqref{F2} yields the instanton-like expansion of the genus two contribution to the Yang-Mills free energy:
\be
{\cal F}_2(\lambda)\,=\,\lambda^2\,\left(8q^2\,+\, 64 q^3\ldots\right)\,+\,\lambda\left(2q\,+\,12q^2\ldots\right)\,+\,\frac{1}{12\lambda}\,.\label{F2YM}
\ee
Here $q=e^{2i\pi\tau}=e^{-\lambda}$.
The coefficient of $\lambda^2$ counts the number of maps with two branch points whilst the 
terms proportional to $\lambda$ are associated to maps with one collapsed handle. The quasi-modular property of ${\cal F}_2$ produces a similar expansion in powers of $e^{-1/\lambda}$ at high temperature, but additional new terms appear which ensure the correct high temperature thermodynamics. These additional terms in the high temperature expansion, which is a weak coupling expansion in the language of YM$_2$, can also be given a geometric interpretation as in \cite{Griguolo:2004uz}.

As already alluded to above, an important general property of ${\cal F}_g(\lambda)$ is that it must be a quasi-modular form with modular weight $6g-6$. In particular, the $g=2$ contribution which we have evaluated has modular weight $6$. We expect therefore that when we compute the entanglement entropy of a spatial interval of length $\Delta$ in the free fermion theory, it will yield a function of $\Delta$ with modular weight 6 at order $\mu^2$. On general grounds we also expect this function to be a quasi-elliptic function of $\Delta$. It would be extremely interesting to understand if the EE contributions can also be given a precise geometric interpretation in terms of maps from genus $g$ Riemann surfaces to a torus with two marked points.
We note that the R\'enyi entropies for YM$_2$ on the torus have already been investigated in \cite{Velytsky:2008rs} and \cite{Gromov:2014kia} and, due to the topological nature of 2D Yang-Mills theory, have been argued to be independent of the interval lengths. This is certainly not what we expect from the free fermion theory as we see below. It would be worth investigating the reason for this fundamental difference between the two theories, given that the two systems are closely related (on the torus).

\section{Thermal correction for ${\cal W}_\infty[1]$ or free bosons }
\label{five}

It is interesting to compare and contrast the thermal correction at order $\mu^2$ for different CFTs on the torus. We can perform this excercise quite easily for $M$ free complex bosons ${\Phi_a}$, $(a=1,2,\ldots M)$. This theory has central charge $2M$.
The spin-three current in the free boson theory, adopting the normalizations of \cite{Kraus:2011ds}, is
\bea
W(z)\,=\,  \frac{\gamma }{\sqrt{2}}\sum_{a=1}^M\left( \pd \bar \Phi_a \, \pd ^2 \Phi^a  - \pd ^2 \bar\Phi_a\, \pd \Phi^a   \right)\,,\qquad \gamma\,=\,i\sqrt{\frac{5}{6\pi^2}}\,.
\eea
Applying Wick contractions, and using the Green's function for free bosons on the torus we obtain
\bea
\langle W(z_1) W(z_2)  \rangle \,=\, M \gamma^2  \left[  - \wp'(z_{12})^2  
+\left( \wp(z_{12})\,+\,\frac{\pi^2}{3L^2} \widehat E_2 (\tau,\bar\tau)  \right)\wp'' (z_{12}) \right]\,,
\eea
where $z_{12}=z_1-z_2$. Since this is an elliptic function, we can express it in terms of the Weierstrass $\zeta$-function and its derivatives
\bea
\langle W(z_1) W(z_2)  \rangle\,=\,M\gamma^2 \left( C - c_2 \zeta ' (z) - \frac{c_4}{3!} \zeta^{(3)} (z)- \frac{c_6}{5!} \zeta^{(5)} (z)\right)\,,
\eea
with coefficients
\bea
C = \frac{8 \pi ^6 }{45 L^6} {E}_6(\tau )
,\,\qquad c_2 = \frac{8\pi^4}{15L^4} E_4 (\tau)\,, \quad c_4 = \frac{\pi^2}{L^2} \widehat E_2\,, \quad c_6 \,=\,1\,. 
\eea
Integrating the two-point function over the torus ${\mathbb T}^2$ using prescription-1 and including the contributions of both holomorphic and antiholomorphic sectors, we find the order $\mu^2$ correction to the free energy ,
\bea
-\beta F^{(2)} 
&&=\, (2M)\mu^2\gamma^2\frac{8\pi^6 \beta^2 }{45L^4} \left(E_6\,-\, E_2\,E_4\right)\label{freebosonans}\\\nonumber\\\nonumber&&=\,
(2M)\mu^2\gamma^2\frac{4\pi^5 \beta^2}{15L^4}\,\,i \frac{d E_4(\tau)}{d\tau}\,.
\eea
Calculating the integrals using prescription-2 leads to the same result with $E_2$ replaced by $\widehat E_2$. 

As in the case of the free fermion theory we can compare this result with the canonical partition function at order $\mu^2$ from the Hamiltonian approach of \cite{Kraus:2011ds}. Accordingly, the partition function of the free boson CFT with a spin three chemical potential (including both holomorphic and anti-holomorphic sectors) is
\bea
&&\ln{\cal Z}\,=\,-\,M\sum_{m=1}^\infty\left[\ln\left(1\,-\,e^{2\pi i\tau m\, +\,  b \mu m^2}\right)\,+\,\ln\left(1-e^{2\pi i\tau m\, -\,  b \mu m^2}\right)\right]\,,\\\nonumber 
&&b\,=\,\frac{16\pi^3}{L}\gamma\tau\,.
\eea
Expanding this exact partition sum to quadratic order in $\mu$ we have
\be
\ln{\cal Z}\,=\,-\,M \ln|\eta(\tau)|^2\,-\,(2M)\mu^2\gamma^2\frac{128\pi^6\beta^2}{L^4}\sum_{m=1}^\infty\frac{m^4 \,q^m}{(1-q^m)^2}\,+\ldots\,.
\ee
From the $q$-expansions of the Eisenstein series \eqref{qexpansion}, we deduce
\be
\sum_{m=1}^\infty\frac{m^4 \,q^m}{(1-q^m)^2}\,=\,-\frac{1}{720}(E_6\,-\,E_2 E_4)\,
\ee
leading to precise agreement between the grand canonical partition sum and eq.\eqref{freebosonans} obtained by the integration method using prescription-1.
Therefore the correction to the free energy is an anomalous modular form of weight six. The leading high temperature  asymptotics, obtained after performing a modular transformation, matches the expected universal answer for ${\cal W}_\infty[\lambda]$ theories \cite{paper1, paper2}, with exponentially subleading corrections,
\be
-\beta F^{(2)}\big|_{\beta/L \to 0}\,=\,(2M)\mu^2\frac{8\pi^3 L}{9\beta^3}\left(1\,+\,{\cal O}(e^{-2\pi L/\beta})\right)\,.
\ee
Note that the central charge of the theory is $2M$. The coefficients of the low temperature  $q$-expansion are non-negative integers, and constitute predictions for solutions with spin-three hair in hs$[\lambda=0]$ theory:
\bea
-\beta F^{(2)}\,=\,-\mu^2\gamma^2M\frac{128\pi^6 \beta^2 }{L^4}\left(q\,+\,18q^2\,+\,84 q^3\,+\ldots\right)\,.
\eea

\section{Single interval R\'enyi entropies}
\label{six}

We will now adapt the prescriptions and method for evaluating the integrals over holomorphic current correlators to compute the singe interval R\'enyi and entanglement entropies in the free fermion theory on the torus. The standard route for this calculation is via the implementation of the replica trick using branch point twist fields 
\cite{Calabrese:2004eu, Calabrese:2005in, Calabrese:2009qy}.  The partition function of the QFT is evaluated on an $n$-sheeted Riemann surface ${\cal R}_n$, branched along the interval, first yielding the so-called R\'enyi entropies $S_{\rm RE}(\Delta, n)$, and subsequently taking the $n\to 1$ limit to obtain the entanglement entropy,
\be
S_{\rm RE}(\Delta,n)\,=\, \frac{1}{1-n}\ln\left[\frac{{\cal Z}^{(n)}}{{\cal Z}^n}\right]\,,\qquad S_{\rm EE}\,=\,\lim_{n\to 1}S_{\rm RE}(\Delta, n)\,.\label{RE}
\ee
Here ${\cal Z}^{(n)}$ and ${\cal Z}$ are the QFT partition functions on the branched Riemann surface and the torus respectively. The R\'enyi and entanglement entropies are formally related to the reduced density matrix $\rho_\Delta$ for a spatial interval of length $\Delta$ as,
\bea
S_{\rm RE}(\Delta,n)\,=\,\frac{1}{1-n}\ln{\rm Tr}\rho_\Delta^n\,,\qquad S_{\rm EE}\,=\,-{\rm Tr}{\rm\rho_\Delta }\ln\rho_\Delta\,.
\eea
As pointed out in \cite{Calabrese:2004eu}, the calculation of ${\cal Z}^{(n)}$ may be achieved by evaluating the partition function in the presence of {\em twist} and {\em antitwist} fields ${\sigma}_n$ and $\overline{\sigma}_n$ inserted at the end-points of the interval, so that in a perturbative expansion in $\mu$  we have
\bea
&& S_{\rm RE}(\Delta, n)\,=\,\frac{1}{1-n}\,\left[\ln \left({\cal Z}_{\rm CFT}^{(n)}\,+\,\mu\int d^2z \langle\bar\sigma_n(y_2)\,W(z)\,\sigma_n(y_1)\rangle\right.\right.\label{perturbativeS}\\\nonumber 
&&\left.\left.\,+\,\frac{\mu^2}{2}\int d^2 z_1 d^2 z_2\, \langle\overline{\sigma}_n(y_2)W(z_1)W(z_2){\sigma}_n(y_1)\rangle\,+\, {\rm h.c.}\ldots\right)\,-\,n\ln{\cal Z}\right]
\eea
with $y_1, y_2 \in {\mathbb R}$  and $|y_2-y_1|=\Delta$. Note that ${\cal Z}$ is the partition function of the {\em deformed} CFT
 and can itself be expressed in a perturbative expansion in $\mu$.
Therefore, to compute the R\'enyi and entanglement entropies at order $\mu^2$ we need the three- and four-point functions involving a twist-antitwist pair and one or two spin-three current insertions, on the $n$-fold branched cover of the torus.

To implement the replica trick we take both the CFT Lagrangian density and the chemical potential deformation as a sum over $n$ independent copies
\be
{\cal L}\,=\,{\cal L}_{\rm CFT}\,-\,\mu W\,=\,\sum_{i=1}^n\left({\cal L}_{\rm CFT}^{(i)}\,-\,\mu{W}^{(i)}\,+\,{\rm h.c.}\right)\,. 
\ee
Note that the deforming operator is also replicated\footnote{R\'enyi entropies for perturbed CFTs have been studied in \cite{Cardy:2010zs}.}. Focussing on the system of $M$ fermions $\{\psi_a\}_{a=1,\dots M}$ for concreteness,
the action of the twist fields on the replicas is via cyclic permutation  of the $n$-copies or branches $\{\psi_{a}^{(i)}\}_{i=1,\ldots n}$. The different sheets or branches are glued along a branch cut that coincides with the entangling interval. 

In general, one is then faced with the difficult task of calculating the partition function of the QFT on the $n$-sheeted Riemann surface. For CFT's with generic deformations turned on this can be complicated. However, in the case at hand, a crucial simplification results from the fact that all the higher spin currents in the free fermion theory, including the spin-three operator $W$, are bilinears in the fields \cite{Bergshoeff:1990yd, Pope:1991ig, paper1}. The deformed theory continues to be Gaussian and therefore it is natural to transform to a Fourier basis which diagonalizes the permutation action of the twist fields:
\be
\tilde \psi_{a,k}\,=\sum_{\ell=1}^{n} e^{2\pi i(\ell-1)\, \frac{k}{n}}\,\psi_{a}^{(\ell)}\,\qquad k\,=\,-(n-1)/2, (n-3)/2,\ldots (n-1)/2\,.
\ee
The action of the twist fields on  $\tilde\psi_{a,k}$ is given by a phase factor $e^{\pm 2\pi i k/n}$ acquired  when taken around one of the two branch points. The theory on the $n$-sheeted surface is then recast in terms of $n$ decoupled free fermions, and the QFT partition function is expressed as the correlation function of the product  of $n$-pairs of twist-antitwist fields,
\be
{\cal Z}^{(n)}\,=\,\prod_{k}\langle\sigma_{k,n}(y_1)\bar\sigma_{k,n}(y_2)\rangle\,.
\ee
In the presence of a deformation away from the conformal point, evaluation of the partition function in conformal perturbation theory is easier when there is an explicit representation for the twist operators. This is possible in the free fermion theory, upon bosonization. On the other hand, for the free boson CFT, there is no such explicit representation and correlators of twist operators with other currents have to be reconstructed from their short distance properties as was done on the cylinder in \cite{paper1}. The situation on the torus is more involved and so we will content ourselves with studying the simpler case of free fermions in this paper.

\subsection{Bosonization}
In order to work with an explicit representation of the twist fields for the R\'enyi entropy calculation, we will use the bosonized representation of  free fermions in terms of $M$ free chiral bosons $\{\varphi_a\}$,
\be
\psi^a(z)\,=\,:e^{i\varphi_a(z)}:
\ee
satisfying the operator product expansion (OPE)
\be
\varphi_a(z)\,\varphi_b(0)\,\sim\,-\,\delta_{a,b}\,\ln z\,.
\ee  
Crucially, the  $\{\varphi_a\}$ must each be viewed as a compact chiral boson with unit radius. The sum over the non-trivial winding sectors is then responsible for reproducing the correct fermion partition function and spin structure \cite{DiFrancesco:1997nk}. We  proceed to outline the key ingredients required for the calculation in the bosonized language, accompanied by consistency checks where necessary.

In the free fermion theory, each replica field $\tilde \psi_{a,k}$ (in the Fourier basis) corresponds to a compact chiral boson $\varphi_{a,k}$, and the twist fields can be explicitly represented as
\be
\sigma_{k,n}\,=\,\prod_{a=1}^M:e^{i\frac{k}{n}\left(\varphi_{a,k}(z)-\bar\varphi_{a,k}(\bar z)\right)}:\,\qquad \bar\sigma_{k,n}\,=\,\prod_{a=1}^M:e^{-i\frac{k}{n}\left(\varphi_{a,k}(z)-\bar\varphi_{a,k}(\bar z)\right)}:
\ee
with $k= -\tfrac{1}{2}(n-1)\,,-\tfrac{1}{2}(n-3)\,,\ldots \tfrac{1}{2}(n-1)$.
Now, the corrections to the partition function (as a power series in $\mu$) can be computed once the correlators involving these twist fields with any number of current insertions are known. 

On the plane and the cylinder such correlators follow immediately from Wick rules, but the torus is more subtle and involves extra contributions from `instanton' sectors as we discuss below. We explain the resulting modified Wick rules in appendix \eqref{appendix:JJVV}.

\subsection{Twist field correlator on ${\mathbb T}^2$}
In the absence of any other operator insertions, the two-point function of a twist-antitwist pair on the square torus with $\tau\,=\,i\beta/L$ and $|y_2-y_1| =\Delta\in {\mathbb R}$, in the undeformed theory, is given by 
\bea
&&\langle\bar\sigma_{k,n}(y_2)\sigma_{k,n}(y_1)\rangle_{\rm CFT}\,=\,
\left|\frac{\pi}{L}\frac{\vartheta_1'(0)}{\vartheta_1\left(\frac{\pi\Delta}{L}\right)}\right|^{2M\frac{k^2}{n^2}}\,\left|\frac{\vartheta_\nu\left(\frac{k}{n}\frac{\pi\Delta}{L}\right)}{\vartheta_\nu(0)}\right|^{2M}\,,\label{twistCFT}
\\\nonumber
&&\Delta\,=\,|y_2-y_1|\,,
\eea
so that
\be
%{\cal Z}^{(n)}_{\rm CFT}\,=
\langle\overline\sigma_\n(y_2)\sigma_n(y_1)\rangle\,=\,\prod_{k=-\frac{n-1}{2}}^{\frac{n-1}{2}}\left|\frac{\pi}{L}\frac{\vartheta_1'(0)}{\vartheta_1\left(\frac{\pi\Delta}{L}\right)}\right|^{2M\frac{k^2}{n^2}}\,\left|\frac{\vartheta_\nu\left(\frac{k}{n}\frac{\pi\Delta}{L}\right)}{\vartheta_\nu(0)}\right|^{2M}\,.\label{zncft}
\ee
The normalisations have been fixed so that in the short distance limit $(\Delta\ll L,\beta)$ the correlator approaches $|y_1-y_2|^{2Mk^2/n^2}$.
Depending on whether the fermions have anti-periodic (NS) or periodic (R) boundary conditions around the spatial circle, $\nu$ takes the values $3$ and $2$ respectively.
A derivation of the twist correlator in the bosonized description is presented in appendix \ref{appendix:JJVV}.
 It has already been demonstrated in \cite{Azeyanagi:2007bj} that the correlator \eqref{twistCFT} eventually leads to the single interval RE and EE in the free fermion CFT and, in particular, satisfies the non-trivial requirement that $S_{\rm EE}(\Delta=L)$ should match the thermal entropy of the CFT.

\subsection{Free boson propagator on ${\mathbb T}^2$}
To calculate the effect of operator insertions in the presence of twist fields, we need to understand both quantum and classical or nonperturbative contributions to the correlations functions.
The quantum pieces are encoded in the free boson propagator on the torus which is one of the main ingredients in the calculation. Recall that the Green's function for a free (real) boson $\Phi(z,\bar z)\equiv \varphi(z)\,+\,\bar\varphi(\bar z)$ on the complex plane
\be
G(z_1,z_2)\,=\,\langle\Phi(z_1,\bar z_1)\,\Phi(z_2,\bar z_2)\rangle\,=\,-\ln|z_1-z_2|^2\,.
\ee
Naively, the Green's function on the torus ${\mathbb C/\Gamma}$, $\Gamma\simeq 2\omega_1{\mathbb Z}\oplus 2\omega_2 {\mathbb Z}$, should be obtained by a sum over images 
\be
\langle\Phi(z_1,\bar z_1)\,\Phi(z_2,\bar z_2)\rangle\,\to\,-\,\ln\prod_{n,m\in{\mathbb Z}}|z_{12}-\Omega_{n,m}|^2\,,
\ee
where $\Omega_{m,n}\,\equiv\, 2n\omega_1 + 2m\omega_2$, with $2\omega_1\,=\,L$ and $2\omega_2\,=\,i\beta$ being the periods of the torus.  However, this is not the complete result since the torus is compact and we cannot have a source for the Green's function equation without a sink, so we need the following modification
\be
\partial\bar\partial\, G_{{\mathbb T}^2}(z_1,z_2)\,=\,\pi\left(\delta^2(z_1-z_2)\,-\,\frac{1}{(2\omega_1)^2\,{\rm Im}(\tau)} \right)\,,\label{gfsink}
\ee
where the second term on the right hand side can be viewed as a negative unit charge distributed uniformly on the torus with area $(2\omega_1)^2\,{\rm Im}(\tau)\,=\,\beta L$.
Integrating the left hand side gives zero on ${\mathbb T}^2$ and the same now holds on the right hand side. Including the effect of the sink term above, up to an additive constant, the  free boson Green's function is
\bea
&&G_{{\mathbb T}^2}(z_{12})\,=\,-\ln\left|\frac{\vartheta_1\left(\frac{\pi\,z_{12}}{L}\right)}{\vartheta_1'(0)}\right|^2\,+\,2\pi\,\frac{\left({\rm Im}\,z_{12}\right)^2}{L^2\,{\rm Im}(\tau)}\,,\label{Phiprop}\\\nonumber
%+\,\frac{\pi^2}{6 L^2}\,E_2(\tau)\,(z_{12})^2\,
&& z_{12}\,=\,z_1-z_2\,.
\eea
This is a standard result (e.g. \cite{itzykson, douglas}). In particular this implies that the correlator of the holomorphic current $\partial\varphi$, up to an additive constant, is
\bea
\langle\partial\varphi(z_1)\partial\varphi(z_2)\rangle\,=\,&&-\wp(z_1-z_2)-\frac{\pi^2}{3L^2}\left(E_2-\frac{3}{{\pi\rm Im }\tau}\right)\,,\label{jjpert}\\\nonumber
=\,&& {\cal H}(z_1-z_2)\,+\,\frac{\pi}{L^2\,{\rm Im}\,\tau}\,.
\eea
The function ${\cal H}$ was  defined in eq.\eqref{jj} for the current correlator in the fermionic picture. Comparing with \eqref{jj} we see two potential sources of disagreement. The first of these is the dependence on $1/{{\rm Im}\,\tau}$ which is absent in \eqref{jj} and the second is the absence of theta-function dependent terms in eq.\eqref{jjpert}. Both discrepancies are resolved by the inclusion of the winding mode contributions.

\subsection{Deformation by the U(1) current}
Following the route taken in our calculation of thermal corrections, we will begin by examining simple examples of chiral deformations before addressing the spin-three chemical potential.
The simplest case is the spin-1 current which, in the bosonized theory, is given by
\be
J\,=\,i\sum_{a=1}^M\partial\varphi_a\,.
\ee
As a warmup towards understanding the structure of the perturbative corrections to the twist-antitwist correlator due to deformations of the type \eqref{perturbativeS} , we will examine the basic structure of correlation functions involving
insertions of  the holomorphic U(1) current alongside the twist fields.
Although the theory is free, calculation of correlators in the bosonized description is subtle. This is because correlation functions in the bosonized framework receive both perturbative and semiclassical winding mode contributions (see appendix\eqref{appendix:JJVV} and \cite{DiFrancesco:1997nk}). Specifically, in a sector
with winding numbers $w,w' \in \mathbb Z$, we have 
\bea
\partial\varphi_a(z)\big|_{w,w'}\,&&=\,\partial\varphi_{a;\,{\rm cl}}\,+\,\partial\varphi_{a;\,\rm qu}(z)\,,\qquad w,w'\in {\mathbb Z}\,,\\\nonumber\\\nonumber
&&=\,-\frac{i\pi}{L}\left(\frac{w'-w\,\bar\tau}{{\rm Im}\,\tau}\right)\,+\,\partial\varphi_{a;\,\rm qu}\,.
\eea
The quantum contribution $\varphi_{a;\,{\rm qu}}$ is a free field with vanishing 
one-point function. Following the steps laid out in appendix \ref{appendix:JJVV}, the three-point function involving two twist fields and one insertion of the U(1) current is given as,
\bea
&&\langle\bar\sigma_{k,n}(y_2)\sigma_{k,n}(y_1)\,\partial\varphi_{a,k}(z)\rangle_{\rm CFT}\,=\,\langle\bar\sigma_{k,n}(y_2)\sigma_{k,n}(y_1)\rangle_{\rm CFT}\,\left\{\frac{ik}{n}G(z)\,-\,iK^{(1)}\left(\tfrac{k}{n}\tfrac{\pi\Delta}{L}\right)\right\}\,,\nonumber\\\nonumber\\
&&G(z)\,=\,\tilde\zeta(z-y_1)\,-\,\tilde\zeta(z-y_2)\,,\qquad K^{(p)}\,=\,\left(\frac{\pi}{L}\right)^p\,\frac{\vartheta_\nu^{(p)}\left(\tfrac{k}{n}\frac{\pi\Delta}{L}\right)}
{\vartheta_\nu\left(\tfrac{k}{n}\frac{\pi\Delta}{L}\right)}\,.\label{jvv}
\eea
 The function $\tilde \zeta(z)$ is the logarithmic derivative of Jacobi's $\vartheta_1$-function defined in eqs.\eqref{zetatdef},\eqref{logtheta}. It is quasi-elliptic and a simple modification of the Weierstrass $\zeta$-function.
 
Here $G(z)$ is the one-point function for $\partial\varphi_{a,k}$ in the presence of a positive and negative point charge pair on the torus. As explained in appendix \ref{appendix:JJVV}, the twist operator insertions act as point sources for the chiral bosons on the torus giving rise to a classical one-point function for the fields.  
This also follows from straightforward Wick contractions of the current with the twist operators. The constant additive piece $K^{(1)}$ is a classical or instanton contribution from the winding modes of the compact boson $\varphi_{a,k}$. It is interesting to check how these two terms combine and approach the correct high temperature limit $\frac{\beta}{L} \to 0$. As usual, this limit is obtained by performing a modular transformation, and subsequently taking $\frac{L}{\beta}\to \infty$ keeping $\frac{\Delta}{\beta}$ fixed, and we find 
\bea
&&G(z)\to\frac{\pi}{\beta}\left(\coth\tfrac{\pi}{\beta}(z-y_1)\,-\,\coth\tfrac{\pi}{\beta}(z-y_2)
%\,-\,\frac{2\Delta}{L}
\right)\,,\\\nonumber
&&K^{(1)}\rightarrow \,-\frac{2\pi}{L}\frac{k}{n}\frac{\Delta}{\beta}\to 0\,.
%\frac{ik}{n}\frac{\pi}{\beta}\frac{2\Delta}{L}\,.
\eea
This agrees with the correlators evaluated directly on the cylinder in  \cite{paper1}. For the two-point function of two insertions of the operator $\partial\varphi_{a,k}$ in the presence of twist operators we similarly obtain,
\bea
&&\langle\bar\sigma_{k,n}(y_2)\sigma_{k,n}(y_1)\,\partial\varphi_{a,k}(z_1)\partial\varphi_{a,k}(z_2)\rangle_{\rm CFT}\,=\,\langle\bar\sigma_{k,n}(y_2)\sigma_{k,n}(y_1)\rangle_{\rm CFT}\,\times\\\nonumber\\\nonumber
&&\qquad\qquad\left\{{\cal H}(z_1-z_2)\,+\,\vartheta_\nu(x)^{-1\,}{\cal G}_k(z_1)\,{\cal G}_k(z_2)\vartheta_\nu(x)\big|_{x\,=\,\frac{k}{n}(\pi\Delta/L)}
%\left(\frac{ik}{n}\right)^2G(z_1)G(z_2)\,+\,\frac{ik}{n}\left(G(z_1)+G(z_2)\right)\,K^{(1)}\,+\,K^{(2)}
\right\},\\\nonumber\\
&&\qquad\qquad{\cal G}_k(z):=\,\frac{ik}{n}G(z)\,-\,\frac{i\pi}{L}\partial_x\,.
\eea
Here ${\cal H}(z_1-z_2)$ is the function defined in \eqref{jj} in the absence of any twist fields. When $k=0$, we recover the U(1) current-correlator encountered in \eqref{jj} within the fermionic formulation. Therefore the factors of $1/{{\rm Im}\,\tau}$ we saw in eq.\eqref{jjpert} cancelled against similar contributions from the winding modes, which also introduced new theta-function dependent terms that contain information on the spin structure.
The expression\footnote{The results can be reproduced by acting with derivatives on the four-point correlators of vertex operators on the torus \cite{Herzog:2013py} and taking two of the momenta to vanish.} above
 illustrates precisely how the Wick contraction rules are modified by winding mode contributions as explained in appendix \ref{appendix:JJVV}. 
 
\subsubsection{R\'enyi entropies for spin-1 deformation}
As a warmup excercise we will now calculate single interval R\'enyi entropy for the free fermion CFT with a chemical potential for the U(1) charge. The R\'enyi entropy in the presence of twist fields and U(1) current perturbation is
\bea
&&S_{\rm RE}(\Delta, n)\,=\\\nonumber
&&\frac{1}{1-n}\left[\ln{{\cal Z}^{(n)}_{\rm CFT}}\,-\,n\ln{\cal Z}\,+\,\varepsilon^2\int d^2z_1\int d^2z_2\,
\frac{\langle\overline\sigma_n(y_2)\,J(z_1)\,J(z_2)\sigma_n(y_1)\rangle}{\langle\overline\sigma_n(y_2)\sigma_n(y_1)\rangle}\,\ldots\label{jpertRE}
\right]\,.
\eea
The CFT partition function in the absence of deformations is simply the twist-antitwist correlator, ${\cal Z}^{(n)}_{\rm CFT}\,=\,\langle\overline\sigma_n(y_2)\,\sigma_n(y_1)\rangle$.  At order $\varepsilon^0$, it has already been explicitly verified in \cite{Azeyanagi:2007bj}, that the twist field correlators satisfy all necessary checks required of the entanglement entropy of the CFT on the torus.
In particular, the CFT partition function ${\cal Z}^{(n)}_{\rm CFT}$ on the $n$-fold cover of the torus is given explicitly by the twist-field correlation function \eqref{zncft}. Upon taking the limit $n\to 1$, this yields the entanglement entropy, $S_{\rm EE}(\Delta)$, which was shown in \cite{Azeyanagi:2007bj} to have the property that it 
yields the thermal entropy of the free fermion CFT when $\Delta=L$.

We have omitted from eq.\eqref{jpertRE} a putative one-point function for $J$ in the presence of the twist fields, because it vanishes identically at the conformal point. The definition of the current operator involves a sum over both the fermion flavour index and the replica Fourier index:
\be
J\,=\,i\sum_{a=1}^M\sum_{k}\partial\varphi_{a,\,k}\,.
\ee
Recall from our previous discussion that the Fourier label ranges from $k\,=\,-(n-1)/2$ to $(n-1)/2$. Explicitly, the current-current correlator on the $n$-sheeted Riemann surface is
\bea
\frac{\langle \overline\sigma_n(y_2)J(z_1)\, J(z_2)\sigma_n(y_1)\rangle}{\langle \overline\sigma_n(y_2)\sigma_n(y_1)\rangle}\,=\,
-\,M\sum_k\left[{\cal H}(z_{12})
%\,+\,\left(\frac{ik}{n}\right)^2 G(z_1)G(z_2)\right.\\\nonumber\\\nonumber
%&&\left.+\,\frac{ik}{n}\left\{G(z_1)\,+\,G(z_2)\right\}\left(\frac{-i\pi}{L}\right)\frac{\vartheta_\nu'(x)}{\vartheta_\nu(x)}
\,-\,\frac{\pi^2}{L^2}\left[\ln\vartheta_\nu(x)\right]''
%\frac{\vartheta_\nu''(x)}{\vartheta_\nu(x)}\,-\,\left(\frac{\vartheta_\nu'(x)}{\vartheta_\nu(x)}\right)^2\right\}
\right]_{x=\frac{k\pi\Delta}{nL}}
\eea
Any terms involving the functions $G(z)$ vanish before integration due to independent sums over a pair of  replica Fourier indices. In addition, the integration prescription we have adopted straightforwardly implies
\be
\int d^2z\, {\cal H}(z)\,=\,0\,,\qquad\qquad\int d^2z \,{G}(z)\,=\,0\,.\label{hgint}
\ee 
Therefore, the order $\varepsilon^2$ correction to the R\'enyi entropy is 
\be
S^{(2)}_{\rm RE}(\Delta,n)\,=\,\frac{\varepsilon^2\pi^2\,M\beta^2}{(1-n)}\left[\sum_k
\left[\ln\vartheta_\nu(x)\right]''_{x=\frac{k\pi\Delta}{nL}}
%\left\{\frac{\vartheta_\nu''(x)}{\vartheta_\nu(x)}\,-\,\left(\frac{\vartheta_\nu'(x)}{\vartheta_\nu(x)}\right)^2\right\}_\,
\,-\,n\,\frac{\vartheta_\nu''}{\vartheta_\nu}\right]\,.\label{renyi2}
\ee
There are various checks that this formula for the R\'enyi entropy satisfies. It vanishes identically when the interval length $\Delta$ shrinks i.e.in the limit 
$\Delta/L\to 0$. Applying modular transformation properties of the Jacobi-theta functions, for  fixed $\Delta/L$, we find that in the high-temperature limit, $\beta/L \to 0$, the RE correction vanishes  exponentially $\sim e^{-\pi L/\beta}$. Therefore, at high temperatures, the  entanglement entropy correction $S_{\rm EE}^{(2)}(\Delta)$ also vanishes exponentially with temperature. This is consistent with the behaviour of the thermal entropy correction at this order. The U(1) chemical potential does not alter the partition sum \eqref{u1partition} at high temperatures. The most non-trivial check of the RE correction evaluated above, comes from the requirement that when $\Delta=L$, the entanglement entropy should reduce to the thermal entropy for the deformed theory. For the undeformed theory ($\varepsilon=0$), this check has already been done in \cite{Azeyanagi:2007bj}. We need only to verify that it is also satisfied by the correction at second order in $\varepsilon$. The thermal entropy correction at order $\varepsilon^2$ is
\be
S_{\rm thermal}^{(2)}\,=\,M\pi^2\varepsilon^2\beta^2\,\partial_\beta\left(\beta\,\frac{\vartheta_\nu''}{\vartheta_\nu}\right)\,.
\ee
To show that the R\'enyi entropy correction \eqref{renyi2} yields this result in the limit $n\to1$ with $\Delta=L$, we first need to pick a specific spin structure, say $\nu=3$ for simplicity. Then we use the identity
\be
\ln\left[\frac{\vartheta_3(z|\tau)}{\eta(\tau)}\right]\,=\,2\sum_{m=1}^\infty\frac{(-1)^{m+1}}{m}\,q^{\frac{1}{2}m}\frac{\cos 2m z}{1-q^m}\,,\qquad q\,=\,e^{2\pi i \tau}\,.\label{logtheta3}
\ee
Although one may be tempted to set $z= k\pi/n$ at this stage, the sum over $k$ is not well defined in the limit $n\to 1$. To circumvent this, we first perform a modular transformation $\vartheta_3(z|\tau)\,=\,{(-i\tau)^{-1/2}}\,\exp(-i z^2/\pi\tau)\,\vartheta_3(z/\tau|-1/\tau)$, then make use of the identity \eqref{logtheta3} and finally perform the sum over $k$. The exercise is straightforward and explicitly leads to the required condition:
\be
S^{(2)}_{\rm EE}(\Delta=L)\,=\,S^{(2)}_{\rm thermal}\,.
\ee

\subsection{Stress tensor deformation}

The second example we touch upon briefly is the deformation of the CFT by the stress tensor as in eq.\eqref{stressdef}. The corrections to the R\'enyi entropies at order $\varepsilon^2$ are determined by both the one- and two-point functions of the stress tensor on the $n$-sheeted cover of the torus. For simplicity, it suffices to focus explicit attention on the first order correction:
\bea
&&S_{\rm RE}(\Delta, n)\,=\,\frac{1}{1-n}\times\\\nonumber
&&\left[\ln{{\cal Z}^{(n)}_{\rm CFT}}\,-\,n\ln{\cal Z}_{\rm CFT}\,+\,
2\varepsilon\int d^2z\,\left(\frac{\langle\overline\sigma_n(y_2)\,T(z)\,\sigma_n(y_1)\rangle}{\langle\overline\sigma_n(y_2)\sigma_n(y_1)\rangle}\,-\,n\,\langle T(z)\rangle_{{\mathbb T}^2}\right)\ldots
%\right.
%\\\nonumber\\\nonumber
%&&\left.+\,\varepsilon^2\int d^2z_1\int d^2z_2\,
%\left(\frac{\langle\overline\sigma_n(y_2)\,T(z_1)\,T(z_2)\sigma_n(y_1)\rangle}{\langle\overline\sigma_n(y_2)\sigma_n(y_1)\rangle}\,-\,n\,\langle T(z_1) T(z_2)\rangle_{{\mathbb T}^2}\right)\,\ldots
\right]\,.
\eea
Here we have also expanded the deformed CFT partition function on the torus ${\cal Z}$ in powers of the parameter $\varepsilon$ (see eq.\eqref{stressperts}).
In the bosonized, replicated theory, the stress tensor operator is 
\be
T\,=\,-\,\frac{1}{2}\sum_k\sum_{a=1}^M:(\partial\varphi_{a,k})^2:\,\,\,.
\ee 
The quickest way to obtain the one-point function of the stress tensor in the presence of branch-point twist fields is to take the coincidence limit $(z_1\to z_2)$ of the correlator of two U(1)-currents $\partial\varphi_{a,k}(z_1)$ and $\partial\varphi_{a,k}(z_2)$ given in \eqref{jjtwist}. Integration of the first order correction over the torus is achieved by making use of eqs.\eqref{hgint} and \eqref{i21} and the order $\varepsilon$ correction the R\'enyi entropy is 
\bea
&&S^{(1)}_{\rm RE}(\Delta,n)\,=\\\nonumber
&&-\,\frac{\varepsilon M\, (\beta L)\pi^2}{(1-n)L^2}\left[-\frac{(n^2-1)}{12 \,n}\left(\frac{\vartheta_1''\left(\frac{\pi \Delta}{L}\right)}{\vartheta_1\left(\frac{\pi \Delta}{L}\right)}\,+\,E_2\right)
\,+\,\sum_k\left(\frac{\vartheta_\nu''(x)}{\vartheta_\nu(x)}-
\frac{\vartheta_\nu''}{\vartheta_\nu}\right)_{x=\frac{k\pi\Delta}{nL}}\right]
\eea
The terms within the first set of parentheses are periodic under $\Delta \to \Delta +L$  due to the periodicity properties of $\vartheta_1$ whilst this is not the case for the Jacobi-theta functions with $\nu=2,3$.
The stress tensor perturbation should directly correspond to a rescaling of the 
temperature. Indeed, using the heat equation for theta functions, we see that
\be
S^{(1)}_{\rm RE}(\Delta,n)\,=\,-\,(2\varepsilon\pi)\frac{i\beta}{L}\,\partial_\tau\ln\,\langle\overline\sigma_n(y_2)\sigma_n(y_1)\rangle\,.
\ee 
This confirms that the correlators we have computed in the bosonized picture yield the expected results. Since our correction to the R\'enyi entropy is purely the effect of a temperature rescaling in the CFT, the resulting entanglement entropy $(n\to 1)$ is guaranteed to satisfy all necessary checks. In particular it was shown in \cite{Azeyanagi:2007bj} that the CFT partition function ${\cal Z}^{(n)}_{\rm CFT}$ in the presence of the branch point twist fields on the torus reproduces the thermal entropy of the free fermion CFT when $\Delta =L$ and for any temperature. Therefore any rescalings of the temperature (infinitesimal or otherwise) will preserve this relationship. 

The order $\varepsilon^2$ correction to the R\'enyi entropy requires the use of the stress tensor correlator in the presence of twist fields \eqref{stressrenyi} and integrating over the torus. We will not present any further details of this calculation as it is straightforward and the corrections are once again perfectly consistent with the effect of rescaling the temperature at this order.

\section{R\'enyi entropies with spin-three chemical potential}
\label{seven}

We finally turn to the evaluation of the R\'enyi and entanglement entropies for the ${\cal W}_{1+\infty}$ theory with a chemical potential for the spin-three current.
After bosonization, the spin three current in the free fermion CFT is \cite{Pope:1991ig}
\be
W(z)\,=\,-{\tilde \gamma}\,\sum_{a=1}^M:(\partial\varphi_a)^3:\,,\qquad {\tilde \gamma}\,=\,\frac{\sqrt 5}{6\pi}\,,\label{wff}
\ee
in terms of the chiral bosons $\varphi_a$, where the normalization $\tilde\gamma$ has been fixed to match the conventions of \cite{Kraus:2011ds, paper1} for the leading singularity in the $WW$ OPE \eqref{opew3}.

\subsection{The ${\cal O}(\mu^2)$ correction to R\'enyi entropy}
For the calculation of the R\'enyi entropy, the spin-three current in the replica theory is 
\be
W(z)\,=\,-\tilde\gamma\sum_{a=1}^M\sum_{k}:\left(\partial\varphi_{a,k}(z)\right)^3:\,\,.
\ee
Formally, the first correction to the partition sum on the $n$-sheeted torus is set by the one-point function for $W$ i.e. $\langle\overline \sigma_n(y_2) W(z)\sigma_n(y_1)\rangle$. Unlike the stress tensor, however, the operator $W(z)$ does not acquire a one-point function in the presence of twist fields. While there is a non-vanishing correlator $\langle\overline\sigma_{k,n}(y_2)W_k(z)\sigma_{k,n}(y_1)\rangle$ in each sector labelled by the integer $k$, it is an odd function of $k$ and vanishes upon performing the sum over $k$.

The first non-vanishing correction to the R\'enyi and entanglement entropies of the deformed theory appears at order $\mu^2$. This is determined by the correlation function \eqref{WWVV} involving two spin-three current insertions in the presence of the branch point twist fields, integrated over the torus. The steps leading to \eqref{WWVV} are involved and explained in in appendix \ref{appendix:JJVV}. Despite the  lengthy final form of eq.\eqref{WWVV}, a number of terms do not contribute upon integrating over the torus due to the vanishing integrals
\be
\int_{{\mathbb T}^2} d^2 z \, {\cal H}(z)\,=\,0\,,\qquad\qquad\int_{{\mathbb T}^2}d^2z\,G(z)\,=\,0\,.
\ee
Therefore, introducing the shorthand notation,
\be
K^{(p)}(x)\,=\,\left(\frac{\pi}{L}\right)^p\,\frac{\vartheta_\nu^{(p)}(x)}{\vartheta_\nu(x)}\,,
\ee
we can write down all the non-vanishing contributions to the ${\cal O}(\mu^2)$ correction for the R\'enyi entropy:
\bea
&&S_{\rm RE}(\Delta, n)\,=\,\frac{1}{1-n}\left[\ln{\cal Z}^{(n)}_{\rm CFT}-n\ln{\cal Z}\,+\,M\tilde\gamma^2\mu^2\sum_{k}\left\{\int d^2z_1\int d^2z_2
\left(6{\cal H}(z_{12})^3\,+\,
\right.\right.\right.\nonumber\\
\label{w3renyi}\\\nonumber
&&\left.\left.\left.-\,18{\cal H}(z_{12})^2\,\left[\tfrac{k^2}{n^2}\,G(z_1)G(z_2)\,+\,K^{(2)}(x)\right]
\,+\,9{\cal H}(z_{12})\left[\tfrac{k^4}{n^4}\,G(z_1)^2 G(z_2)^2\,+\right.\right.\right.\right.\\\nonumber\\\nonumber
&&\left.\left.\left.\left.+\,\tfrac{2\,k^3}{n^3}\,G(z_1)G(z_2)(G(z_1)\,+\,G(z_2))\,K^{(1)}(x)\,+\,\tfrac{4\,k^2}{n^2}\,G(z_1)G(z_2)\,K^{(2)}(x)\right]\right.\right.\right.\\\nonumber\\\nonumber
&&\left.\left.\left.
%+\,\tfrac{k^4}{n^4}\left(\tfrac{\pi
%^4}{9L^4}E_2^2\,+
%\tfrac{2\pi^2}{3L^2}E_2\,K^{(2)}(x)\,+\,K^{(4)}(x)\right)\right]
\,-\,\tfrac{9\,k^4}{n^4}G(z_1)^2\,G(z_2)^2\,\tfrac{\pi^2}{L^2}\left(\ln\vartheta_\nu(x)\right)''\,+\,\tfrac{3\,k^2}{n^2}\left[G(z_1)^2+G(z_2)^2\right]\times\right.\right.\right.\\\nonumber\\\nonumber
&&\left.\left.\left.\left[K^{(3)}(x)K^{(1)}(x)-K^{(4)}(x)-\tfrac{\pi^4}{L^4}E_2\left\{\ln\vartheta_\nu(x)\right\}''\right]\,+\,\left[K^{(3)}(x)^2 \,-\,K^{(6)}(x)\right]\right.\right.\right.\\\nonumber\\\nonumber
&&\left.\left.\left.+\,\tfrac{2\,\pi^2}{L^2}E_2\left[K^{(3)}(x)K^{(1)}(x)\,-\,K^{(4)}(x)\right]\,-\,\tfrac{\pi^6}{L^6}E_2^2\left(\ln\vartheta_\nu(x)\right)''
\right)\right\}\,+\,{\cal O}(\mu^3)\right]_{x\,=\,\frac{k}{n}(\frac{\pi\Delta}{L})}
\eea
The summation ranges over $k\,=\,-(n-1)/2,-(n-3)/2, \ldots (n-1)/2$.
The first term $\sim H^3$ in the integral above is purely a correction to the thermal partition function at this order and cancels against an identical contribution to $n\ln {\cal Z}$ in the first line. All thermal corrections can  be removed by subtracting out the $WW$ correlator \eqref{ww}. 
 The integrals necessary for the explicit form of this expression have been evaluated in appendix \ref{integrals}. Schematically, there are five distinct integrals that appear above:
\bea
&&{\cal I}_1\,=\,\int_1\int_2 {\cal H}(z_{12})^2\,G(z_1)\,G(z_2)\,,\qquad \I_2\,=\,\int_1\int_2 {\cal H}(z_{12})\,G(z_1)^2\,G(z_2)^2\,,\nonumber\\\\\nonumber
&&
\I_3\,=\int_1\int_2{\cal H}(z_{12})G(z_1)^2G(z_2)\,,\quad
\I_4=\int_1\int_2\H(z_{12})\,G(z_1)\,G(z_2)\,,\quad \I_5=\int G(z)^2\,.
\eea
Of these, ${\cal I}_1$ and $\I_2$ are directly related to the high temperature cylinder limit. They are generalizations to the torus of the integrals encountered in \cite{paper1, paper2} which are responsible for the high temperature universality of the entanglement entropy for ${\cal W}_{\infty}[\lambda]$ theories, at this order in $\mu$. The remaining integrals ${\cal I}_3, {\cal I}_4$ and ${\cal I}_5$ are accompanied by coefficients whose origin can be traced to the winding mode sectors in the bosonized framework. This is also the origin of all the constant terms in the $\langle WW\rangle$ correlator. Technically, the evaluation of ${\cal I}_1$ and ${\I_2}$ is the most involved.
The integrals ${\I_3}$, ${\cal I}_4$ and ${\cal I}_5$ are already contained in ${\cal I}_1$, $\I_2$. The final results for $\I_3$, $\I_4$ and ${\cal I}_5$ are written out in \eqref{I3}, \eqref{I4} and \eqref{I5}.

While the complete expression for the R\'enyi entropy is not particularly illuminating, we write down the explicit forms of the two terms ${\cal I}_1$ and ${\cal I}_2$ which are directly connected to the high temperature cylinder limit:
\bea
{\cal I}_1(\Delta)&&\,=\,\\\nonumber
&&\frac{1}{6}(\beta L)^2\left[\tfrac{1}{20}\wp^{(4)}(\Delta)\,+\,\tfrac{1}{2}\wp^{(3)}(\Delta)\tilde\zeta(\Delta)\,+\,\wp''(\Delta)\,\tilde\zeta^2(\Delta)\,+\,
\tfrac{2\pi^2}{L^2}\,E_2\,\left\{\wp'(\Delta)\,\tilde\zeta(\Delta)\right.\right.\\\nonumber\\\nonumber
&&\left.\left.+\,2\tilde\zeta^2(\Delta)\,\left(\wp(\Delta)\,+\,\tfrac{\pi^2}{3 L^2}E_2\right)\right\}\,-\,\tfrac{2\pi^4}{15 L^4}\,\wp(\Delta)\,\left(5E_2^2\,+\,E_4\right)\,+\right.\\\nonumber\\\nonumber
&&\left.+\tfrac{4\pi^6}{45 L^6}(5E_2^3\,-\,5E_2E_4\,-\,3E_6)\right]
%\,-\,(2\pi {\cal A})\,\left[\tfrac{1}{6}\wp''(\Delta)\,+\,\wp'(\Delta)\,\tilde\zeta(\Delta)\right.\\\nonumber\\\nonumber
%&&\left.+\,\wp(\Delta)\,\tilde\zeta^2(\Delta)\,+\,\tfrac{\pi^2}{3L^2}E_2\,\left(\tilde\zeta^2(\Delta)\,-\,\wp(\Delta)\right)\,-\,\tfrac{\pi^4}{9L^4}\left(2E_4\,-\,E_2^2\right)\right]
\eea
and
\bea
{\cal I}_2(\Delta)&&\,=\,\\\nonumber
&&{(\beta L)}^2\left[-\tfrac{1}{20}\wp^{(4)}(\Delta)\,-\,\tfrac{1}{6}\wp^{(3)}(\Delta)\tilde\zeta(\Delta)\,+\,\wp''(\Delta)\left(\tilde\zeta(\Delta)^2\,+\,\tfrac{2\pi^2}{3 L^2}E_2\right)\right.\\\nonumber\\\nonumber
&&\left.-\,\tfrac{2\pi^2}{3L^2}\,E_2\,\wp'(\Delta)\,\tilde\zeta(\Delta)\,+\,\wp(\Delta)\left\{-4\tilde\zeta(\Delta)^4 \,-\,\tfrac{4\pi^2}{L^2}E_2\tilde\zeta(\Delta)^2-\tfrac{2\pi^4}{15L^4}(5E_2^2-E_4)\right\}\right.\\\nonumber\\\nonumber
&&\left.-\,\tfrac{4\pi^2}{3L^2}E_2\left(\tilde \zeta(\Delta)^4\,+\,\tfrac{\pi^2}{L^2}E_2\tilde\zeta(\Delta)^2\right)\,-\,\tfrac{4\pi^6}{135 L^6}\left(5E_2^3+E_6\right)\right]\,.
%-\,(2\pi{\cal A})\left[-\,\tfrac{1}{12}\wp''(\Delta)\right.
%\\\nonumber\\\nonumber
%&&\left.+\wp(\Delta)\left(\tilde\zeta(\Delta)^2\,+\,\tfrac{2\pi^2}{3L^2}E_2\right)\,-\,\tfrac{1}{2}\tilde\zeta(\Delta)^4\, -\,\tfrac{2\pi^2}{3L^2}E_2\tilde\zeta(\Delta)^2\,-\,\tfrac{2\pi^4}{9L^4}\left(E_2^2\,+\,\tfrac{1}{4}E_4\right)\right]\,.
\eea
These are quasi-elliptic functions of $\Delta$ with modular weight six. The functions $\tilde\zeta(\Delta)$ are only quasi-periodic under $\Delta\to \Delta +i\beta$. Interestingly, they are singly periodic functions of $\Delta$, under the shift $\Delta \to \Delta + L$. .
Therefore $\I_1$ and $\I_2$ vanish both at $\Delta=0$ and $\Delta=L$. The winding mode contributions are responsible for spoiling this periodicity and reproducing the correct physics of entanglement entropy, as we will see later.

%\subsection{Complete expression for RE correction}
%We quote here the complete expression for the ${\cal O}(\mu^2)$ correction to the RE for completeness:
%\bea
%S^{(2)}_{\rm RE}(\Delta, n)\,=\,\frac{1}{(n-1)}\frac{5M}{36\pi^2}\sum_k\left[-18\tfrac{k^2}{n^2}{\I_1}\right]
%\eea

\subsection{High temperature cylinder limits}
The first test that we will subject our result for the 
R\'enyi entropy \eqref{w3renyi} to,  is to take the high temperature limit. The high temperature calculations in \cite{paper1}, later shown to be universal in  \cite{paper2}, demonstrated that the order $\mu^2$ 
R\'enyi entropy corrections are universal for all theories with a ${\cal W}_\infty[\lambda]$ symmetry (for any $\lambda$). The free fermion theory with ${\cal W}_{1+\infty}$ symmetry is somewhat of an exception to this due to the presence of the additional U(1) current. In particular, we saw in section \ref{section:hightw3} that the high temperature thermal corrections in this theory can be empirically separated out into a ${\cal W}_\infty [0]$ piece and the U(1)-current contribution. Our task is to repeat this exercise for the much more complicated R\'enyi entropies.

The R\'enyi entropy correction at order $\mu^2$ is obtained by first removing the $k=0$ piece in \eqref{w3renyi} which is the thermal correction to the partition sum at this order. To extract the high temperature limit $\beta/L\to 0$  we must perform a modular transformation first and then take $L/\beta \to \infty$. Careful examination of this limit reveals that only three terms survive when $L/\beta$ is taken to infinity:
\bea
S^{(2)}_{\rm RE}(\Delta, n)\big|_{\frac{\beta}{L}\to 0}\,=\,\frac{1}{(1-n)}\frac{5M\mu^2}{36\pi^2}&&\sum_k\left[-18\left(\frac{k^2}{n^2}\right)\,\I_1(\Delta)\,+\,9\,\left(\frac{k^4}{n^4}\right)\,\I_2(\Delta)\right.\nonumber\\\\\nonumber
&&\left.-\,6\left(\frac{k^2}{n^2}\right)\,\beta L\,\,\I_5(\Delta)\,
%K^{(3)}(x)K^{(1)}(x) - K^{(4)}(x)
\tfrac{\pi^4}{L^4}E_2\left(\ln\vartheta_\nu(x)\right)''\right]_{\frac{L}{\beta}\to\infty}\,,
\eea
where $x= \frac{k\pi}{n}\frac{\Delta}{L}$. The high temperature limiting forms of the three terms are
\bea
\I_1(\Delta)\to &&\frac{4\pi^4}{3\beta^2}\left(\frac{4\pi\Delta}{\beta}\coth\left(\tfrac{\pi\Delta}{\beta}\right)\,-\,1\right)\,+\,\\\nonumber
&&\,\frac{4\pi^4}{\beta^2}\sinh^{-2}\left(\tfrac{\pi\Delta}{\beta}\right)\left\{\left(1\,-\,\frac{\pi\Delta}{\beta}\coth\left(\tfrac{\pi\Delta}{\beta}\right)\right)\,-\,\left(\tfrac{\pi\Delta}{\beta}\right)^2\right\}
\eea
and
\bea
\I_2(\Delta)\to &&-\,\frac{8\pi^4}{\beta^2}\left(5\,-\,\frac{4\pi\Delta}{\beta}\coth\left(\tfrac{\pi\Delta}{\beta}\right)\right)\,-\,\\\nonumber
&&\,\frac{72\pi^4}{\beta^2}\sinh^{-2}\left(\tfrac{\pi\Delta}{\beta}\right)\left\{\left(1\,-\,\frac{\pi\Delta}{\beta}\coth\left(\tfrac{\pi\Delta}{\beta}\right)\right)\,-\,\frac{1}{9}\left(\tfrac{\pi\Delta}{\beta}\right)^2\right\}\,,
\eea
while the final term surviving in this limit becomes
\bea
(\beta L)\,\I_5(\Delta)\,\tfrac{\pi^4}{L^4}E_2(\ln\vartheta_\nu)''\to
-\frac{8\pi^4}{\beta^2}\left(1\,-\,\frac{\pi\Delta}{\beta}\coth\left(\tfrac{\pi\Delta}{\beta}\right)\right)\,.\label{u1limit}
\eea
The non-trivial limits for $\I_1$ and $\I_2$ coincide\footnote{The definition of $\I_2$ in this paper differs by a sign from that in \cite{paper1}. This is because the quantum propagator ${\cal H}$ as defined in this paper contains an extra minus sign.} with the corresponding objects found directly on the high temperature cylinder  in \cite{paper1, paper2}. This should be viewed as a stringent test of our integration prescription on the torus. In addition to these two terms which lead to the universal entanglement entropy correction in the limit $n\to 1$, we have a new contribution from \eqref{u1limit} whose origin lies in a sum over winding modes of the compact boson.

In this limit, the sum over the replica Fourier index $k$ is rendered trivial. Using the identities
\be
\sum_k\,\frac{k^2}{n^2}\,=\,\frac{n^2-1}{12\,n}\,,
\qquad\sum_k\frac{k^4}{n^4}\,=\,\frac{(n^2-1)(3n^2-7)}{240 \,n^3}\,,
\ee
and taking the limit $n\to 1$ to get the entanglement entropy, we can view the high temperature correction to the entanglement entropy as the sum of two parts,
\be
\lim_{\frac{\beta}{L}\to 0}\,\,S^{(2)}_{\rm EE}\,\Big|_{{\cal W}_{1+\infty}}\,\,=\,\,
S^{(2)}_{\rm EE}\,\Big|_{{\cal W}_\infty}\,+\,\left(\frac{5M\mu^2}{36\pi^2}\right)\frac{8\pi^4}{\beta^2}\left(\frac{\pi\Delta}{\beta}\coth\left(\tfrac{\pi\Delta}{\beta}\right)\,-\,1\right)\,.
\ee
The first term on the right hand side, labelled as the ${\cal W}_\infty$ portion, is universal \cite{paper1, paper2}:
\be
S^{(2)}_{\rm EE}\,\Big|_{{\cal W}_\infty}\,=\,
\frac{5M\mu^2}{12\pi^2}\left[\I_1(\Delta)\,+\,\frac{1}{10}\I_2(\Delta)\right]_{\frac{\beta}{L}\to 0}\,\,.
\ee
The high temperature entanglement entropy needs to satisfy another important test. For large interval size (relative to the inverse temperature), the entanglement entropy must be extensive in the interval length $\Delta$, and approach the thermal entropy. The high temperature limit of the thermal entropy correction for free fermions inferred from eq.\eqref{F2ff} is,
\be
S^{(2)}_{\rm thermal}\Big|_{{\cal W}_{1+\infty}}\,\to\,
M\mu^2\frac{14\pi^3\,L}{3\,\beta^3}\,(1\,+\,{\O}\left(e^{-2\pi L/\beta}\right))\,.
\ee
The corresponding large interval limit, $\Delta/\beta\gg 1$, for the entanglement entropies yields,
\be
S^{(2)}_{\rm EE}\,\Big|_{{\cal W}_{1+\infty}}\,\to\,
M\mu^2\frac{14\pi^3\,\Delta}{3\,\beta^3}\,,\qquad\qquad
S^{(2)}_{\rm EE}\,\Big|_{{\cal W}_{\infty}}\,\to\,
M\mu^2\frac{32\pi^3\,\Delta}{9\,\beta^3}\,.
\ee
This precisely mirrors the behaviour of the thermal entropies from \eqref{winftyvs}. It is quite remarkable that {\it a priori} very different and complicated looking contributions, from the perturbative and winding sectors in the bosonized theory, conspire to give rise to the correct high temperature behaviour as expected on physical grounds. 

Equally noteworthy is the fact that, unlike the thermal calculation we have seen earlier, for the high temperature RE/EE  on the torus it is not so straightforward to interpret the difference between ${\cal W}_{1+\infty}$ and ${\cal W}_\infty$ as a shift involving the U(1)-current. Such an interpretation may still possible given the origin of the term \eqref{u1limit} as it can be related to an operator of the type $\sim (\partial\varphi_{k,\,{\rm cl}})^2 \langle T_k\rangle^2$, but it is unclear at the moment how this connects up with a direct calculation on the high temperature cylinder. In particular, if we had taken the cylinder limit {\it before} doing the integrals, at the level of the correlation function of the $W$-currents \eqref{WWVV}, we would have found all winding mode contributions to be suppressed leaving behind terms of the type, ${\cal H}(z_{12})^2 G(z_1) G(z_2)$, ${\cal H}(z_{12})G^2(z_1) G^2(z_2)$ and ${\cal H}(z_{12}) G^2(z_1)$. The first two are the result of standard Wick contraction rules (on the plane or the cylinder) between two $W$-currents and twist fields considered in \cite{paper1}. The third contribution $\sim{\cal H}G^2$ can be related to the correlator of the spin-three and spin-one currents $\langle W J\rangle$ in the ${\cal W}_{1+\infty}$ theory. Such a cross term arises naturally and is the only new contribution to the R\'enyi entropy at this order in $\mu$ if the spin-three current is shifted in accordance with  
eq.\eqref{Wtransform}.  Indeed, this interpretation works in a very precise sense and calculation of the relevant integrals on the cylinder yields exactly \eqref{u1limit} as the new contribution to EE/RE. This is remarkable because the integral of the term ${\cal H} G^2$ on the torus vanishes identically. Thus there is a subtle sense in which the large $L$ (or high temperature) limit does not commute with integration, and yet gives rise to the same final result. It would be worth understanding this somewhat mysterious behaviour better.

\subsection{Matching thermal entropy at $\Delta=L$}
Perhaps the strongest test of the RE/EE calculation comes from the requirement that when the entangling interval spans the entire system, the entanglement entropy should precisely equal the thermal entropy. It is already known to be true explicitly at the conformal point \cite{Azeyanagi:2007bj}. Verifying this property for the higher spin correction \eqref{w3renyi} appears to be an arduous task at first sight. However, we encounter a significant simplification due the fact that when $y_2-y_1\,=L$, all terms in the correlator that depend on the function $G(z)$ vanish identically since
\be
G(z)\Big|_{y_2=y_1+L}\,=\,\tilde\zeta(z-y_1)\,-\,\tilde\zeta(z-y_1-L)\,=\,0.
\ee
The R\'enyi entropy correction at order $\mu^2$, with 
$\Delta =L$, can then be represented relatively compactly after integration on the torus:
\bea
&&S^{(2)}_{\rm RE}(\Delta,n)\Big|_{\Delta=L}\,=\,\frac{\mu^2 M}{(1-n)}\,\frac{80\pi^4 \beta^2}{L^4}\,\sum_{k}\,
\left[\frac{1}{2^5\cdot 3^2}\left(E_2^2-E_4\right)\left(\frac{\vartheta_\nu''(x)}{\vartheta_\nu(x)}\,-\,\frac{\vartheta_\nu''}{\vartheta_\nu}\right)\right.\nonumber\\\
\label{renyiL}
\\\nonumber
&&\left.-\,\frac{1}{2^6\cdot 3^2}\left\{E_2^2\left(\left[\ln\vartheta_\nu(x)\right]'' \,-\,\frac{\vartheta_\nu''}{\vartheta_\nu}\right)
\,+\,2E_2\left(\frac{\vartheta_\nu^{(4)}(x)}{\vartheta_\nu(x)}\,-\,\frac{\vartheta_\nu^{(3)}(x)}{\vartheta_\nu(x)}\frac{\vartheta_\nu'(x)}{\vartheta_\nu(x)}
\,-\,\frac{\vartheta_\nu^{(4)}}{\vartheta_\nu}\right)
\right.\right.\\\nonumber\\\nonumber
&&\left.\left.\,+\,\frac{\vartheta_\nu^{(6)}(x)}{\vartheta_\nu(x)}\,-\,\left(\frac{\vartheta_\nu^{(3)}(x)}{\vartheta_\nu(x)}\right)^2 \,-\,\frac{\vartheta_\nu^{(6)}}{\vartheta_\nu}\right\}\right]_{x=\frac{k\pi}{n}}\,.
\eea
Note that Jacobi-theta functions without a specified argument should be understood as being evaluated at vanishing argument. We emphasize that the RE at $\Delta=L$ is completely determined by  contributions from the winding mode sectors of the compact chiral bosons. Their presence, captured by the theta-functions, is precisely what guarantees that the RE (for any $\Delta$) is {\em not} a periodic function of $\Delta$ under $\Delta \to \Delta +L$.
Had we not included the winding modes we would have found vanishing R\'enyi entropy and entanglement entropy at $\Delta = L$ which would not be consistent with thermodynamical expectations. 

The thermal entropy correction can be obtained from the free energy correction eq.\eqref{fermionF} or from its elegant $q$-expansion \eqref{qexpff1} as
\be
S^{(2)}_{\rm thermal}\,=\,\beta^2\frac{\partial F^{(2)}}{\partial\beta}\,,\qquad -\beta F^{(2)}\,=\, 160 M\mu^2\beta^2\frac{\pi^4}{L^4}\sum_{m=1}^\infty \frac{m^4 q^m}{(1+q^m)^2}\,.
\ee
Demonstrating the matching of the entire $q$-expansion with the $n\to 1$ limit of the R\'enyi entropy at $\Delta=L$ is an interesting and challenging exercise that we leave for the future. We can nevertheless demonstrate the extremely non-trivial agreement  between $S^{(2)}_{\rm thermal}$ and $S^{(2)}_{\rm EE}(\Delta=L)$ at high temperature.

The high temperature limit of eq.\eqref{renyiL} is obtained as usual after a modular transform and then taking $L/\beta\to \infty$. Up to exponentially suppressed terms, we find
\be
S^{(2)}_{\rm RE}(L,n)\Big|_{\frac{\beta}{L}\to 0}
\,=\,M\mu^2\,\frac{7\pi^3\,L}{6 \beta^3}\frac{(1-n^4)}{(1-n)n^3}\,.
\ee
In the limit $n\to 1$ we then have
\be
S^{(2)}_{\rm EE}(L)\Big|_{\frac{\beta}{L}\to 0}
\,=\,M\mu^2\,\frac{14\pi^3\,L}{3 \beta^3}
\ee
which agrees precisely with the high temperature thermal entropy of the ${\cal W}_{1+\infty}$ theory.

\section{Discussion}
\label{eight}

In this paper we have performed an analytical computation of the perturbative corrections to the single interval 
R\'enyi and entanglement entropies (RE/EE) for the free fermion CFT on the torus, deformed by a chemical potential for the spin-three current. At quadratic order in the chemical potential our computation captures all finite size corrections to the higher spin entanglement entropy for this CFT. The RE and EE are controlled by quasi-elliptic functions of the interval length, which we have determined in closed form. 
The two main ingredients that made the calculation possible are -- the spin-three current correlator in the presence of branch-point twist fields on the torus, and a prescription for integrating holomorphic correlators over the torus. Both these ingredients have been shown to satisfy several consistency checks.

A significant outcome of our calculation is a detailed understanding of the intricate interplay between perturbative and winding mode contributions to higher point correlation functions (involving twist fields) in the bosonized description of free fermions. In fact, we have seen that internal consistency requires inclusion of both contributions. Discarding the winding modes on the torus leads to a formula for EE that does not reduce to the thermal entropy when the interval in question spans the system. It would be interesting to understand whether this feature manifests itself within the context of large-$N$ Yang-Mills theory on the torus whose large-$N$ expansion has been argued to be equivalent to the  ${\cal W}_{1+\infty}$ theory with spin-three deformation, but without inclusion of winding modes \cite{douglas,dijkgraaf}.

Our calculation was motivated primarily by higher spin holographic duality, and existing proposals for entanglement entropies therein, between ${\cal W}$-algebra CFTs and Vasiliev's hs$[\lambda]$ theories. Since higher spin holography is a duality between systems that can both be tractable (or weakly coupled) simultaneously, perturbative results in the the QFT are intrinsically valuable.

Our explicit results for thermal corrections in the Lagrangian formulation for the free boson theory (on the torus) with spin-three deformation agree with the results of \cite{Kraus:2011ds}. These can be viewed as predictions of finite size effects for black hole solutions carrying spin three charge in hs$[1]$ theory. Furthermore, it should be possible to carry out RE/EE calculations for the free boson theory by extending the results of \cite{Atick:1987kd,Datta:2013hba} to obtain the two point function of spin-3 currents in presence of twist operators on the torus for this
theory. Then  performing the integrals over the torus  on this correlator  would lead to results for the R\'{e}nyi entropy for the $\mathcal{W}_{\infty}[1]$  theory. 

For the free fermions on the torus,  understanding how to extract ${\cal W}_\infty[0]$ answers from the ${\cal W}_{1+\infty}$ results on the torus holds the key to making predictions for holographically dual higher spin black holes. At present, we have been able to see how to do this at high temperatures (the cylinder limit) for both EE and thermal entropy allowing us to match with the universal high temperature results for ${\cal W}_\infty[\lambda]$ theories in \cite{paper1, paper2}. It should be possible to do this consistently at any temperature.

Finally, both spin-one (the U(1)-current) and spin-two (stress tensor) chemical potentials have  natural interpretations in terms of transformations acting on the original partition function -- the first as a type of `twisting' or modification of boundary conditions around the thermal circle, and the second as a temperature rescaling. It would be interesting to see if our results for the RE and thermal entropy can be interpreted as some transformation on the respective partition functions due to the spin three chemical
potential.

\acknowledgments

We thank Michael Ferlaino for discussions and collaboration during the early stages of the project. We would also like to thank Tim Hollowood for enjoyable discussions and insights on various aspects of this work.
JRD thanks the warm hospitality of the High Energy Section, ICTP, Trieste
during
which part of this work was done.
He also thanks
Dieter Lust and the string theory group at  Max-Planck-Institut f\"{u}r
Physik, Munich
for hospitality and the stimulating environment during which this work was
completed.
 SPK acknowledges financial support from U.K. Science and Technology Facilities Council (STFC) under the grants ST/J000043/1 and ST/L000369/1. The work of JRD is partially supported by the
Ramanujan fellowship DST-SR/S2/RJN-59/2009.

\newpage
\appendix
\section*{Appendix}

\section{${\cal W}$-algebra OPEs} 
\label{walgebra}
We list OPEs involving the stress tensor $T$ and the spin-three current $W$, which are universal for any ${\cal W}$-algebra. These particular OPEs (importantly, the $WW$ OPE) are independent of $\lambda$ for ${\cal W_\infty[\lambda]}$ (see e.g. appendix D of 
\cite{Gaberdiel:2012yb})
\bea
&&T(z)T(w)\sim \frac{c/2}{(z-w)^4}+\frac{2 T(w)}{(z-w)^2}+\frac{T'(w)}{(z-w)}\,,\nonumber\\\nonumber
&& T(z)W(w)\sim\frac{3}{(z-w)^2}W(w)+\frac{1}{z-w}W'(w)\,,\\\nonumber
&&W(z)W(w)\sim\frac{2c/3}{(z-w)^6}+\frac{4T(w)}{(z-w)^4}+\frac{2T'(w)}{(z-w)^3}+\frac{4U(w)+\tfrac{3}{5}T''(w)}{(z-w)^2}+\\
&&\qquad+\frac{2U'(w)+
\tfrac{2}{15}T'''(w)}{(z-w)}\,.\label{opew3}
\eea
Here $U$ is the spin-4 current. For the ${\cal W}_3$ algebra, $U$ is replaced by a composite operator since there is no spin-4 current:
\be
U\to\frac{16}{22+5c}\left(:TT: -\tfrac{3}{10}\,T''\right)\,.
\ee

\section{Some properties of elliptic functions and modular forms}
\label{app:elliptic}
We provide some details of the properties of elliptic, quasi-elliptic functions and modular forms that are useful for our calculations. For a more complete treatment we refer the reader to \cite{ww} and \cite{koblitz}. An elliptic function is a function on the complex plane, periodic with two periods $2\omega_1$ and $2\omega_2$. Defining the lattice $\Gamma\,=\,2\omega_1{\mathbb Z}\oplus2\omega_2{\mathbb Z}$ and the basic period parallelogram as
\be
{\cal D}\,=\,\left\{z=2\mu\omega_1+2\nu\omega_2\,|\,\mu,\nu\in[0,1)\right\}\,,
\ee
the Weierstrass $\wp$-function is analytic in ${\cal D}$ except at $z=0$, where it has a double pole. ${\wp}(z)$ is an even function of $z$, defined via the sum
\be
\wp(z;\omega_1,\omega_2)\,=\,\frac{1}{z^2}+\sum_{(m,n)\neq(0,0)}\left\{\frac{1}{(z-2m\omega_1-2n\omega_2)^2}-\frac{1}{(2m\omega_1+2n\omega_2)^2}\right\}\,.
\ee
\begin{itemize}
\item{The Weierstrass $\wp$-function satisfies
\be
\wp^{\prime}(z)^2\,=\,4\wp(z)^3-g_2\wp(z)-g_3\,;\qquad
g_2\,=\,60\sum\Omega_{mn}^{-4}\,,\quad
g_3\,=\,140\sum\Omega_{mn}^{-6}\,,
\ee
where $\Omega_{mn}=2m\omega_1+2n\omega_2$ with $(m,n)\neq (0,0)$ and $g_2, g_3$ are the {\em Weierstrass invariants}.} 

\item{Under a modular transformation, of the complex structure parameter $\tau\,=\,\omega_2/\omega_1$ of the torus, the $\wp$-functions transforms with weight two:
\be
\wp\left(\frac{z}{\tau};\, \omega_1,\,-\frac{\omega_1}{\tau}\right)\,=\,
\tau^2\,\wp(z;\, \omega_1,\,\tau\omega_1)\,.\label{pmodular}
\ee}
\item{The following formula can be used to infer the behaviour of $\wp(z)$ in the limit
${\rm Im}(\tau)\gg 1$ \cite{ww}
\be
\wp(z;\omega_1,\omega_2)\,=\,\left(\tfrac{\pi}{2\omega_1}\right)^2\left[-\tfrac{1}{3}\,+\,\sum_{n=-\infty}^\infty{\rm csc}^2\left(\tfrac{z-2n\omega_2}{2\omega_1}\pi\right)\,-\,\sum_{n=-\infty}^{\infty\prime}{\rm csc}^2\left(\tfrac{n\omega_2}{\omega_1}\pi\right)\right].\label{plowt}
\ee}
\item{The Weiserstrass ${\wp}$-function has a Laurent series expansion around $z=0$ with modular forms as coefficients:
\be
{\wp}(z)\,=\,\frac{1}{z^2}\,+\,2\sum_{k=1}^\infty (2k+1)\,\frac{\zeta_{\rm R}(2k+2)\,E_{2k+2}(\tau)}{(2\omega_1)^{2k+2}}\,z^{2k}\,,\label{laurent}
\ee
where $\zeta_R$ is the Riemann-zeta function.
}
\item{The Weierstrass $\zeta$-function is a quasi-elliptic function, analytic in ${\cal D}$ but with a simple pole at $z=0$ 
\bea
&&\wp(z)\,=\,-\zeta'(z)\,,\qquad\qquad\zeta(z+2\omega_{1,2})\,=\,\zeta(z)+2\zeta(\omega_{1,2})\label{zetadef}\label{zetasigmadef}\\\nonumber\\\nonumber
&&\zeta(z)\,=\,\frac{d}{dz}\ln\sigma(z)\,,\qquad\qquad\sigma(z+2\omega_{1,2})\,=\,-\sigma(z)\,e^{2\zeta(\omega_{1,2})(z+2\omega_{1,2})}
\eea
The $\sigma$-function is also only quasiperiodic, with a simple zero at $z=0$. Both $\sigma(z)$ and $\zeta(z)$ are odd under $z\to -z$.}

\item{A slight modification renders the Weierstrass $\zeta$-function periodic along one of the periods of the torus ${\mathbb C}/\Gamma$
\be
\tilde\zeta(z)\,=\,\zeta(z)\,-\,\frac{\zeta(\omega_1)}{\omega_1}\,z\,.
\label{zetatdef}
\ee
The new function is periodic along the A-cycle, but only quasiperiodic along the B-cycle.
\be
\tilde\zeta(z+2\omega_1)\,=\,\zeta(z)\,,\qquad\zeta(z+2\omega_2)\,=\,
\zeta(z)-\frac{i\pi}{\omega_1},
\ee
where we have used the identity
\be
\omega_2\zeta(\omega_1)-\omega_1\zeta(\omega_2)\,=\,i\pi\,.
\ee}
\item{The quasiperiodic function $\tilde\zeta(z)$ is the logarithmic derivative of the Jacobi theta function
\be
\tilde\zeta(z)\,=\,\frac{\pi}{2\omega_1}\,\frac{\vartheta_1'\left(\frac{\pi z}{2\omega_1}\right)}{\vartheta_1\left(\frac{\pi z}{2\omega_1}\right)}\,.
\label{logtheta}
\ee
}
\item{The Jacobi-theta functions satisfy the heat equation:
\be
\frac{\partial^2\vartheta_\nu(u|\tau)}{\partial u^2}\,=\,-\frac{4}{i\pi}\frac{\partial\vartheta_\nu(u|\tau)}{\partial\tau}\,.\label{heateq}
\ee
}
\item{A useful relation between the Weierstrass $\wp$-function and Jacobi theta functions can be obtained using
\be
\wp(z)\,-\,e_r\,=\,\frac{\sigma_r(z)^2}{\sigma(z)^2}\,,\qquad e_{r}\,=\,\wp(\omega_r)\,,\qquad\sigma_r(z)\,=\,\frac{e^{-\eta_r z}\sigma(z+\omega_r)}{\sigma(\omega_r)}\,,
\label{wpandtheta}
\ee
where $r=1,2,3$ and $\omega_3 = \omega_1+\omega_2$ and $\eta_r=\zeta(\omega_r)$\,.
}
\end{itemize}
{\em Representation of any elliptic function in terms of the $\zeta$-function and its derivatives \cite{ww}:}
Let $f(z)$ be any elliptic function on ${\cal D}$. Let $a_1, a_2\ldots a_n$ be isolated poles of the function in ${\cal D}$ and let the singular terms near $z=a_k$ be
\be
f(z)\,\simeq \,\sum_{r=1}^{r_k}\frac{c_{k,r}}{(z-a_k)^r}\,+\,\ldots
\ee
Then it can be shown that
\bea
f(z)\,=\,C\,+\,\sum_{k=1}^n\left(\sum_{s=1}^{r_k}\frac{(-1)^{s-1}\,c_{k,s}\,\zeta^{(s-1)}(z-a_k)}{(s-1)!}\right)\,,\label{zetaexp}
\eea
provided also that $\sum_k c_{k,1}\,=0\,$ i.e. the sum of simple pole residues must vanish.

This is a particularly useful theorem which allows us to integrate elliptic functions along the cycles of the torus ${\mathbb C}/\Gamma$. The theorem can be used to deduce the following set of identities:
\bea
&&\wp(z)^2\,=\,\tfrac{1}{6}\wp''(z)\,+\,\tfrac{g_2}{12}\,,
\label{psquared}\\\nonumber\\\
&&\wp(z-a)\,\wp(z-b)\,=\,\wp(b-a)\,\left\{\wp(z-a)+\wp(z-b)-\wp(a)-\wp(b)\right\}\,+\label{wpproduct1}\\\nonumber
&&\qquad\qquad\,\wp'(b-a)\,\left\{\zeta(z-b)-\zeta(z-a)+\zeta(b)-\zeta(a)\right\}\,+\,\wp(a)\wp(b)\,.\\\nonumber\\
&&\wp(z)\wp(z-a)\,=\,\wp(a)\left\{\wp(z)+\wp(z-a)\right\}\,+\,\wp'(a)\{\zeta(z-a)-\zeta(z-b)\}+\label{wpproduct2}\nonumber\\
&&\qquad\qquad\qquad\quad\zeta(a)\wp'(a)+\tfrac{1}{3}\wp''(a)-\tfrac{\pi^4}{9L^4}E_4(\tau)\,,\\\nonumber\\\nonumber
&&\wp(z)\left(\zeta(z-a)-\zeta(z-b)\right)\,=\,\wp(b)\left\{\zeta(z)-\zeta(z-b)\right\}+\wp(a)\left\{\zeta(z-a)-\zeta(z)\right\}
\label{pzetaproduct}\\\nonumber
&&+\wp(z)\{\zeta(b)-\zeta(a)\}+\wp(a)\zeta(a)-\wp(b)\zeta(b)+
\tfrac{1}{2}\left\{\wp'(a)-\wp'(b)\right\}\,.
\eea
\subsection{The Eisenstein series }
There are a number of ways to introduce the Eisenstein series (see \cite{koblitz})
\bea
E_k(\tau)\,=\,\frac{1}{2}\sum_{\substack{m,n \in{\mathbb Z}\\(m,n)=1}}\frac{1}{(m\tau+n)^k}\,,\qquad\qquad{\tau}=\frac{\omega_2}{\omega_1}
\eea
where $\tau$ is the complex structure parameter of the torus defined by ${\mathbb C}/\Gamma$ and $(m,n)$ denotes the greatest common divisor. Each series has a $q$-expansion
\bea
&&E_2(\tau)\,=\,1-24\sum_{n=1}^\infty\sigma_1(n)\,q^n\,,\qquad q=e^{2\pi i \tau}
\label{qexpansion}\\\nonumber
&&E_4(\tau)\,=\,1+240\sum_{n=1}^\infty\sigma_3(n)\,q^n\,,\\\nonumber
&&E_6(\tau)\,=\,1-504\sum_{n=1}^\infty\sigma_5(n)\,q^n\,,
\eea
where $\sigma_j(n)$ is a sum over each positive integral divisor of $n$ raised to the $j^{\rm th}$ power. Under the S-transformation $\tau\to -1/\tau$, the modular forms with the exception of $E_2(\tau)$, transform covariantly with a specific modular weight
\be
E_k(-1/\tau)=\tau^k\,E_k(\tau)\qquad k\geq 4\,,\qquad E_2(-1/\tau)=\tau^2E_2(\tau) +\frac{6\tau}{i\pi}\,.\label{modularE2}
\ee
The anomalous transformation of $E_2(\tau)$ can be fixed by a shift.
\be
\widehat E_2(\tau,\bar\tau) = E_2(\tau)-\frac{3}{\pi {\rm Im}(\tau)}\,.
\ee
This is a modular form of weight two, although it is not holomorphic. Further useful relations include
\be
\zeta(\omega_1)=\frac{\pi^2}{12\omega_1} E_2(\tau)\,,\qquad
g_2=\frac{4\pi^4}{3(2\omega_1)^4}E_4(\tau)\,,\qquad
g_3\,=\,\frac{8\pi^6}{27 (2\omega_1)^6}\,E_6(\tau)\,.
\ee
\section{Twist field correlators on the torus}
\label{appendix:JJVV}
Correlation function involving twist fields on the torus are somewhat subtle due to the presence of winding mode sectors \cite{DiFrancesco:1997nk}. The reason for this is that bosonization relates a free fermion field $\psi$ to a compact chiral boson $\varphi$ with unit radius in our convention:
\be
\psi\,=\,:e^{i\varphi}:
\ee
The free boson partition function in the presence of twist and antitwist fields, can be formally written  as
\bea
{\cal Z}\,=\,\int [d\varphi][d\bar\varphi]\,\exp\left({-S[\varphi,\bar\varphi]\,+\,\frac{ik}{n}\left(\varphi(y_1)-\varphi(y_2)\right)\,-\,\frac{ik}{n}\left(\bar\varphi(\bar y_1)-\bar\varphi(\bar y_2)\right)}\right)\label{twistaction}
\eea
The chiral and anti-chiral bosons are together  packaged into the boson $\Phi$:
\be
\Phi(z,\bar z)\,=\,\left(\varphi(z)\,-\bar\varphi(\bar z)\right)\,,\qquad
S\,=\,\frac{1}{2\pi}\int d^2z\,\partial\Phi\bar\partial\Phi\,.
\ee
Since $\Phi$ is compact, it contains a non-perturbative classical piece characterized by the winding number around the target space circle. Taking $z=x+iy$ and the torus periods to be $2\omega_1= L$ and $2\omega_2=i\beta$ with $\tau=i\beta/L$, we can split the field into a classical and quantum part. The classical portion contains the winding modes and the background field generated by the point charge sources on the torus due to the insertion of the twist operators:
\bea
&&\Phi(z,\bar z)\,=\,\Phi_{\rm cl}\,+\,\Phi_{\rm qu}\,,\qquad
\Phi_{\rm cl}\,=\,\Phi_{w,w'}\,+\,\Phi_{\rm source}\\\nonumber\\
&&\Phi_{w,w'}\,=\,\frac{2\pi}{L}w\,x\,+\,\frac{2\pi}{\beta}w'\,y\,,\qquad w,w'\in{\mathbb Z}\\\nonumber\\
&&\Phi_{\rm source}(z)\,=\,\frac{ik}{n}\,\ln\left|\frac{\vartheta_1\left(\tfrac{\pi}{L}(z-y_1)\right)}{\vartheta_1\left(\tfrac{\pi}{L}(z-y_2)\right)}\right|^2\,.
\eea
Substituting the classical solutions into the action and summing over the winding numbers $w,w'$ (after careful Poisson resummation over $w'$), the twist field correlator on the square torus is 
\bea
&&\sigma_{k,n}\,=\,:e^{ik\Phi/n}:\,,\qquad \bar\sigma_{k,n}\,=\,:e^{-ik\Phi/n}:\\\nonumber
&&\langle\sigma_{k,n}(y_1)\,\bar\sigma_{k,n}(y_2)\rangle\,=\,\left|\frac{\vartheta_1'(0)}{\vartheta_1\left(\frac{\pi\Delta}{L}\right)}\right|^{2k^2/n^2}\,\left|\frac{\vartheta_\nu\left(\frac{k}{n}\frac{\pi\Delta}{L}\right)}{\vartheta_\nu(0)}\right|^2\,.
\eea
When the fermions are periodic around the spatial circle we take $\nu=2$, whilst when they satisfy anti-periodic boundary conditions we must set $\nu=3$. 

The primary object of our interest is the correlator of currents in the free fermion theory in the presence of twist fields. The correlators of holomorphic currents in the bosonized theory are built from the two-point function of the U(1) current operator $\partial\varphi_k$, which can be split in the manner indicated above, into three pieces
\be
\partial\varphi_k(z)\,=\,\partial\varphi_{k;{w,w'}}\,+\,\partial\varphi_{k,{\rm source}}\,+\,\partial\varphi_{k,{\rm qu}}\,.\label{expansion1}
\ee 
The quantum fluctuation $\varphi_{k,qu}$ has vanishing one-point function. Its two-point function for operators 
located at two different points $z_1, z_2$ is  given 
by 
\be\label{qcore}
\langle\partial\varphi_{k, {\rm qu}} (z_1)\partial\varphi_{k, {\rm qu} } (z_2) \rangle
=  {\cal H} (z_1 - z_2)  + \frac{\pi}{L^2 {\rm Im} \tau } 
\ee
Note here the presence of the $1/{\rm Im} \tau$ term, which originates from the background negative charge on the 
torus (\ref{gfsink}).  
However if the location of the operators $\varphi_{k,qu}$ are coincident one has to allow for self contractions. 
This is due to the fact that the normal ordering constant does not vanish on the torus and must be taken into account. 
Therefore we have
\be\label{qcore1}
\langle\partial\varphi_{k, {\rm qu}} (z_1)\partial\varphi_{k, {\rm qu} } (z_1) \rangle
= -\frac{\pi^2 E_2}{3L^2} + \frac{\pi}{L^2 {\rm Im} \tau}
\ee
Note that this is the constant term that occurs in the two point function in (\ref{qcore}). 
Self contractions do not have any singular terms since those have been already subtracted while 
defining the operator. 
The first two pieces  in (\ref{expansion1})
are entirely classical in the sense described previously, and are given by 
\bea
&&\partial\varphi_{k,{\rm source}}(z)\,=\,\frac{ik}{n}\partial_z\left(\ln\frac{\vartheta_1\left(\tfrac{\pi}{L}(z-y_1)\right)}{\vartheta_1\left(\tfrac{\pi}{L}(z-y_2)\right)}\right)\,=\,\frac{ik}{n}\left(\tilde\zeta(z-y_1)-\tilde\zeta(z-y_2)\right)\nonumber\\\nonumber\\
&&\partial\varphi_{k;w,w'}\,=\,-i\frac{\pi}{L}\left(\frac{w'-w\bar \tau}{{\rm Im}(\tau)}\right) \,,
\eea
In the winding number sector with integers $(w,w')$, the classical effective action in the presence of twist operators (the exponent of eq.\eqref{twistaction})
\bea
&&\tilde S\,=\,-\frac{1}{2{\rm Im}(\tau)}|w'-w\tau|^2\,+\,\frac{2\pi k}{n}{\rm Re}\left(y_{12}\frac{w'-w\bar\tau}{{\rm Im}\tau}\right)
-\,\frac{k^2}{n^2}\ln\left|\frac{\vartheta_1(y_{12})}{\vartheta_1'(0)}\right|^2\\\nonumber
&&-\,\frac{2\pi\, k^2}{n^2\, {\rm Im}\tau}\left({\rm Re}\, y_{12}\right)^2\,.
\eea
where $y_{12}\,=\, y_1-y_2$.
The insertion of $\partial\varphi_{k;w,w'}$, into the partition can be re-expressed in terms of a derivative of this classical action (with respect to $y_{12}$), so that  the one-point function of the full operator $\partial\varphi_k$ in the presence of the twist fields is  
\be
\frac{\langle\partial\varphi_k(z)\,\sigma_{k,n}(y_1)\,\bar\sigma_{k,n}(y_2)\rangle}{\langle\sigma_{k,n}(y_1)\,\bar\sigma_{k,n}(y_2)\rangle}\,=\,\frac{ik}{n}G(z)\,-\,\frac{i\pi}{L}
\frac{\vartheta_\nu'\left(x\right)}{\vartheta_\nu(x)}\Big|_{x=\frac{k}{n}\frac{\pi\Delta}{L}}\,.
\ee
This expression is obtained after performing the sum over all winding number sectors (accompanied by Poisson resummation as an intermediate step).
Here 
\be
G(z)\,=\,\tilde\zeta(z-y_1)\,-\tilde\zeta(z-y_2)
\ee
which is periodic under the shift $y_2 \to y_2 + 2\omega_1$ with $2\omega_1=L$.

The two-point correlator of two U(1) currents is also computed similarly by splitting each current insertion into quantum and classical pieces and then performing the sum over the winding number sectors. We find,
\bea
&&\frac{\langle\partial\varphi_k(z_1)\partial\varphi_k(z_2)\,\sigma_{k,n}(y_1)\,
\bar\sigma_{k,n}(y_2)\rangle}{\langle\sigma_{k,n}(y_1)\,\bar\sigma_{k,n}(y_2)\rangle}\,=\,\left({\cal H}(z_{12})\, + \frac{\pi}{L^2 {\rm Im} \tau} \right. \nonumber
+\,\left(\frac{ik}{n}\right)^2\,G(z_1)\,G(z_2)\nonumber\\\nonumber\\
&&\left.\,+\,\frac{ik}{n}(G(z_1)+G(z_2))\left(\frac{-i\pi}{L}\right)\frac{\vartheta_\nu'\left(x\right)}{\vartheta_\nu(x)}\,+\,\left(\frac{-i\pi}{L}\right)^2\frac{\vartheta_\nu''\left(x\right)}{\vartheta_\nu(x)} - \frac{\pi}{L^2 {\rm Im} \tau} \right)_{x=k\pi\Delta/nL}  \,. 
\eea
A remarkable feature of this result is that all  factors of $1/{\rm Im}\,\tau$ cancel. 
The first line of the above equation has  such a factor due to 
 correlator for the quantum fluctuations  given in (\ref{qcore}). 
 The last line also contains a factor of $1/{\rm Im}\,\tau$ which arises  when terms of the kind 
 $(\partial\varphi_{k; w, w'} )^2 $ are re-expressed in terms of the derivatives of the classical action due to the winding
 modes. 
 It is noteworthy that such factors of $1/{\rm Im}\tau$ consistently cancel between the quantum contributions and the classical winding number sectors. We will continue to observe such cancellations for the two point functions of the 
 stress tensor as well as that of the spin-3 current in presences of the twists.  We observe that the above correlator 
 at $k=0$ reduces to the two point function of the $U(1)$ currents  evaluated using the free fermion description of the 
 theory  given in (\ref{jj})  as expected .
 We can use this result to evaluate 
 the U(1) correlator in the presence of twist operators which appears in the  correction to the R\'{e}nyi entropy
 due to a perturbation by a $U(1)$ current. 
 This is given by 
 \bea
&&\frac{ \langle 
 \sum_{k, k'} \left( \partial\varphi_k(z_1)\partial\varphi_k'(z_2)\right) \, \prod_{k, k'} \sigma_{k,n}(y_1)\,\bar\sigma_{k',n}(y_2)\rangle}{\prod_{k,k'}  \langle\sigma_{k,n}(y_1)\,\bar\sigma_{k',n}(y_2)\rangle}\,= \label{jjtwist}\\ \nonumber
 &&\hspace{2.5in}\sum_k \left[  {\cal H}(z_{12})   - \frac{\pi^2}{L^2} \left\{ \ln\vartheta_\nu(x) \right\}''_{x=\frac{k\pi\Delta}{nL}}
 \right]
 \eea
 Here the sum over $k, k'$ runs over the values $k, k' = -\frac{1}{2}(n-1) , - \frac{1}{2}( n-3) , \cdots \frac{1}{2}(n-1) $. 
 In obtaining the above answer we have also set all the terms which involve sums of odd functions in $k$ to zero. 
 Note that the last term in the above equation arises from a careful consideration of the $k=k'$ term in the double sum. 

 We now consider the two point function of the stress tensor in presence of the twist operators. Defining
 \be
\widehat {\cal H }(z)\,:=\,{\cal H}( z )\,  +\, \frac{\pi}{ L^2 {\rm Im} \tau}\,, 
 \ee 
 for this we obtain,
 \begin{eqnarray}
&&\frac{\langle\partial\varphi_k(z_1)^2\partial\varphi_k(z_2)^2\,\sigma_{k,n}(y_1)\,
\bar\sigma_{k,n}(y_2)\rangle}{\langle\sigma_{k,n}(y_1)\,\bar\sigma_{k,n}(y_2)\rangle}\,=\,2 \widehat{\cal H}( z_{12} )^2 
 -4 \widehat{\cal H}( z_{12} )\times\nonumber\\ \nonumber
& &
\times\left\{ \left( \frac{\pi}{L} \right)^2   \frac{\vartheta_\nu^{(2)} (x)}{\vartheta_\nu(x) }
+ \frac{\pi}{L^2 {\rm Im}{\tau} }
+  \frac{\pi}{L} \frac{\vartheta_\nu^{(1)} (x)}{\vartheta_\nu(x) }
\left( G(z_1) + G(z_2) \right)  + G(z_1) G(z_2)  \right\}\\\nonumber
& &+\, \left( \frac{\pi}{L} \right)^4 \frac{\vartheta_\nu^{(4)} (x)}{\vartheta_\nu(x) }
+ 6 \left( \frac{\pi}{L} \right)^2  \frac{\vartheta_\nu^{(2)} (x)}{\vartheta_\nu(x) } \frac{\pi}{L^2 {\rm Im}{\tau} }
+ 3 \left(  \frac{\pi}{L^2 {\rm Im}{\tau}}\right)^2 
\\ \nonumber
& & + 2 \left[ \left( \frac{\pi}{L} \right)^3  \frac{\vartheta_\nu^{(3)} (x)}{\vartheta_\nu(x) }
+ 3 \frac{\pi}{L} \frac{\vartheta_\nu^{(1)} (x)}{\vartheta_\nu(x) }\frac{\pi}{L^2 {\rm Im}{\tau}}
\right]
\left[ \frac{k}{n} \left( G(z_2) + G(z_1) \right) \right] \\ \nonumber
& & + \left[  \left( \frac{\pi}{L} \right)^2  \frac{\vartheta_\nu^{(2)} (x)}{\vartheta_\nu(x) }
+ \frac{\pi}{L^2 {\rm Im}{\tau} } \right]
\left[ \left( \frac{k}{n} \right)^2   \left( G(z_2)^2 + G(z_1)^2 + 4 G(z_2) G(z_1) \right) \right] 
\\ \nonumber
& & + 2 \frac{\pi}{L} \frac{\vartheta_\nu^{(1)} (x)}{\vartheta_\nu(x) }
\left( \frac{k}{n} \right)^3 
\left(  G(z_1) G(z_2) ( G(z_1) + G(z_2)  \right) 
+ \left( \frac{k}{n} \right)^4 ( G(z_1) G(z_2) )^2 \\ \nonumber
& & +\,\left( - \frac{\pi^2 E_2}{3L^2} + \frac{\pi}{L^2 {\rm Im }\tau} \right)^2 
\,-\,\left( - \frac{\pi^2 E_2}{3L^2} + \frac{\pi}{L^2 {\rm Im }\tau} \right)\times
\\\nonumber
& &
\times\left\{
 2 \left[ \left( \frac{\pi}{L} \right)^2  \frac{\vartheta_\nu^{(2)} (x)}{\vartheta_\nu(x) }
+ \frac{\pi}{L^2 {\rm Im}{\tau} } \right]  \right. \\ \nonumber
& & \left. 
\qquad\qquad\qquad\qquad + 2 \frac{\pi}{L} \frac{\vartheta_\nu^{(1)} (x)}{\vartheta_\nu(x) }
( G(z_1) + G(z_2) )  + G(z_1)^2 + G(z_2)^2 
\right\} 
%\\\nonumber
\end{eqnarray}
Note that all terms with  the $1/{\rm{Im} \tau}$  cancel and the terms due to the self contractions
contribute in this cancellation.  In the above formula $x = \frac{k}{n} \frac{\pi \Delta}{L}$. 
On simplification we see that the correlator can be written as
\bea
&&\frac{\langle\partial\varphi_k(z_1)^2\partial\varphi_k(z_2)^2\,\sigma_{k,n}(y_1)\,
\bar\sigma_{k,n}(y_2)\rangle}{\langle\sigma_{k,n}(y_1)\,\bar\sigma_{k,n}(y_2)\rangle}\,= \\ \nonumber
& & \left( \frac{\pi}{L} \right)^4
\left[  \frac{\vartheta_\nu^{(4)} (x)}{\vartheta_\nu(x) }
- \left( \frac{\vartheta_\nu^{(2)} (x)}{\vartheta_\nu(x) } \right)^2 \right]
 + 2 \left( \frac{\pi}{L} \right)^3  \frac{\vartheta_\nu^{(3)} (x)}{\vartheta_\nu(x) }
\left[ \frac{k}{n} \left( G(z_2) + G(z_1) \right) \right]  \\ \nonumber
& & +
 \left( \frac{\pi}{L} \right)^2  \frac{\vartheta_\nu^{(2)} (x)}{\vartheta_\nu(x) }
 \left[ \left( \frac{k}{n} \right)^2   \left( G(z_2)^2 + G(z_1)^2 + 4 G(z_2) G(z_1) \right) \right] 
 \\ \nonumber
 & & + 2 \frac{\pi}{L} \frac{\vartheta_\nu^{(1)} (x)}{\vartheta_\nu(x) }
\left( \frac{k}{n} \right)^3 
\left(  G(z_1) G(z_2) ( G(z_1) + G(z_2)  \right) 
+ \left( \frac{k}{n} \right)^4 ( G(z_1) G(z_2) )^2 \\ \nonumber
&&+ \frac{\pi^2 E_2}{3L^2}  \left[
 2 \frac{\pi}{L} \frac{\vartheta_\nu^{(1)} (x)}{\vartheta_\nu(x) }
( G(z_1) + G(z_2) )  + G(z_1)^2 + G(z_2)^2  \right] \\ \nonumber
&& -4  {\cal H}( z_{12} )
\left[ \left( \frac{\pi}{L} \right)^2   \frac{\vartheta_\nu^{(2)} (x)}{\vartheta_\nu(x) }
+  \frac{\pi}{L} \frac{k}{n} \frac{\vartheta_\nu^{(1)} (x)}{\vartheta_\nu(x) }
\left( G(z_1) + G(z_2) \right)  + \left( \frac{k}{n}\right)^2 G(z_1) G(z_2)  \right]  \\ \nonumber
& &+ 2 \left( {\cal H}( z_{12} )    \right)^2 +  \left(  \frac{\pi^2 E_2}{3L^2} + \left(\frac{\pi}{L} \right)^2  
\frac{\vartheta_\nu^{(2)} (x)}{\vartheta_\nu(x) }  \right)^2
\eea
The last term in the above expression can be seen to arise from the disconnected 
diagram $ \langle\partial\varphi_k(z_1)^2 \rangle \langle \partial\varphi_k(z_2)^2\, \rangle $. 
We see that on subtracting this term,  the $k=0$ limit of the above correlator agrees precisely 
with that obtained in the free fermion description of the theory in (\ref{tt}) after 
taking into account the factor of $4$ which appears in the definition of the stress tensor. 
We now quote the result  for the correlator which appears in the 
correction to the R\'{e}nyi entropy on deformation by the stress tensor. 
\bea
&&\frac{\langle \sum_{k, k'} 
\left( \partial\varphi_k(z_1)^2\partial\varphi_{k'}(z_2)^2\, \right) \prod_{k, k'} \sigma_{k,n}(y_1)\,
\bar\sigma_{k',n}(y_2)\rangle}{\langle 
\prod_{k,k'} \sigma_{k,n}(y_1)\,\bar\sigma_{k',n}(y_2)\rangle}\,=\,
\sum_{k} \left\{2 \left( {\cal H}( z_{12} )    \right)^2\,+\, \right.\label{stressrenyi}\\ \nonumber
& & \left.
+\,\left( \frac{\pi}{L} \right)^4
\left[  \frac{\vartheta_\nu^{(4)} (x)}{\vartheta_\nu(x) }
- \left( \frac{\vartheta_\nu^{(2)} (x)}{\vartheta_\nu(x) } \right)^2 \right]
\,+\, \frac{2k}{n} \left( G(z_2) + G(z_1) \right)\times \right. \\ \nonumber
&& \times  \left( \frac{\pi}{L} \right)^3  \left[ \frac{\vartheta_\nu^{(3)} (x)}{\vartheta_\nu(x) }
-  \frac{\vartheta_\nu^{(2)} (x)}{\vartheta_\nu(x) }  \frac{\vartheta_\nu^{(1)} (x)}{\vartheta_\nu(x) }
\right]\,+\,4 \left( \frac{\pi}{L} \right)^2   \left( \frac{k}{n} \right)^2\,G(z_1) G(z_2) 
\left[\ln \vartheta_\nu(x)\right]''\\ \nonumber 
&&  \left. -4  {\cal H}( z_{12} )
\left[ \left( \frac{\pi}{L} \right)^2   \frac{\vartheta_\nu^{(2)} (x)}{\vartheta_\nu(x) }
+  \frac{\pi}{L}\frac{k}{n}  \frac{\vartheta_\nu^{(1)} (x)}{\vartheta_\nu(x) }
\left( G(z_1) + G(z_2) \right)  +  \left(\frac{k}{n}\right)^2G(z_1) G(z_2)  \right]  
\right\} \\ \nonumber
& & 
\,+\,\sum_{k, k'} 
\left[  \frac{\pi^2 E_2}{3L^2} + \left(\frac{\pi}{L} \right)^2  
\frac{\vartheta_\nu^{(2)} (x)}{\vartheta_\nu(x) }   + 2 \frac{k}{n}
\frac{\pi}{L} \frac{\vartheta_\nu^{(1)} (x)}{\vartheta_\nu(x) }    G(z_1) 
+ \left( \frac{k}{n} \right)^2  G(z_1)^2    \right] \\ \nonumber
& &  \qquad\times
\left[  \frac{\pi^2 E_2}{3L^2} + \left(\frac{\pi}{L} \right)^2  
\frac{\vartheta_\nu^{(2)} (x')}{\vartheta_\nu(x') }   + 2 \frac{k}{n}
\frac{\pi}{L} \frac{\vartheta_\nu^{(1)} (x')}{\vartheta_\nu(x') }    G(z_1) 
+ \left( \frac{k}{n} \right)^2  G(z_1)^2    \right]
\eea
The terms in the  last lines which contains a double sum over $k, k'$ is due to the 
disconnected contributions to the stress tensor correlator with $x ' = \frac{k'}{n} \frac{\pi \Delta}{L}$.  In evaluating the R\'{e}nyi 
entropy these terms should simply be removed. 

Finally, we write down the correlator that appears in the 
correction to the R\'{e}nyi entropy for a spin three deformation. 
\bea
&&\frac{\langle \sum_{k, k'} 
\left( \partial\varphi_k(z_1)^3\partial\varphi_{k'}(z_2)^3\, \right) \prod_{k, k'} \sigma_{k,n}(y_1)\,
\bar\sigma_{k',n}(y_2)\rangle}{\langle 
\prod_{k,k'} \sigma_{k,n}(y_1)\,\bar\sigma_{k',n}(y_2)\rangle}\,= 
\label{WWVV}\\ \nonumber
& &\sum_k \left\{ 
- \left( \tfrac{\pi}{L} \right)^6  \left[
 \tfrac{\vartheta_\nu^{(6)} (x)}{\vartheta_\nu(x) } 
 +2  E_2  \tfrac{\vartheta_\nu^{(4)} (x)}{\vartheta_\nu(x) } 
 +E_2^2  \tfrac{\vartheta_\nu^{(2)} (x)}{\vartheta_\nu(x) } \right] \right.\\\nonumber
 &&+ \left( \tfrac{\pi}{L} \right)^6 \left[
\left( \tfrac{\vartheta_\nu^{(3)} (x)}{\vartheta_\nu(x) } \right)^2 
+ 2 E_2 \tfrac{\vartheta_\nu^{(3)} (x)}{\vartheta_\nu(x) } \tfrac{\vartheta_\nu^{(1)} (x)}{\vartheta_\nu(x) }
+ E_2^2 \left(\tfrac{\vartheta_\nu^{(1)} (x)}{\vartheta_\nu(x) } \right)^2  \right] +\,\\\nonumber
& &
+\tfrac{3k}{n}  \left( \tfrac{\pi}{L} \right)^5 ( G(z_1) + G(z_2) )\left[ - \tfrac{\vartheta_\nu^{(5)} (x)}{\vartheta_\nu(x) }  + 
 \tfrac{\vartheta_\nu^{(3)} (x)}{\vartheta_\nu(x) } \tfrac{\vartheta_\nu^{(2)} (x)}{\vartheta_\nu(x) } 
\,-\,E_2 \left( \tfrac{\vartheta_\nu^{(3)} (x)}{\vartheta_\nu(x) } 
- \tfrac{\vartheta_\nu^{(2)} (x)}{\vartheta_\nu(x) } \tfrac{\vartheta_\nu^{(1)} (x)}{\vartheta_\nu(x) }  \right)
\right] \\ \nonumber
& &  + 3 \left( \tfrac{k}{n}\right)^2 \left( \tfrac{\pi}{L} \right)^4 
( G(z_1)^2 + G(z_2)^2  ) 
\left[ - \tfrac{\vartheta_\nu^{(4)} (x)}{\vartheta_\nu(x) }  + 
 \tfrac{\vartheta_\nu^{(3)} (x)}{\vartheta_\nu(x) } \tfrac{\vartheta_\nu^{(1)} (x)}{\vartheta_\nu(x) } 
- E_2  \left( \tfrac{\vartheta_\nu^{(2)} (x)}{\vartheta_\nu(x) } 
-\left( \tfrac{\vartheta_\nu^{(1)} (x)}{\vartheta_\nu(x) } \right)^2  \right)\right] \\ \nonumber
& & + 9 \left( \tfrac{k}{n}\right)^2 \left( \tfrac{\pi}{L} \right)^4  G(z_1) G(z_2) 
\left[ -  \tfrac{\vartheta_\nu^{(4)} (x)}{\vartheta_\nu(x) } 
+  \left( \tfrac{\vartheta_\nu^{(2)} (x)}{\vartheta_\nu(x) } \right)^2 \right] \\ \nonumber
& &+9 \left( \tfrac{k}{n}\right)^3 \left( \tfrac{\pi}{L} \right)^3  G(z_1) G(z_2) 
( G(z_1) + G(z_2) ) \left[
- \tfrac{\vartheta_\nu^{(3)} (x)}{\vartheta_\nu(x) }   +
\tfrac{\vartheta_\nu^{(2)} (x)}{\vartheta_\nu(x) } \tfrac{\vartheta_\nu^{(1)} (x)}{\vartheta_\nu(x) } 
\right] \\ \nonumber
& & +9 \left( \tfrac{k}{n}\right)^4 \left( \tfrac{\pi}{L} \right)^2(  G(z_1) G(z_2) )^2 
\left[ - \tfrac{\vartheta_\nu^{(2)} (x)}{\vartheta_\nu(x) }  
+\left(  \tfrac{\vartheta_\nu^{(1)} (x)}{\vartheta_\nu(x) }   \right)^2 \right] \\ \nonumber
& & + 9 {\cal H}(z_{12})  \left[  \left( \tfrac{\pi}{L} \right)^4   
 \left( \tfrac{\vartheta_\nu^{(4)} (x)}{\vartheta_\nu(x) }  
+ \frac{2}{3} E_2 \tfrac{\vartheta_\nu^{(2)} (x)}{\vartheta_\nu(x) }
+ \frac{1}{9} E_2^2  \right)  \right.  \\ \nonumber
& & \qquad\qquad + 2\, \tfrac{k}{n} \left( \frac{\pi}{L}\right)^3  ( G(z_1) + G(z_2) ) \left( 
\tfrac{\vartheta_\nu^{(3)} (x)}{\vartheta_\nu(x) } + \frac{E_2}{3} 
\tfrac{\vartheta_\nu^{(1)} (x)}{\vartheta_\nu(x) } \right)  \\ \nonumber
& &  \qquad\quad + \left( \tfrac{k}{n} \right)^2 \left( \tfrac{\pi}{L} \right)^2  \left(
\left( G(z_1)^2  + G(z_2)^2  \right) \left(
 \tfrac{\vartheta_\nu^{(2)} (x)}{\vartheta_\nu(x) } + \frac{E_2}{3}  
 \right) \, +\, 4 G(z_1) G(z_2)\, \tfrac{\vartheta_\nu^{(2)} (x)}{\vartheta_\nu(x) } \right) \\ \nonumber
 & & \qquad\quad\left.+\, 2\, \left( \tfrac{k}{n}\right)^3 \tfrac{\pi}{L} \left(  ( G(z_1) + G(z_2) ) G(z_1) G(z_2) 
 \tfrac{\vartheta_\nu^{(1)} (x)}{\vartheta_\nu(x) }   \right) +\left( \tfrac{k}{n}\right)^4 ( G(z_1) G(z_2))^2
 \right] \\ \nonumber
 & & -18 ( {\cal H}(z_{12} )^2  \left[ 
 \left( \tfrac{\pi}{L}\right)^2 \tfrac{\vartheta_\nu^{(2)} (x)}{\vartheta_\nu(x) }
 + \tfrac{k}{n} \tfrac{\pi}{L} ( G(z_1) + G(z_2) ) \tfrac{\vartheta_\nu^{(1)} (x)}{\vartheta_\nu(x) }
 + \left( \tfrac{k}{n} \right)^2 G(z_1) G(z_2) \right] \\ \nonumber
 & &\left.   + \,6 ( {\cal H}(z_{12} )^3 \right\}\,.
\eea
Note that if one just retains the $k=0$ term in the above expression, we obtain the 
two point function of the spin-three currents in the fermionic description given in 
eq.(\ref{ww}) as expected.  This completes the evaluation of all the correlators required to compute the deformation of the  R\'{e}nyi  entropies  by the chemical potentials considered in this paper. 

\section{Integrals on the torus}
\label{integrals}
In the following, we take the two periods of the torus to be
\be
2\omega_1\,=\,L\,,\qquad 2\omega_2\,=\,i\beta\,,
\ee
the spatial extent of the system and the inverse temperature respectively.
The first two non-trivial  integrals relevant for the order $\mu^2$ correction to the R\'enyi entropies are
\bea
&&{\I}_1(y_1,y_2)\,=\,\int d^2 z_2 \int d^2 z_1\,{\H}(z_1-z_2)^2\,G(z_1)\,G(z_2)\,,\\\nonumber
&&{\I}_2(y_1,y_2)\,=\,\int d^2 z_2 \int d^2 z_1\,\H(z_1-z_2)\,G(z_1)^2\,G(z_2)^2\,,\\\nonumber
&&\H(z)\,=\,-{\wp}(z)\,+\,\kappa_1\,;\qquad \kappa_1\,=\,-\frac{\pi^2}{3L^2}\,E_2(\tau)\,,\\\nonumber
&& G(z)\,=\,\tilde\zeta(z-y_2)\,-\,\tilde\zeta(z-y_1)\,.
\eea
For the purposes of integration and keeping track of different contributions we find it useful to split $G(z)$ as
\bea
&&G(z)\,=\,g(z)\,+\kappa_2\,,\\\nonumber
&& g(z)\,=\,\zeta(z-y_2)\,-\,\zeta(z-y_1)\,;\qquad
\kappa_2\,=\,\frac{\pi^2}{3L^2}\,E_2(\tau)\,(y_2-y_1)\,.
\eea
The integrals are evaluated on the torus ${\mathbb C}/\Gamma$ with the lattice $\Gamma\simeq \omega_1{\mathbb Z}\oplus\omega_2{\mathbb Z}$. Although the computation of the integrals is tedious, it is rendered relatively straightforward by the fact that the integrands are elliptic (doubly periodic) functions of the integration variables $z_1$ and $z_2$.

Formal manipulations, namely, the exchange operation $y_1\leftrightarrow y_2$ accompanied by the shifts $z_{1,2}\to z_{1,2}+y_1$ inside the integral sign can be used to show that the integrals must be even functions of the variable $\Delta = (y_2-y_1)$:
\be
{\cal I}_{1,2}\,=\,{\cal I}_{1,2}(\Delta)\,,\qquad \Delta= (y_2-y_1).
\ee

\subsection{Evaluation of ${\cal I}_1(\Delta\,|\tau)$}
The integral ${\cal I}_1$ can be calculated by repeated application of theorem \eqref{zetaexp} on the integrand which is an elliptic function of both $z_1$ and $z_2$. We first  break up the expression into three separate pieces:
\be
{\cal I}_1\,=\,{\cal I}_{1,1}\,+\,{\cal I}_{1,2}\,+\,{\cal I}_{1,3}\,
\ee
where
\bea
&&{\cal I}_{1,1}\,=\,\kappa_2^2\int d^2z_2\int\, d^2 z_1\, \H(z_1-z_2)^2\\\nonumber
&&{\cal I}_{1,2}\,=\,\kappa_2\int d^2z_2\int\, d^2 z_1\, \H(z_1-z_2)^2\,\left\{g(z_1)\,+\,g(z_2)\right\}\\\nonumber
&&{\cal I}_{1,3}\,=\,\int d^2z_2\int d^2 z_1\,\H(z_1-z_2)^2\,g(z_1)\,g(z_2)\,.
\eea
Using \eqref{zetadef}, \eqref{psquared} and our prescription for calculating the integrals of holomorphic functions, we find that 
\bea
&&\boxed{{\cal I}_{1,2}\,=\,-\,2\,{\cal I}_{1,1}}\\
&&\boxed{{\cal I}_{1,1}
\,=\,\frac{\pi^8\,\beta^2}{81\,L^6}\,\Delta^2\,E_2^2(\tau)\,\left(E_4(\tau)\,-\,E_2(\tau)^2\right)}
\eea
We are then left with the computation of ${\cal I}_{1,3}$ which is more involved. This can be split further into  `easy' and `difficult' pieces:
\bea
&&{{\cal I}_{1,3}}\,=\,{\cal I}_{1,3a}\,+\,{\cal I}_{1,3b}\,+\,{\cal I}_{1,3c}\\\nonumber
&&{\cal I}_{1,3a}\,=\,\int_{{{\rm\mathbb T}}^2}\int_{{{\rm\mathbb T}}^2}\,\tfrac{1}{6}\wp''(z_1-z_2)\,g(z_1)\,g(z_2)\\\nonumber
&&{\cal I}_{1,3b}\,=\,-2\kappa_1\int_{{{\rm\mathbb T}}^2}\int_{{{\rm\mathbb T}}^2}\,
\wp(z_1-z_2)\,g(z_1)\,g(z_2)\\\nonumber
&& {\cal I}_{1,3c}\,=\,\left(\kappa_1^2+\tfrac{g_2}{12}\right)\int_{{{\rm\mathbb T}}^2}\int_{{{\rm\mathbb T}}^2}\,\,g(z_1)\,g(z_2)\,.
\eea
Of these, the last one is easiest to perform 
\be
\boxed{{\cal I}_{1,3c}\,=\,\frac{\pi^8\,\beta^2}{81\,L^6}\,\Delta^2\,E_2^2(\tau)\,\left(E_4(\tau)\,+\,E_2^2(\tau)\right)\,}.
\ee
We now turn to ${\cal I}_{1,3a}$. Using the periodicity of the integrand we can integrate by parts without acquiring any boundary terms so that
\be
{\cal I}_{1,3a}\,=\,-\tfrac{1}{6}\int_{{{\rm\mathbb T}}^2}\int_{{{\rm\mathbb T}}^2}\wp(z_1-z_2)\,\left[\wp(z_1-y_2)-\wp(z_1-y_1)\right]
\left[\wp(z_2-y_2)-\wp(z_2-y_1)\right]\,.\nonumber
\ee
We then make repeated use of the product formulae \eqref{wpproduct1}, 
\eqref{wpproduct2}  to write the products of Weierstrass-$\wp$ functions as a sum of $\zeta$ and $\wp$ functions which can be integrated and if necessary simplified using \eqref{pzetaproduct}.
Along the way we make use of the following integrals:
\bea
&&\int_{{\rm\mathbb T}^2}\wp(z-a)\wp(z-b)\,=\,(\beta L)\,\left[
\wp(b-a)\left\{-\wp(a)-\wp(b)-\tfrac{2\pi^2}{3L^2}E_2(\tau)\right\}
\right.\\\nonumber
&&\left.+\wp'(b-a)\left\{\tilde\zeta(b)-\tilde\zeta(a)\right\} +\wp(a)\wp(b)\right]\,,\\\nonumber
&&\int_{{\rm\mathbb T}^2}\wp(z)\left[\zeta(z-a)-\zeta(z-b)\right]\,=\,(\beta L)\,\left[
\tfrac{1}{2}\left\{\wp'(a)-\wp'(b)+
\wp(a)\tilde\zeta(a)-\wp(b)\tilde\zeta(b)\right\}
\right.\\\nonumber
&&\left.-\tfrac{\pi^2}{3L^2}E_2(\tau)\left\{\tilde\zeta(b)-\tilde\zeta(a)\right\}\right]\,.
\eea
With this procedure we obtain a result for ${\cal I}_{1,3a}$ which is a function only of $\Delta=(y_2-y_1)$. After careful evaluation, we find 
\bea
{\cal I}_{1,3a}(\Delta)&&=\,\tfrac{(\beta L)^2}{6}\left(\tfrac{1}{20}\,\wp^{(4)}(\Delta)\,+\,\tfrac{1}{2}\wp^{(3)}(\Delta)\,\tilde\zeta(\Delta)\,+\,\wp''(\Delta)\left(\tilde\zeta^2(\Delta)-\tfrac{2\pi^2}{3L^2}E_2\right)\right.\\\nonumber\\\nonumber
&&\left.-\tfrac{2\pi^2}{L^2}E_2\,\wp'(\Delta)\,\tilde\zeta(\Delta)\,+\,\tfrac{2\pi^4}{15L^4}\wp(\Delta)\left(5E_2^2-E_4\right)\,+\,\tfrac{4\pi^6}{45L^6}(5E_2E_4\,-\,3 E_6)\,\right).
\eea
It is straightforward to show that
\be
\boxed{\tfrac{1}{6}\partial_{y_1}\partial_{y_2}{\cal I}_{1,3b}(\Delta)\,=\,2k_1\left[{\cal I}_{1,3a}(\Delta)+\tfrac{\pi^6\beta^2}{81 L^4}\left\{2E_6(\tau)\,-\,3E_2(\tau)E_4(\tau)\right\}\right]}\,.\label{abrelation}
\ee
Since all our integrals are only functions of $\Delta$, 
\be
-\partial_{y_1}\partial_{y_2}{\cal I}_{1,3b}\,=\,{\cal I}_{1,3b}''(\Delta)\,.
\ee
Therefore, \eqref{abrelation} along with the requirements
\be
{\cal I}_{1,3b}(0)=0\,,\qquad {\cal I}_{1,3b}(\Delta)={\cal I}_{1,3b}(-\Delta)
\ee
completely determines ${\cal I}_{1,3b}$ upon integrating ${\cal I}_{1,3a}$ twice with respect to $\Delta$. 
We list a set of identities which enable us to integrate ${\cal I}_{1,3a}(\Delta)$ with respect to $\Delta$:
\bea
&&\tilde\zeta(\Delta)\wp'''(\Delta)\,=\\\nonumber
&&\qquad\frac{d}{d\Delta}
\left(\tfrac{1}{20}{\wp'''(\Delta)}\,+\,\wp''(\Delta)\tilde\zeta(\Delta)\,+\,\tfrac{\pi^2}{3L^2}E_2(\tau)\,\wp'(\Delta)-\tfrac{2}{5}g_2\,\zeta(\Delta)\,+\,\tfrac{3}{5}g_3\Delta\right)\nonumber\\\nonumber
&&\wp''(\Delta)\tilde\zeta(\Delta)^2\,=\,\frac{d}{d\Delta}\left\{\tfrac{1}{120}\wp'''+\,\tfrac{1}{6}\wp''\tilde\zeta\,+\,\wp'\tilde\zeta^2\,-\,\tfrac{g_2}{15}\zeta +\tfrac{g_3}{10}\Delta+\tfrac{\pi^2}{6L^2}E_2\left(\wp'+\tfrac{g_2}{3}\Delta+\right.\right.\\
&&\left.\left.\qquad\qquad\qquad\qquad\,+\,4\wp\tilde\zeta\,-\, 8\zeta\tfrac{\pi^2}{6L^2}E_2\right)\right\}\\
&&\wp'(\Delta)\tilde\zeta(\Delta)\,=\,\frac{d}{d\Delta}
\left(\tfrac{1}{6}\wp'+\wp\,\tilde\zeta - \tfrac{\pi^2}{3L^2}E_2\,\zeta\,+\,\tfrac{g_2}{12}\,\Delta\right)\\
&&\wp''(\Delta)\tilde\zeta(\Delta)\,=\,\frac{d}{d\Delta}
\left(\wp'\,\tilde\zeta \,+\,\tfrac{1}{2}{\wp}^2\,+\,
\tfrac{\pi^2}{3 L^2}E_2\,\wp\right)\\\nonumber
&&\wp'(\Delta)\tilde\zeta(\Delta)^2\,=\,\frac{d}{d\Delta}\left[\wp\,\tilde\zeta^2\,+\,\tfrac{1}{3}\wp'\tilde\zeta+\tfrac{1}{36}\wp''+\tfrac{g_2}{72}+\tfrac{2\pi^2}{3L^2}E_2\wp\,+\,\tfrac{2\pi^4}{9L^4}\ln\sigma\,\left(E_4-E_2^2\right)\right.\\
&&\left.-\tfrac{\pi^6}{27L^6}\,\Delta^2\,E_2E_4\,-\,\tfrac{\pi^2}{3L^2}E_2\,\zeta^2\,+\,\tfrac{2\pi^4}{9L^4}\,\Delta\,E_2^2\,\zeta\right]
\eea
We then find the indefinite integral
\bea
&&\hat{\cal I}\,=\,\int{\cal I}_{1,3a}(\Delta)d\Delta\,= \label{firstintegral}\\\nonumber
&&\tfrac{(\beta L)^2}{6}
\left[\tfrac{1}{12}\wp'''+\tfrac{2}{3}\wp''\,\tilde\zeta+\wp'\tilde\zeta^2-\tfrac{2\pi^2}{3 L^2}\wp' E_2-\tfrac{4\pi^2}{3L^2}E_2\,\wp\,\tilde\zeta-\tfrac{2\pi^4}{9L^4}\,\zeta\left(E_4+\,E_2^2\right)\right.\\\nonumber
&&\left.
\,+\,\tfrac{8\pi^6}{27 L^6}\,\Delta\,\left(E_2E_4\,-\,\tfrac{1}{2} E_6\right)\right]
\eea
Now we can integrate this a second time using previous identities and ${\cal I}_{1,3b}$ is completely fixed to be
\bea
&&{\cal I}_{1,3b}(\Delta\,,\tau)\,=\,\\\nonumber
&&\boxed{\tfrac{4\pi^2(\beta L)^2}{6 L^2} E_2\left(\tfrac{1}{6}\wp''+\wp'\tilde\zeta+\wp\tilde \zeta^2-\tfrac{\pi^2}{3L^2}\wp\,E_2\,+\tfrac{\pi^2}{3L^2} E_2\zeta^2-\tfrac{2\pi^4}{9L^4}E_2^2\,\Delta\,\zeta-\tfrac{\pi^4}{9L^4}(2 E_4-E_2^2) \right)}
\eea
This is an even function of $\Delta$, vanishing at $\Delta=0$ and satisfying the differential equation \eqref{abrelation}.

\subsection{Evaluation of ${\cal I}_2(\Delta|\tau)$}

We can also see that the integrals in ${\cal I}_2$ can be related to the ones we have already calculated for ${\cal I}_1$. We recall \be
{\cal I}_2\,=\,\int d^2z_2\int d^2z_1\,\H(z_1-z_2)\,(g(z_1)+\kappa_2)^2\,(g(z_2)+\kappa_2)^2\,.
\ee
Next, we make use of the theorem \eqref{zetaexp} to write
\bea
&&(g(z)+\kappa_2)^2\,=\,\wp(z-y_1)+\wp(z-y_2)-2\tilde\zeta(\Delta)\,g(z)\,+\,\kappa_3\nonumber\\
&&\Delta\,=\,y_2-y_1\,,\qquad \kappa_3\,=\,\left(\wp(\Delta)-\zeta(\Delta)^2+\kappa_2^2\right)
\eea
where we have made use of the fact that $\tilde\zeta(\Delta)\,=\,\zeta(\Delta)-\kappa_2$. 
The integral can be organized into six separate terms:
\be
{\cal I}_{2}\,=\,\sum_{m=1}^6{\cal I}_{2,m}\label{sixparts}
\ee
where 
\bea
&&{\cal I}_{2,1}\,=\,\kappa_1\left(\int d^2z\, (g(z)+\kappa_2)^2\right)^2\,,\qquad
{\cal I}_{2,2}\,=\,-\kappa_3^2\int d^2z_1\int d^2z_2\,\wp(z_1-z_2)\,
\\\nonumber
&&{\cal I}_{2,3}\,=\,-2\kappa_3\int d^2 z_1\int d^2 z_2\,\left(\wp(z_2-y_1)+\wp(z_2-y_2)-2\tilde\zeta(\Delta)g(z_2)\right)\,\wp(z_1-z_2)\\\nonumber
&&{\cal I}_{2,4}\,=\,-4\tilde\zeta(\Delta)^2\int d^2z_1\int d^2z_2\,\wp(z_1-z_2)\,g(z_1)\,g(z_2)\,,\\\nonumber
&&{\cal I}_{2,5}\,=\,4\tilde\zeta(\Delta)\int d^2z_1\int d^2z_2\,\wp(z_1-z_2)\,g(z_1)\,\left(\wp(z_2-y_1)+\wp(z_2-y_2)\right)\\\nonumber
&&{\cal I}_{2,6}\,=\,-\int d^2z_1\int d^2z_2\wp(z_1-z_2)\left[\wp(z_1-y_1)+\wp(z_1-y_2)\right]\left[\wp(z_2-y_1)+\wp(z_2-y_2)\right]
\eea
Both ${\cal I}_{2,1}$ and ${\cal I}_{2,2}$ are trivially evaluated. The integral ${\cal I}_{2,3}$ is also rendered simple if one performs the $z_1$-integration first. Furthermore ${\cal I}_{2,4}$ is identical to ${\cal I}_{1,3b}$ up to the constant of proportionality. Finally, the calculation of ${\cal I}_{2,5}$ and ${\cal I}_{2,6}$ is slightly long winded but straightforward and  closely follows the procedure we have used for ${\cal I}_{1,3a}$. 
We obtain,
\bea
&&{\cal I}_{2,1}\,=\,-\tfrac{\pi^2 (L\beta)^2}{3L^2}E_2(\tau)\,\left(\wp(\Delta)-\tilde\zeta(\Delta)^2-\tfrac{2\pi^2}{3 L^2}E_2(\tau)\right)^2,
\label{i21}\\
&&{\cal I}_{2,2}\,=\,(\beta\,L)^2\tfrac{\pi^2}{3 L^2}E_2(\tau)\,\left(\wp(\Delta)-\zeta(\Delta)^2+\tfrac{\pi^4}{9L^4}\Delta^2\,E_2(\tau)^2\right)^2\\\
&&{\cal I}_{2,3}\,=\,\tfrac{4\pi^4(\beta L)^2}{9L^4}\,E_2(\tau)^2\left(\wp(\Delta)-\zeta(\Delta)^2+\tfrac{\pi^4}{9L^4}\Delta^2\,E_2(\tau)^2\right)\left(\Delta\tilde\zeta(\Delta)-1\right)\\
&&{\cal I}_{2,4}\,=\,\frac{2\tilde\zeta(\Delta)^2}{\kappa_1}\,{\cal I}_{1,3b}(\Delta)\\
&&{\cal I}_{2,6}\,=\,-\,6\, {\cal I}_{1,3a}(\Delta)\,-\,(\beta L)^2\tfrac{4\pi^6}{27 L^6}
\,\left(2\,E_6(\tau)\,-\,3\, E_2(\tau) E_4(\tau)\right)\,.
\eea
Finally ${\cal I}_{2,5}$ can be expressed in terms of the function $\hat {\cal I }$ in \eqref{firstintegral} obtained by integrating ${\cal I}_{1,3a}$ once with respect to $\Delta$:
\be
{\cal I}_{2,5}\,=\, 4\tilde\zeta(\Delta)\,\left(6\hat{\cal I}(\Delta)\,+\,(\beta L)^2\Delta\tfrac{4\pi^6}{27L^6}(E_6-\tfrac{3}{2}E_2 E_4)\right)\,.
\ee
\subsection{$\I_3(\Delta|\tau)$}
The third major integral we encounter in the correlation function is 
\be
\I_3\,=\,\int d^2z_1\int d^2 z_2\, {\cal H}(z_{12})\, G(z_1)^2\, G(z_2)\,.
\ee
It is easy to show, using the properties $\int H = \int G =0$, that this integral can be written in terms of those which we have already evaluated above. In particular
\be
\I_3\,=\,-\frac{\I_{2,5}}{4\tilde\zeta(\Delta)}\,-\,\frac{\I_{2,4}}{2\tilde\zeta(\Delta)}\,+\,(\beta L)^2\frac{2\pi^6}{27 L^6}E_2^3(\Delta\tilde\zeta(\Delta)-1)\Delta\,.
\ee
Inserting the complete expressions for these integrals we obtain
\bea
&&\I_3\,=\,(\beta L)^2\,\left[-\tfrac{1}{12}\wp'''(\Delta)-\tfrac{1}{3}\wp''(\Delta)\tilde\zeta(\Delta)\,+\,\wp'(\Delta)\tilde\zeta(\Delta)^2\,+\,2\left[\wp(\Delta)+\tfrac{\pi^2}{3L^2}E_2\right]\tilde\zeta(\Delta)^3\right.\nonumber\\
&&\left.+\tfrac{2\pi^2}{3 L^2}\wp'(\Delta)E_2\,+\,\tfrac{2\pi^2}{3L^2}\wp(\Delta)E_2\tilde\zeta(\Delta)\,+\,\tfrac{2\pi^4}{9L^4}\tilde\zeta(\Delta)(2E_2^2-E_4)\right]\,.\label{I3}
\eea
In the R\'enyi entropy this term is accompanied by a factor of $\vartheta_\nu'/\vartheta_\nu(x)$ which yields  (quasi)modular weight six for this contribution. A non-trivial consistency check satisfied by this expression is that it is periodic under $\Delta \to \Delta+ L$. This periodicity is necessarily spoilt by the Jacobi-theta functions multiplying it in \eqref{w3renyi}.

\subsection{$\I_4(\Delta|\tau)$}
It is easily shown that the integral
\be
\I_4\,=\,\int d^2z_1\int d^2z_2\,\H(z_1-z_2)\,G(z_1)\,G(z_2)\,,
\ee
is determined in terms of integrals that we have already evaluated. In particular
\be
\I_4\,=\,\frac{\I_{1,3b}}{2\kappa_1}\,-\,\kappa_2^2(\beta L)^2\,\frac{\pi^2}{3L^2}E_2\,.
\ee
Explicitly, we find,
\bea
&&\I_4\,=\,-(\beta L)^2 \label{I4}\\\nonumber
&&\left[\tfrac{1}{6}\wp''(\Delta)\,+\,\wp'(\Delta)\,\tilde\zeta(\Delta)\,+\,\wp(\Delta)\tilde\zeta(\Delta)^2\,-\,\tfrac{\pi^2}{3L^2}E_2\,\left(\wp(\Delta)-\tilde\zeta(\Delta)^2\right)\,-\,\tfrac{\pi^4}{9L^4}(2E_4-E_2^2)\right]\,.
\eea
\subsection{$\I_5(\Delta|\tau)$}
This is  among the simplest integrals on the torus
\be
{\I}_5\,=\,\int d^2z\,G(z)^2\,.
\ee
We find
\be
{\I}_5\,=\,(\beta L)\left(\wp(\Delta)-\tilde\zeta(\Delta)^2 -\tfrac{2\pi^2}{3L^2}E_2\right)\,.\label{I5}
\ee


\newpage

\begin{thebibliography}{99} 

%\cite{Maldacena:1997re}
\bibitem{maldacena} 
  J.~M.~Maldacena,
  ``The Large N limit of superconformal field theories and supergravity,''
  Adv.\ Theor.\ Math.\ Phys.\  {\bf 2}, 231 (1998)
  [hep-th/9711200].
  %%CITATION = HEP-TH/9711200;%%
  %9555 citations counted in INSPIRE as of 27 Jan 2014

%\cite{Witten:1998qj}
\bibitem{witten} 
  E.~Witten,
  ``Anti-de Sitter space and holography,''
  Adv.\ Theor.\ Math.\ Phys.\  {\bf 2}, 253 (1998)
  [hep-th/9802150].
  %%CITATION = HEP-TH/9802150;%%
  %6419 citations counted in INSPIRE as of 27 Jan 2014

%\cite{Ryu:2006bv}
\bibitem{Ryu:2006bv} 
  S.~Ryu and T.~Takayanagi,
  ``Holographic derivation of entanglement entropy from AdS/CFT,''
  Phys.\ Rev.\ Lett.\  {\bf 96}, 181602 (2006)
  [hep-th/0603001].
  %%CITATION = HEP-TH/0603001;%%
  %418 citations counted in INSPIRE as of 27 Jan 2014
 
%\cite{Ryu:2006ef}
\bibitem{Ryu:2006ef} 
  S.~Ryu and T.~Takayanagi,
  ``Aspects of Holographic Entanglement Entropy,''
  JHEP {\bf 0608}, 045 (2006)
  [hep-th/0605073].
  %%CITATION = HEP-TH/0605073;%%
  %308 citations counted in INSPIRE as of 27 Jan 2014


%\cite{Casini:2009sr}
\bibitem{Casini:2009sr} 
  H.~Casini and M.~Huerta,
  ``Entanglement entropy in free quantum field theory,''
  J.\ Phys.\ A {\bf 42}, 504007 (2009)
  [arXiv:0905.2562 [hep-th]].
  %%CITATION = ARXIV:0905.2562;%%
  %120 citations counted in INSPIRE as of 10 Dec 2014


%\cite{Holzhey:1994we}
\bibitem{Holzhey:1994we} 
  C.~Holzhey, F.~Larsen and F.~Wilczek,
  ``Geometric and renormalized entropy in conformal field theory,''
  Nucl.\ Phys.\ B {\bf 424}, 443 (1994)
  [hep-th/9403108].
  %%CITATION = HEP-TH/9403108;%%
  %352 citations counted in INSPIRE as of 10 Dec 2014
  
%\cite{Calabrese:2004eu}
\bibitem{Calabrese:2004eu} 
  P.~Calabrese and J.~L.~Cardy,
  ``Entanglement entropy and quantum field theory,''
  J.\ Stat.\ Mech.\  {\bf 0406}, P06002 (2004)
  [hep-th/0405152].
  %%CITATION = HEP-TH/0405152;%%
  %231 citations counted in INSPIRE as of 09 Jan 2014  

%\cite{Calabrese:2005in}
\bibitem{Calabrese:2005in} 
  P.~Calabrese and J.~L.~Cardy,
  ``Evolution of entanglement entropy in one-dimensional systems,''
  J.\ Stat.\ Mech.\  {\bf 0504}, P04010 (2005)
  [cond-mat/0503393].
  %%CITATION = COND-MAT/0503393;%%
  %69 citations counted in INSPIRE as of 27 Jan 2014

 %\cite{Calabrese:2009qy}
\bibitem{Calabrese:2009qy} 
  P.~Calabrese and J.~Cardy,
  ``Entanglement entropy and conformal field theory,''
  J.\ Phys.\ A {\bf 42}, 504005 (2009)
  [arXiv:0905.4013 [cond-mat.stat-mech]].
  %%CITATION = ARXIV:0905.4013;%%
  %113 citations counted in INSPIRE as of 09 Jan 2014  

%\cite{Cardy:2010zs}
\bibitem{Cardy:2010zs} 
  J.~Cardy and P.~Calabrese,
  ``Unusual Corrections to Scaling in Entanglement Entropy,''
  J.\ Stat.\ Mech.\  {\bf 1004}, P04023 (2010)
  [arXiv:1002.4353 [cond-mat.stat-mech]].
  %%CITATION = ARXIV:1002.4353;%%
  %17 citations counted in INSPIRE as of 30 Sep 2014

  %\cite{Douglas:1993wy}
\bibitem{douglas} 
  M.~R.~Douglas,
  ``Conformal field theory techniques in large N Yang-Mills theory,''
  hep-th/9311130.
  %%CITATION = HEP-TH/9311130;%%
  %56 citations counted in INSPIRE as of 02 Jun 2014
  


  %\cite{Dijkgraaf:1996iy}
\bibitem{dijkgraaf} 
  R.~Dijkgraaf,
  ``Chiral deformations of conformal field theories,''
  Nucl.\ Phys.\ B {\bf 493}, 588 (1997)
  [hep-th/9609022].
  %%CITATION = HEP-TH/9609022;%%
  %29 citations counted in INSPIRE as of 02 Jun 2014
  
%\cite{Gutperle:2011kf}
\bibitem{Gutperle:2011kf} 
  M.~Gutperle and P.~Kraus,
  ``Higher Spin Black Holes,''
  JHEP {\bf 1105}, 022 (2011)
  [arXiv:1103.4304 [hep-th]];
  %%CITATION = ARXIV:1103.4304;%%
  %84 citations counted in INSPIRE as of 02 Jan 2014
%\cite{Ammon:2011nk}
%\bibitem{Ammon:2011nk} 
  M.~Ammon, M.~Gutperle, P.~Kraus and E.~Perlmutter,
  ``Spacetime Geometry in Higher Spin Gravity,''
  JHEP {\bf 1110}, 053 (2011)
  [arXiv:1106.4788 [hep-th]].
  %%CITATION = ARXIV:1106.4788;%%
  %70 citations counted in INSPIRE as of 06 Jan 2014  
  
  
  %\cite{Kraus:2011ds}
\bibitem{Kraus:2011ds} 
  P.~Kraus and E.~Perlmutter,
  ``Partition functions of higher spin black holes and their CFT duals,''
  JHEP {\bf 1111}, 061 (2011)
  [arXiv:1108.2567 [hep-th]].
  %%CITATION = ARXIV:1108.2567;%%
  %66 citations counted in INSPIRE as of 06 Jan 2014
 
 %\cite{Gaberdiel:2012yb}
\bibitem{Gaberdiel:2012yb} 
  M.~R.~Gaberdiel, T.~Hartman and K.~Jin,
  ``Higher Spin Black Holes from CFT,''
  JHEP {\bf 1204}, 103 (2012)
  [arXiv:1203.0015 [hep-th]].
  %%CITATION = ARXIV:1203.0015;%%
  %53 citations counted in INSPIRE as of 06 Jan 2014  
  



%\cite{Ammon:2012wc}
\bibitem{Ammon:2012wc} 
  M.~Ammon, M.~Gutperle, P.~Kraus and E.~Perlmutter,
  ``Black holes in three dimensional higher spin gravity: A review,''
  J.\ Phys.\ A {\bf 46}, 214001 (2013)
  [arXiv:1208.5182 [hep-th]].
  %%CITATION = ARXIV:1208.5182;%%
  %72 citations counted in INSPIRE as of 10 Dec 2014

  %\cite{Gaberdiel:2013jca}
\bibitem{Gaberdiel:2013jca} 
  M.~R.~Gaberdiel, K.~Jin and E.~Perlmutter,
  ``Probing higher spin black holes from CFT,''
  JHEP {\bf 1310}, 045 (2013)
  [arXiv:1307.2221 [hep-th]].
  %%CITATION = ARXIV:1307.2221;%%
  %6 citations counted in INSPIRE as of 06 Jan 2014




   %\cite{Prokushkin:1998bq}
\bibitem{Prokushkin:1998bq} 
  S.~F.~Prokushkin and M.~A.~Vasiliev,
  {``Higher spin gauge interactions for massive matter fields in 3-D AdS space-time,''}
  Nucl.\ Phys.\ B {\bf 545}, 385 (1999)
  [hep-th/9806236].
  %%CITATION = HEP-TH/9806236;%%
  %90 citations counted in INSPIRE as of 05 May 2013

 

  
  %\cite{Klebanov:2002ja}
\bibitem{Klebanov:2002ja} 
  I.~R.~Klebanov and A.~M.~Polyakov,
  ``AdS dual of the critical O(N) vector model,''
  Phys.\ Lett.\ B {\bf 550}, 213 (2002)
  [hep-th/0210114].
  %%CITATION = HEP-TH/0210114;%%
  %306 citations counted in INSPIRE as of 28 Jan 2014 
  
  %\cite{Giombi:2009wh}
 \bibitem{Giombi:2009wh} 
  S.~Giombi and X.~Yin,
  {``Higher Spin Gauge Theory and Holography: The Three-Point Functions,''}
  JHEP {\bf 1009}, 115 (2010)
  [arXiv:0912.3462 [hep-th]].
  %%CITATION = ARXIV:0912.3462;%% 
%\cite{Giombi:2010vg}
%\bibitem{Giombi:2010vg} 
  S.~Giombi and X.~Yin,
  {``Higher Spins in AdS and Twistorial Holography,''}
  JHEP {\bf 1104}, 086 (2011)
  [arXiv:1004.3736 [hep-th]].
  %%CITATION = ARXIV:1004.3736;%%

%\cite{Gaberdiel:2010pz}
\bibitem{gg} 
  M.~R.~Gaberdiel and R.~Gopakumar,
  {``An AdS3 Dual for Minimal Model CFTs,''}
  Phys.\ Rev.\ D {\bf 83}, 066007 (2011)
  [arXiv:1011.2986 [hep-th]].
  %%CITATION = ARXIV:1011.2986;%%



%\cite{Gaberdiel:2012uj}
\bibitem{Gaberdiel:2012uj} 
  M.~R.~Gaberdiel and R.~Gopakumar,
  ``Minimal Model Holography,''
  J.\ Phys.\ A {\bf 46}, 214002 (2013)
  [arXiv:1207.6697 [hep-th]].
  %%CITATION = ARXIV:1207.6697;%%
  %113 citations counted in INSPIRE as of 11 Dec 2014
  


  %\cite{Datta:2014ska}
  \bibitem{paper1}
  S.~Datta, J.~R.~David, M.~Ferlaino and S.~P.~Kumar,
  ``Higher spin entanglement entropy from CFT,''
  JHEP {\bf 1406} (2014) 096
  [arXiv:1402.0007 [hep-th]].

  %%CITATION = ARXIV:1402.0007;%%
  %4 citations counted in INSPIRE as of 03 Jun 2014
  
  
  %\cite{Datta:2014uxa}
\bibitem{paper2} 
S.~Datta, J.~R.~David, M.~Ferlaino and S.~P.~Kumar,
  ``Universal correction to higher spin entanglement entropy,''
  Phys.\ Rev.\ D {\bf 90} (2014) 4,  041903
  [arXiv:1405.0015 [hep-th]].
  %%CITATION = ARXIV:1405.0015;%%
  %1 citations counted in INSPIRE as of 03 Jun 2014



%\cite{deBoer:2013vca}
\bibitem{deBoer:2013vca} 
  J.~de Boer and J.~I.~Jottar,
  ``Entanglement Entropy and Higher Spin Holography in AdS$_3$,''
  JHEP {\bf 1404}, 089 (2014)
  [arXiv:1306.4347 [hep-th]].
  %%CITATION = ARXIV:1306.4347;%%
  %31 citations counted in INSPIRE as of 11 Dec 2014

 
 %\cite{Ammon:2013hba}
\bibitem{Ammon:2013hba} 
  M.~Ammon, A.~Castro and N.~Iqbal,
  ``Wilson Lines and Entanglement Entropy in Higher Spin Gravity,''
  JHEP {\bf 1310}, 110 (2013)
  [arXiv:1306.4338 [hep-th]].
  %%CITATION = ARXIV:1306.4338;%%
  %10 citations counted in INSPIRE as of 14 Jan 2014

\bibitem{Datta:2014ypa}
  S.~Datta,
  ``Relative entropy in higher spin holography,''
  arXiv:1406.0520 [hep-th].

\bibitem{Castro:2014mza}
  A.~Castro and E.~Llabr\'es,
  ``Unravelling Holographic Entanglement Entropy in Higher Spin Theories,''
   arXiv:1410.2870 [hep-th]

%\cite{Long:2014oxa}
\bibitem{Long:2014oxa} 
  J.~Long,
  ``Higher Spin Entanglement Entropy,''
  arXiv:1408.1298 [hep-th].
  %%CITATION = ARXIV:1408.1298;%%
  %2 citations counted in INSPIRE as of 10 Dec 2014


%\cite{Chen:2013dxa}
\bibitem{Chen:2013dxa} 
  B.~Chen, J.~Long and J.~j.~Zhang,
  ``Holographic RŽnyi entropy for CFT with W symmetry,''
  JHEP {\bf 1404}, 041 (2014)
  [arXiv:1312.5510 [hep-th]].
  %%CITATION = ARXIV:1312.5510;%%
  %12 citations counted in INSPIRE as of 11 Dec 2014



%\cite{Pope:1989ew}
\bibitem{Pope:1989ew} 
  C.~N.~Pope, L.~J.~Romans and X.~Shen,
  ``The Complete Structure of W(Infinity),''
  Phys.\ Lett.\ B {\bf 236}, 173 (1990).
  %%CITATION = PHLTA,B236,173;%%
  %176 citations counted in INSPIRE as of 10 Jan 2014

%\cite{Bergshoeff:1990yd}
\bibitem{Bergshoeff:1990yd} 
  E.~Bergshoeff, C.~N.~Pope, L.~J.~Romans, E.~Sezgin and X.~Shen,
  ``The Super $W$(infinity) Algebra,''
  Phys.\ Lett.\ B {\bf 245}, 447 (1990).
  %%CITATION = PHLTA,B245,447;%%
  %79 citations counted in INSPIRE as of 10 Jan 2014  
    
  %\cite{Pope:1991ig}
\bibitem{Pope:1991ig} 
  C.~N.~Pope,
  ``Lectures on W algebras and W gravity,''
  hep-th/9112076.
  %%CITATION = HEP-TH/9112076;%%
  %9 citations counted in INSPIRE as of 12 Jan 2014

%\cite{Bakas:1990ry}
\bibitem{Bakas:1990ry} 
  I.~Bakas and E.~Kiritsis,
  ``Bosonic Realization of a Universal $W$ Algebra and $Z$(infinity) Parafermions,''
  Nucl.\ Phys.\ B {\bf 343}, 185 (1990)
  [Erratum-ibid.\ B {\bf 350}, 512 (1991)].
  %%CITATION = NUPHA,B343,185;%%
  %68 citations counted in INSPIRE as of 27 Jan 2014

%\cite{deBoer:2014fra}
\bibitem{deBoer:2014fra} 
  J.~de Boer and J.~I.~Jottar,
  ``Boundary Conditions and Partition Functions in Higher Spin AdS$_3$/CFT$_2$,''
  arXiv:1407.3844 [hep-th].
  %%CITATION = ARXIV:1407.3844;%%
  %6 citations counted in INSPIRE as of 10 Dec 2014
  
 %\cite{Henneaux:2010xg}
\bibitem{Henneaux:2010xg} 
  M.~Henneaux and S.~-J.~Rey,
  {``Nonlinear $W_{infinity}$ as Asymptotic Symmetry of Three-Dimensional Higher Spin Anti-de Sitter Gravity,''}
  JHEP {\bf 1012}, 007 (2010)
  [arXiv:1008.4579 [hep-th]].
  %%CITATION = ARXIV:1008.4579;%%



%\cite{Campoleoni:2010zq}
\bibitem{campoleoni} 
  A.~Campoleoni, S.~Fredenhagen, S.~Pfenninger and S.~Theisen,
  { ``Asymptotic symmetries of three-dimensional gravity coupled to higher-spin fields,''}
  JHEP {\bf 1011}, 007 (2010)
  [arXiv:1008.4744 [hep-th]].
  %%CITATION = ARXIV:1008.4744;%%
%\cite{Campoleoni:2011hg}
%\bibitem{Campoleoni:2011hg} 
  A.~Campoleoni, S.~Fredenhagen and S.~Pfenninger,
  { ``Asymptotic W-symmetries in three-dimensional higher-spin gauge theories,''}
  JHEP {\bf 1109}, 113 (2011)
  [arXiv:1107.0290 [hep-th]].
  %%CITATION = ARXIV:1107.0290;%%

%\cite{Afshar:2012hc}
\bibitem{Afshar:2012hc} 
  H.~Afshar, M.~Gary, D.~Grumiller, R.~Rashkov and M.~Riegler,
  ``Semi-classical unitarity in 3-dimensional higher-spin gravity for non-principal embeddings,''
  Class.\ Quant.\ Grav.\  {\bf 30}, 104004 (2013)
  [arXiv:1211.4454 [hep-th]].
  %%CITATION = ARXIV:1211.4454;%%
  %10 citations counted in INSPIRE as of 12 Dec 2014
  
%\cite{David:2012iu}  
\bibitem{D-F-K} 
J.~R.~David, M.~Ferlaino and S.~P.~Kumar,
``Thermodynamics of higher spin black holes in 3D,''
JHEP {\bf 1211}, 135 (2012)
[arXiv:1210.0284 [hep-th]].
%%CITATION = ARXIV:1210.0284;%%
%24 citations counted in INSPIRE as of 29 Jan 2014


%\cite{Ferlaino:2013vga}
\bibitem{Ferlaino:2013vga} 
  M.~Ferlaino, T.~Hollowood and S.~P.~Kumar,
  ``Asymptotic symmetries and thermodynamics of higher spin black holes in AdS3,''
  Phys.\ Rev.\ D {\bf 88}, 066010 (2013)
  [arXiv:1305.2011 [hep-th]].
  %%CITATION = ARXIV:1305.2011;%%
  %14 citations counted in INSPIRE as of 10 Dec 2014
  
%\cite{Compere:2013nba}
\bibitem{Compere:2013nba} 
  G.~Comp\'ere, J.~I.~Jottar and W.~Song,
  ``Observables and Microscopic Entropy of Higher Spin Black Holes,''
  JHEP {\bf 1311}, 054 (2013)
  [arXiv:1308.2175 [hep-th]];
  %%CITATION = ARXIV:1308.2175;%%
  %8 citations counted in INSPIRE as of 31 Jan 2014
  %\cite{Compere:2013gja}
  %\bibitem{Compere:2013gja} 
  G.~Comp\'ere and W.~Song,
  ``$\mathcal{W}$ symmetry and integrability of higher spin black holes,''
  JHEP {\bf 1309}, 144 (2013)
  [arXiv:1306.0014 [hep-th]].
  %%CITATION = ARXIV:1306.0014;%%
  %8 citations counted in INSPIRE as of 06 Jan 2014

 %\cite{Perez:2012cf}
\bibitem{Perez:2012cf} 
  A.~Perez, D.~Tempo and R.~Troncoso,
  ``Higher spin gravity in 3D: Black holes, global charges and thermodynamics,''
  Phys.\ Lett.\ B {\bf 726}, 444 (2013)
  [arXiv:1207.2844 [hep-th]];
  %%CITATION = ARXIV:1207.2844;%%
  %26 citations counted in INSPIRE as of 06 Jan 2014
  %\cite{Perez:2013xi}
%\bibitem{Perez:2013xi} 
  A.~Perez, D.~Tempo and R.~Troncoso,
  ``Higher spin black hole entropy in three dimensions,''
  JHEP {\bf 1304}, 143 (2013)
  [arXiv:1301.0847 [hep-th]];
  %%CITATION = ARXIV:1301.0847;%%
  %21 citations counted in INSPIRE as of 06 Jan 2014
%\cite{Henneaux:2013dra}
%\bibitem{Henneaux:2013dra} 
  M.~Henneaux, A.~Perez, D.~Tempo and R.~Troncoso,
  ``Chemical potentials in three-dimensional higher spin anti-de Sitter gravity,''
  JHEP {\bf 1312}, 048 (2013)
  [arXiv:1309.4362 [hep-th]].
  %%CITATION = ARXIV:1309.4362;%%
  %2 citations counted in INSPIRE as of 06 Jan 2014


%\cite{deBoer:2013gz}
\bibitem{deBoer:2013gz} 
  J.~de Boer and J.~I.~Jottar,
  ``Thermodynamics of higher spin black holes in $AdS_3$,''
  JHEP {\bf 1401}, 023 (2014)
  [arXiv:1302.0816 [hep-th]].
  %%CITATION = ARXIV:1302.0816;%%
  %41 citations counted in INSPIRE as of 11 Dec 2014




  %\cite{Faulkner:2013yia}
  \bibitem{Faulkner:2013yia}
    T.~Faulkner,
    ``The Entanglement Renyi Entropies of Disjoint Intervals in AdS/CFT,''
    arXiv:1303.7221 [hep-th].
  
 
 
 \bibitem{Barrella:2013wja}
   T.~Barrella, X.~Dong, S.~A.~Hartnoll and V.~L.~Martin,
   ``Holographic entanglement beyond classical gravity,''
   JHEP {\bf 1309} (2013) 109
   [arXiv:1306.4682 [hep-th]].
 
 
 %\cite{Datta:2013hba}
 \bibitem{Datta:2013hba} 
   S.~Datta and J.~R.~David,
   ``R\'{e}nyi entropies of free bosons on the torus and holography,''
   JHEP {\bf 1404}, 081 (2014)
   [arXiv:1311.1218 [hep-th]].
   %%CITATION = ARXIV:1311.1218;%%
   %9 citations counted in INSP
 

%\cite{DiFrancesco:1997nk}
\bibitem{DiFrancesco:1997nk} 
  P.~Di Francesco, P.~Mathieu and D.~Senechal,
  ``Conformal field theory,''
  New York, USA: Springer (1997) 890 p
  %42 citations counted in INSPIRE as of 06 Oct 2014

\bibitem{Iles:2014gra}
  N.~J.~Iles and G.~M.~T.~Watts,
  ``Modular properties of characters of the W3 algebra,''
  arXiv:1411.4039 [hep-th].
  %%CITATION = ARXIV:1411.4039;%%

    \bibitem{itzykson} C. Itzykson and J.-M. Drouffe, ``Statistical Field Theory,''  Cambridge, 1989.


  
  
  %\cite{Gaberdiel:2013jpa}
\bibitem{Gaberdiel:2013jpa} 
  M.~R.~Gaberdiel, K.~Jin and W.~Li,
  ``Perturbations of W(infinity) CFTs,''
  JHEP {\bf 1310}, 162 (2013)
  [arXiv:1307.4087].
  %%CITATION = ARXIV:1307.4087;%%
  %2 citations counted in INSPIRE as of 10 Jan 2014
  
  

  
  \bibitem{ww} 
  E. T. Whittaker and G. N. Watson, 
  `` A Course of Modern Analysis,'' 
  Cambridge University Press, 1927.
  
 

%\cite{Griguolo:2004uz}
\bibitem{Griguolo:2004uz} 
  L.~Griguolo, D.~Seminara and R.~J.~Szabo,
  ``Two-dimensional Yang-Mills theory and moduli spaces of holomorphic differentials,''
  Phys.\ Lett.\ B {\bf 600}, 275 (2004)
  [hep-th/0408055].
  %%CITATION = HEP-TH/0408055;%%
  %2 citations counted in INSPIRE as of 15 Sep 2014
%46

%\cite{Gross:1992tu}
\bibitem{grosstaylor} 
  D.~J.~Gross,
  ``Two-dimensional QCD as a string theory,''
  Nucl.\ Phys.\ B {\bf 400}, 161 (1993)
  [hep-th/9212149];
  %%CITATION = HEP-TH/9212149;%%
  %211 citations counted in INSPIRE as of 16 Sep 2014
%\cite{Gross:1993hu}
%\bibitem{Gross:1993hu}
  D.~J.~Gross and W.~Taylor,
  ``Two-dimensional QCD is a string theory,''
  Nucl.\ Phys.\ B {\bf 400}, 181 (1993)
  [hep-th/9301068];
  %%CITATION = HEP-TH/9301068;%%
  %291 citations counted in INSPIRE as of 16 Sep 2014
%\cite{Gross:1993yt}
%\bibitem{Gross:1993yt} 
  D.~J.~Gross and W.~Taylor,
  ``Twists and Wilson loops in the string theory of two-dimensional QCD,''
  Nucl.\ Phys.\ B {\bf 403}, 395 (1993)
  [hep-th/9303046].
  %%CITATION = HEP-TH/9303046;%%
  %197 citations counted in INSPIRE as of 16 Sep 2014
  %47
  



 
%\cite{Rudd:1994ta}
\bibitem{rudd} 
  R.~E.~Rudd,
  ``The String partition function for QCD on the torus,'' [hep-th/9407176].
  %%CITATION = HEP-TH/9407176;%%
  %29 citations counted in INSPIRE as of 16 Sep 2014

\bibitem{dijkgraaf2}
R. Dijkgraaf,
``Mirror symmetry and elliptic curves'', in {\it The Moduli Space of Curves}, Progress
in Mathematics {\bf 129} (Birkh\"auser, 1995), 149 - 163.

\bibitem{kaneko}
M. Kaneko and D. Zagier, ``A generalized Jacobi theta function and quasimodular forms'',
in {\it The Moduli Space of Curves}, Progress
in Mathematics {\bf 129} (Birkh\"auser, 1995), 165 - 172.





%\cite{Velytsky:2008rs}
\bibitem{Velytsky:2008rs} 
  A.~Velytsky,
  %``Entanglement entropy in d+1 SU(N) gauge theory,''
  Phys.\ Rev.\ D {\bf 77}, 085021 (2008)
  [arXiv:0801.4111 [hep-th]].
  %%CITATION = ARXIV:0801.4111;%%
  %24 citations counted in INSPIRE as of 29 Nov 2014
  

%\cite{Gromov:2014kia}
\bibitem{Gromov:2014kia} 
  A.~Gromov and R.~A.~Santos,
  %``Entanglement Entropy in 2D Non-abelian Pure Gauge Theory,''
  Phys.\ Lett.\ B {\bf 737}, 60 (2014)
  [arXiv:1403.5035 [hep-th]].
  %%CITATION = ARXIV:1403.5035;%%
  %2 citations counted in INSPIRE as of 29 Nov 2014

%\cite{Azeyanagi:2007bj}
\bibitem{Azeyanagi:2007bj} 
  T.~Azeyanagi, T.~Nishioka and T.~Takayanagi,
  ``Near Extremal Black Hole Entropy as Entanglement Entropy via AdS(2)/CFT(1),''
  Phys.\ Rev.\ D {\bf 77}, 064005 (2008)
  [arXiv:0710.2956 [hep-th]].
  %%CITATION = ARXIV:0710.2956;%%
  %60 citations counted in INSPIRE as of 02 Oct 2014

%\cite{Herzog:2013py}
\bibitem{Herzog:2013py} 
  C.~P.~Herzog and T.~Nishioka,
  ``Entanglement Entropy of a Massive Fermion on a Torus,''
  JHEP {\bf 1303}, 077 (2013)
  [arXiv:1301.0336 [hep-th]].
  %%CITATION = ARXIV:1301.0336;%%
  %7 citations counted in INSPIRE as of 04 Sep 2014

%  %\cite{Atick:1987kd}
\bibitem{Atick:1987kd} 
  J.~J.~Atick, L.~J.~Dixon, P.~A.~Griffin and D.~Nemeschansky,
  ``Multiloop Twist Field Correlation Functions for $Z(N$) Orbifolds,''
  Nucl.\ Phys.\ B {\bf 298}, 1 (1988).
  %%CITATION = NUPHA,B298,1;%%
%  %51 citations counted in INSPIRE as of 10 Dec 2014
%
%
%
\bibitem{koblitz} N. Koblitz, `` Introduction to Elliptic Curves and Modular Forms ,'' Springer-Verlag, 1984.  
%
%

\end{thebibliography}
\end{document}